\documentclass{aa}  

\usepackage{graphicx}
\usepackage{natbib}
\bibpunct{(}{)}{;}{a}{}{,} 
\usepackage[varg]{txfonts}
\usepackage{stfloats}
\usepackage{xurl}
\usepackage[colorlinks=true,linkcolor=bluea,allcolors=blue]{hyperref}
\usepackage{multirow}
\usepackage{soul}

\def\o3{[O\textsc{iii}]}
\def\h2{H$_{\rm 2}$}

\def\msol{\,M$_{\rm \odot}$}
\def\mout{\,M$\rm _{\odot}\, yr^{-1}$}
\def\mdot{$\dot{M}_{\rm out}$}
\def\cm3{$\rm cm^{-3}$}

\def\s3{[S\textsc{iii}]}

\newcommand{\kms}{\,\hbox{\hbox{km}\,\hbox{s}$^{-1}$}}
\newcommand{\casa}{\textsc{CASA}}
\newcommand{\barolo}{\textsuperscript{3D}BAROLO}
\newcommand{\pajet}{PA$\rm_{jet}$}

\newcommand{\degree}{\ensuremath{^\circ}}
\newcommand{\ergs}{\,\hbox{\hbox{erg}\,\hbox{s}$^{-1}$}}
\newcommand{\ratio}{\hbox{R$_{\rm 32}$}}

\newcommand{\cra}{{\color{blue} RA22}}
\newcommand{\ane}{{\color{blue} AA23}}

\begin{document} 

   \title{Molecular gas excitation and outflow properties of obscured quasars at $z\sim$0.1}
  \author{A. Audibert \inst{1,2}  \and C. Ramos Almeida \inst{1,2} \and S. Garc{\'i}a-Burillo\inst{3} \and G. Speranza\inst{4} \and I. Lamperti\inst{5,6}  \and M. Pereira-Santaella\inst{4}  \and F. Panessa\inst{7}}

\institute{Instituto de Astrof\' isica de Canarias, Calle V\' ia L\'actea, s/n, E-38205, La Laguna, Tenerife, Spain  \email{anelise.audibert@iac.es} \and
 Departamento de Astrof\' isica, Universidad de La Laguna, E-38206, La Laguna, Tenerife, Spain \and
Observatorio Astron{\'o}mico Nacional (OAN-IGN)-Observatorio de Madrid, Alfonso XII, 3, 28014 Madrid, Spain \and
Instituto de F\'isica Fundamental, CSIC, Calle Serrano 123, 28006 Madrid, Spain \and
Dipartimento di Fisica e Astronomia, Università di Firenze, Via G. Sansone 1, 50019, Sesto F.no (Firenze), Italy \and
INAF - Osservatorio Astrofisico di Arcetri, largo E. Fermi 5, 50127 Firenze, Italy \and
INAF Istituto di Astrofisica e Planetologia Spaziali, Roma, Italy} 

   \date{Received XX XX, 2024}

 
\abstract{

To investigate the impact of winds and low-to-moderate power jets on the cold molecular gas reservoirs of AGN, we present high angular resolution ALMA CO(2-1) and CO(3-2) observations of a sample of six type-2 quasars (QSO2s) at z$\sim$0.1 from the Quasar Feedback (QSOFEED) sample. Spatially resolved molecular line ratio maps, defined as \ratio$\equiv L^\prime_{\rm CO(3-2)}/L^\prime_{\rm CO(2-1)}$, and kinematic modelling were used to constrain changes in gas excitation and to identify gas outflows, respectively. We find that the molecular outflows are co-spatial with regions with \ratio>1, indicating enhanced temperature relative to the discs and the presence of optically thin gas in the outflows. Considering more and less conservative scenarios to measure the outflow properties, we find mass outflow rates of 5$\lesssim$\mdot$\lesssim$150\mout, much lower than those expected from their AGN luminosities of $\sim10^{45.5-46}$\ergs, assuming literature scaling relations. The outflow kinetic energies might be driven by the combined action of jets and winds/radiation pressure, with radiative coupling efficiencies ($\epsilon_{\rm AGN}\equiv \dot{E}_{\rm out}/L_{\rm bol}$) ranging from $10^{-6}<\epsilon_{\rm AGN}<10^{-4}$ and jet coupling efficiencies ($\epsilon_{\rm jet}\equiv \dot{E}_{\rm out}/P_{\rm jet}$) from $10^{-3}<\epsilon_{\rm AGN}<10^{-2}$. 
A linear regression including the six QSO2s follows the locus of $\epsilon_{\rm jet}\sim$0.1\%, although we do not find a strong correlation because of the small number statistics.  
Our results provide further evidence that AGN-driven jets/winds disturb the molecular gas kinematics and excitation within the central kiloparsecs of the galaxies. 
The coupling between compact jets and the interstellar medium might be relevant to AGN feedback, even in the case of radio-quiet galaxies, which are more representative of the AGN population. Finally, we find that the warm (H$_2$) and cold (CO) molecular gas phases seem to be tracing the same outflow, with the main distinction between them being the mass they carry, while the warm ionized outflows ([OIII]) do not seem to be another face of the same outflow, as they show different orientation, velocity, and radius.}

   \keywords{galaxies: active --  galaxies: kinematics and dynamics -- galaxies: jets -- ISM: jets and outflows}

   \maketitle
%

\section{Introduction}


Active Galactic Nuclei (AGN) are nowadays understood as events in the life-cycle of a galaxy \citep{harrison24} due to their short and episodic nature \citep[0.1-100\,Myr,][]{martini04, hickox14, king15, schawinski15}.  The accretion of matter onto the supermassive black holes (SMBHs) in the center of galaxies and subsequent release of radiation and kinetic energy in form of outflows, profoundly influence the surrounding interstellar medium (ISM) and galactic-scale gas reservoirs \citep[e.g.][hereafter \cra]{veilleux20,valentini20,santi21,santi24,mandal21,Ward24,cra22}. AGN feedback plays a pivotal role in shaping the evolution of massive galaxies by regulating the growth of their SMBHs and regulating star formation within their host galaxies \citep[e.g.][]{croton06,dimatteo08,harrison17,nelson19,harrison24}.

It is widely recognized that AGN feedback operates through two primary channels: the "quasar/radiative" mode and the "radio/kinetic" mode \citep{fabian12, dubois16}. The quasar mode 
predominates in luminous AGN with high accretion rates, where radiatively-driven wide-angle winds are launched and commonly manifests in high-redshift, young quasars \citep{king03, zubovas12, bieri17}. The radio mode 
is often associated with collimated relativistic jets, primarily observed in low-luminosity AGN with low accretion rates. The kinetic energy deposit by the jets into the the galaxies delays cooling flows from the intracluster/intragroup medium and prevents the formation of extremely massive galaxies \citep{fabian12, mcnamara12, bourne17,bourne23}.

Despite this dichotomy between feedback modes, low-to-moderate luminosity ($\log P_{\rm1.4GHz}<10^{24}$\,W~Hz$^{-1}$) kiloparsec scale radio jets are starting to be recognised as a potential driver of multi-phase outflows in radiatively efficient ``radio-quiet'' quasars. For example, in 
the case of optically selected type 2 quasars (QSO2s; e.g. \citealt{reyes08}), compact radio jets are associated with morphologically and kinematically distinct features in the ionized and molecular gas, such as outflowing bubbles and increased turbulence, revealing jet-gas interactions on galactic scales \citep[e.g.,][]{mullaney13, vm17,cra22,girdhar22,ane23,venturi23,speranza24,ulivi24}. Simulations 
show that indeed, confined low-power jets ($P_{\rm jet}\sim10^{43-44}$\ergs) are able to induce feedback into the ISM in the form of local turbulence and shocks that alter the physical conditions of the surrounding gas \citep{mukherjee18sim, meenakshi22, talbot22}.

Hosting ubiquitous galactic scale multi-phase outflows, QSO2s offer an excellent opportunity to 
study both quasar and radio mode AGN feedback impact on the host galaxies properties. In particular, the impact of AGN on the molecular gas content of the ISM plays a critical role for galaxy evolution, since in the nearby Universe, the cold molecular gas phase appears to carry the bulk of the outflow mass budget, while the ionized gas phase only carries a small fraction of the total outflowing mass, at least for low-to-intermediate AGN luminosities \citep[$L_{\rm bol}<10^{46}$\ergs][]{fiore17,fluetsch19}.  
However, most of the molecular outflows traced using CO 
were detected in low-luminosity AGN \citep[i.e., Seyfert galaxies; e.g. see][]{combes13,ane19,ane21,santi19,santi21,feruglio20, mavi21} and in ultra-luminous infrared galaxies \citep[ULIRGs; e.g.][]{feruglio10,cicone14,santi15,pereira18,fluetsch19,lamperti22,luke24} and until recently, the high-luminosity regime was scarcely observed in the molecular gas phase. 

In the quasar regime of 
AGN luminosities ($L_{\rm AGN}\gtrsim10^{45-46}$\ergs), where most massive outflows are expected in the molecular phase, recent observational results revealed a 
different scenario. In a sample of seven QSO2s at redshifts $z\sim$0.1 observed with the Atacama Large Millimeter/submillimeter Array (ALMA) in CO(2-1) at 0\farcs2 (400\,pc), \cra~reported
cold molecular outflows in the five QSO2s with CO(2-1) detections. The molecular mass outflow rates (\mdot) are lower than those expected from their AGN luminosities according to empirical scaling relations \citep{cicone14,fiore17,fluetsch19}. The measured 
outflow mass rates and velocities, of \.M$_{\rm out}\sim$8-16\,\mout~and $v_{\rm out}\sim$200-350\kms~(\cra), are intermediate between those of the molecular outflows detected in Seyfert galaxies (\mdot$\sim$0.3-5\,\mout, $v_{\rm out}\sim$90-200\kms) and ULIRGs hosting AGN (\mdot$\sim$60-400\,\mout, $v_{\rm out}\sim$600-1200\kms). See \cra~ for a discussion on the different outflows properties in these systems. 

Additionally, 
molecular outflows with sizes of a few kiloparsecs were detected in four QSO2s with redshifts z<0.2 using CO(3-2) observations at 0\farcs3-0\farcs6 spatial resolution, and in the case of two of the targets molecular gas was detected extending along 5-13\,kpc scale radio lobes of these ``radio-quiet'' sources \citep{girdhar24}.  
In the case of the the ‘Teacup’, a radio-quiet quasar hosting a compact jet almost coplanar with the cold molecular disk, the bulk of the molecular outflow is mostly detected along the direction perpendicular to the compact radio jet \citep[$\sim$1\,kpc in size][hereafter \ane]{ane23}. This is traced by the large CO velocity dispersion and also by a high 
L$^\prime_{\rm CO(3-2)}/L^\prime_{\rm CO(2-1)}$ line ratio (\ratio) perpendicular to the radio jet. A comparison with the tailored simulations of jet-ISM interaction first presented in \citet{mukherjee18sim} and \citet{meenakshi22} supports an scenario where the radio jet compresses and accelerates the molecular gas, driving a lateral outflow that shows enhanced turbulence and gas excitation perpendicular to the radio jet in the Teacup (\ane). Similar enhanced velocity dispersion perpendicular to jets has been observed in CO(3-2) for another QSO2 by \citet{girdhar22} and in a few QSO2s in the ionized gas traced by [OIII] \citep{venturi23,speranza24,ulivi24}.
Altogether, these results suggest that other factors aside from AGN luminosity, such as the jet power, the spatial distribution of the dense gas in the ISM, and the coupling between jets and the dense gas, might also be relevant for driving more or less massive molecular outflows 
(\cra, \ane).

An alternative approach 
for discerning the imprints of AGN feedback on molecular gas involves examining various CO transitions, or CO spectral line energy distributions (SLEDs), which provide insights into the mechanisms driving molecular gas excitation. The CO excitation is suggested to be driven by photodissociation regions (PDRs) and X-ray-dominated regions (XDRs), displaying a wide range of temperature and gas densities \citep{hollenbach99,meijerink07,esposito22,esposito24}. Processes related to AGN activity, such as shocks induced by jets/outflows and by X-ray emission can enhance the gas emission, specially at high J transitions \citep[e.g.,] []{vanderwerf10,mingozi18,vallini19}. Recently, \citet{moly24} analyzed multiple integrated CO transitions in a sample of 17 QSO2s at z<0.2 observed with the  Atacama Compact Array (ACA) or the Atacama Pathfinder EXperiment telescope (APEX). Their CO SLEDs indicate that AGN feedback does not seem to exert a substantial 
influence on the molecular gas content and excitation on global galaxy scales. Instead, they propose that its effects might be 
observed on localized and more nuclear scales that cannot be resolved with the ACA and APEX observations. Indeed, spatially resolved line ratios offer a means to explore these scales and unveil localized alterations in gas excitation attributable to jets and/or 
outflows, as evidenced by elevated values of low-J CO line ratios observed in certain jetted Seyfert/radio galaxies \citep[e.g.,][]{santi14,dasyra16,oosterloo17,oosterloo19, fotopoulou19,ruffa22}. However, investigations into local variations in gas conditions due to compact jets or AGN-driven outflows in``radio-quiet'' quasars utilizing spatially resolved molecular line ratios have thus far been limited to the study of the Teacup galaxy (\ane).

In this work we present an analysis of the molecular gas kinematics and excitation of 
six QSO2s which are part of the Quasar Feedback (\href{http://research.iac.es/galeria/cra/qsofeed/}{QSOFEED}) sample, using ALMA \hbox{CO(3-2)} observations at 0\farcs25$-$0\farcs6 (450$-$1300\,pc) resolution, in addition to the \hbox{CO(2-1)} data at 0\farcs2 resolution first presented in {\color{blue} RA22}. 
The observations used in this work and  the data reduction are described in Section~\ref{sec:data}, and the morphology, kinematics, and CO line ratios 
are presented in Section~\ref{sec:results}. We discuss the dependency of the molecular outflow energetics on the AGN, radio-jet, and host galaxy properties in Section~\ref{sec:discussion}, and the main conclusions are drawn in Section~\ref{sec:sum}. 
We adopt a flat $\Lambda$CDM cosmology with H$_{\rm 0}$=70\kms Mpc$^{-1}$, $\Omega_{\rm M}$=0.3, and $\Omega_{\rm \Lambda}$=0.7.

\section{Sample and observations}\label{sec:data}

This work is part of the QSOFEED project \citep{cra22,cra25,pierce23,bessiere24,speranza24}, whose aim is to
characterise and understand the direct impact of AGN feedback on the central region of nearby quasar host galaxies. The QSOFEED sample was drawn from the compilation of narrow-line AGN by \citet{reyes08} and were selected based on their [OIII] luminosites ($L_{\rm [OIII]}>10^{8.5}L_\odot$) and redshits ($z<$0.14), resulting in a sample of 48 QSO2s with stellar masses ranging from 10$^{10.7}$ to 10$^{11.6}$M$_\odot$. The redshift cut was set to detect 2.12 $\mu$m H$_2$ emission line 
in the near-infrared (NIR; see for instance \citealt{cra17,cra19,speranza22,mavi25}) from the ground. The ionized outflow properties and stellar populations of the full sample are reported in \citet{bessiere24} and the optical morphologies, which are dominated by galaxy interactions, in \citet{pierce23}. 

The subset of six quasars presented here consist of the five QSO2s with CO(2-1) ALMA observations at 0\farcs2 spatial resolution analysed in \cra~plus J1347+1217 (also known as 4C12.50 and F13451+1232), a ULIRG with the largest radio luminosity in the QSOFEED sample. The main properties of the selected six QSO2s are listed in Table~\ref{tab:main}. In this work, we re-analysed the CO(2-1) data of the five QSO2s presented in \cra~together with archival ALMA CO(3-2) observations at 0\farcs25$-$0\farcs6 resolutions (corresponding to physical scales of 450\,pc to 1.1\,kpc at the redshifts of our targets), and in the case of J1347+1217, we use publicly available CO(2-1) and CO(3-2) ALMA observations at similar angular resolution, first presented in  \citet{fotopoulou19} and \citet{lamperti22}. 


\begin{table*}
\caption{Main properties of the quasars.}
\centering
\begin{tabular}{lccccccccccc}
\hline
\hline
SDSS ID  	       &      z  & D$_{\rm L}$ &   Scale   &  log L$_{\rm [OIII]}^{\rm corr}$  & log L$_{\rm bol}$& log L$_{\rm 1.4GHz}$ & log M$_{\rm*}$ &  SFR &  Galaxy \\
                       &    SDSS	      & (Mpc) &(kpc/\arcsec) & (L$_{\rm \sun}$)	  & (erg/s) & (W/Hz) &(M$_{\rm \sun}$) & \mout & Morphology \\
\hline

J101043.36+061201.4   &      0.0977  & 449   &  1.807   &  9.30  &   45.54   & 24.37  &  10.99   &   30    & Interacting ETG  \\
J110012.39+084616.3   &      0.1004  & 462   &  1.851   &  9.60  &   45.84   & 24.18  & 11.02    &   34    & Barred spiral \\
J134733.36+121724.3 &     0.1217$^*$ & 568   &  2.189   &  9.28  &   45.52   & 26.30  & 11.70    & 59$^\dag$    & Merging ETG \\
J135646.10+102609.0   &      0.1232  & 576   &  2.213   &  9.29  &   45.53   & 24.36  & 11.27    &  69    &  Merging ETG  \\
J143029.88+133912.0   &      0.0851  & 388   &  1.597   &  9.58  &   45.82   & 23.67  &  11.15   &  12   & Post-merger ETG  \\
J150904.22+043441.8   &      0.1115  & 517   &  2.028   &  9.79  &   46.03   & 23.81  &  10.94   &   34   & Barred spiral  \\ 
\hline	   					 			    					 			      
\end{tabular}	
\tablefoot{
The values of L$_{\rm bol}$ were derived from the extinction-corrected [OIII] luminosities from \citet{kong18}, listed in column 5, multiplied by the factor of 474 from \citet{lamastra09}.
Rest-frame radio luminosities are calculated from integrated FIRST fluxes \citep{becker95}, assuming a spectral index $\alpha$=-0.7, and the stellar masses and SFRs are from \cra.
$^*$ The redshift measured from the CO(2-1) is computed by \citet{lamperti22}. $^\dag$ SFR from \citet{bessiere24}.}
\label{tab:main}
\end{table*}

The ALMA observations of the \hbox{CO(2-1)} and \hbox{CO(3-2)} emission lines were observed in bands 5, 6, and 7. The \hbox{CO(2-1)} observations and data reduction are described in detail in \cra, and those of J1347+1217 in \citet{lamperti22}. We retrieve the CO(3-2) observations from the ALMA archive for five of the targets, as no CO(3-2) observations are available for J1509+0434. These observations are part of different projects, whose details, including principal investigator (PI) name, project ID, synthesised beam, and root mean square (rms) noise are listed in Table~\ref{tab:obs}.

\begin{table*}
\caption{Properties of the ALMA observations.}
\centering
\begin{tabular}{lcccccccccccc}
\hline
\hline
Short  	 &    CO      & ALMA  &  $t_{\rm int}$ &    beam   & beam & MRS          &    rms  &  $v_{\rm res}$  & PI & project   \\
name         & transition &  band &        (h)     & (\arcsec $\times$ \arcsec) & (pc $\times$ pc) & (\arcsec) &  (mJy) &      (\kms)     & name & ID \\
\hline
J1010   &   (2-1)    & 5   &  0.17  &  0.76$\times$0.67  &  1373 $\times$ 1211 & 7.3 & 0.82  & 11.2  &  C. Ramos Almeida & (1) \\
        &   (3-2)    & 7   &  1.58  &  0.25$\times$0.24  & 452 $\times$ 434 & 8.3 & 0.31  & 15.0  &  A. Thomson       &(2) \\
J1100   &   (2-1)    & 5   &  0.87  &  0.23$\times$0.20  & 426 $\times$ 370  & 7.4 & 0.42  & 11.2  &  C. Ramos Almeida  & (1) \\
        &   (3-2)    & 7   &  1.35  &  0.31$\times$0.24  &  574 $\times$ 444 & 8.1 & 0.34  & 15.0  &  A. Thomson  & (2) \\
J1347   &   (2-1)    & 5   &  1.34  &  0.22$\times$0.17 & 482 $\times$ 372 & 4.2 & 0.74  & 11.4  &  M. Pereira Santaella  & (3) \\
        &   (3-2)    & 7   &  0.08  &  0.61$\times$0.47 & 1335 $\times$ 1029 & 4.3 & 1.39  & 10.0  &  K. Dasyra  & (4) \\
J1356   &   (2-1)    & 5   &  0.99  &  0.24$\times$0.23 & 531 $\times$ 509 & 8.7 &  0.39  & 11.4  &  C. Ramos Almeida  & (1) \\
        &   (3-2)    & 7   &  1.08  &  0.53$\times$0.35 & 1125 $\times$ 743 & 9.4 & 1.30  & 16.0  &  J. Greene  & (5,6) \\
J1430   &   (2-1)    & 6   &  0.56  &  0.24$\times$0.19 & 383 $\times$ 303  & 7.5 & 0.39  & 11.0  &  C. Ramos Almeida  & (1) \\
        &   (3-2)    & 7   &  0.51  &  0.60$\times$0.54  & 958 $\times$ 862  & 5.3 & 1.93  & 10.0  &   G. Lansbury  & (7) \\
J1509   &   (2-1)    & 5   &  0.91  &  0.26$\times$0.23 & 527 $\times$ 466 & 9.0 &  0.42  & 11.3  &  C. Ramos Almeida  & (1) \\ 
\hline	  
\end{tabular}	
\tablefoot{Columns indicate the main observational properties of the sample: name, CO transition, integration time, beam size, maximum recoverable scale (MRS), root mean square (rms), velocity resolution, Principal Investigator (PI) and ALMA Project ID. Project IDs: (1) 2018.1.00870.S, (2): 2018.1.01767.S, (3) 2018.1.00699.S, (4) 2013.1.00180.S, (5) 2011.0.00652.S, (6) 2012.1.00797.S, (7) 2016.1.01535.S}
\label{tab:obs}
\end{table*}

The \hbox{CO(3-2)} data were calibrated using the appropriate \casa~ software version \citep{casa} in the pipeline mode and using the standard flux, bandpass, and phase calibrators.
Once calibrated, the imaging and cleaning were performed with the task \textsc{tclean}. The spectral line map was obtained after subtracting the continuum in the $uv$-plane using the tasks \textsc{uvcontsub} and a 0th order polynomial in the channels free from emission lines. The \hbox{CO(3-2)} data cubes were produced with spectral resolutions listed in Table~\ref{tab:obs} and using Briggs weighting mode and a robust parameter set to 0.5 in order to achieve the best compromise between resolution and sensitivity. In the case of J1347, the CO(3-2) data reduction was performed by \citet{fotopoulou19} using similar \casa~ routines. Finally, the datacubes were corrected for primary beam attenuation. In order to study the line ratios and compare the \hbox{CO(2-1)} to the new \hbox{CO(3-2)} datacube, we regridded the \hbox{CO(2-1)} data to the same pixel scale of the \hbox{CO(3-2)} and then convolved it to the same common beam sizes.

The auxiliary radio data used here are the 6\,GHz Very Large Array (VLA) C-band observations of J1010, J1100, J1356, and J1430 first presented in \citet{jarvis19,jarvis21}. These data were obtained in two different configurations: the high-resolution (0\farcs25, HR) A-array and the low-resolution (1\farcs0, LR) B-array maps. In the case of J1100, we also consider the e-MERLIN observations at 1.5\,GHz obtained at a 0\farcs3 resolution from \citet{jarvis19}. 
In the case of J1347 (4C12.50), we used the 15\,GHz observations from the Very Long Baseline Array (VLBA), retrieved from the NASA/IPAC Extragalactic Database (NED) and presented in \citet{kellermann04}. 
Finally, no VLA observations are yet available for J1509. The main properties of the radio observations used in this work are listed in Table~\ref{tab:radio} of Appendix~\ref{sec:radio}.

\section{Results and analysis}\label{sec:results}

\subsection{Moment maps of the CO(3-2) emission}\label{sec:moms}

\begin{figure*}
    \centering
    \includegraphics[width=17cm]{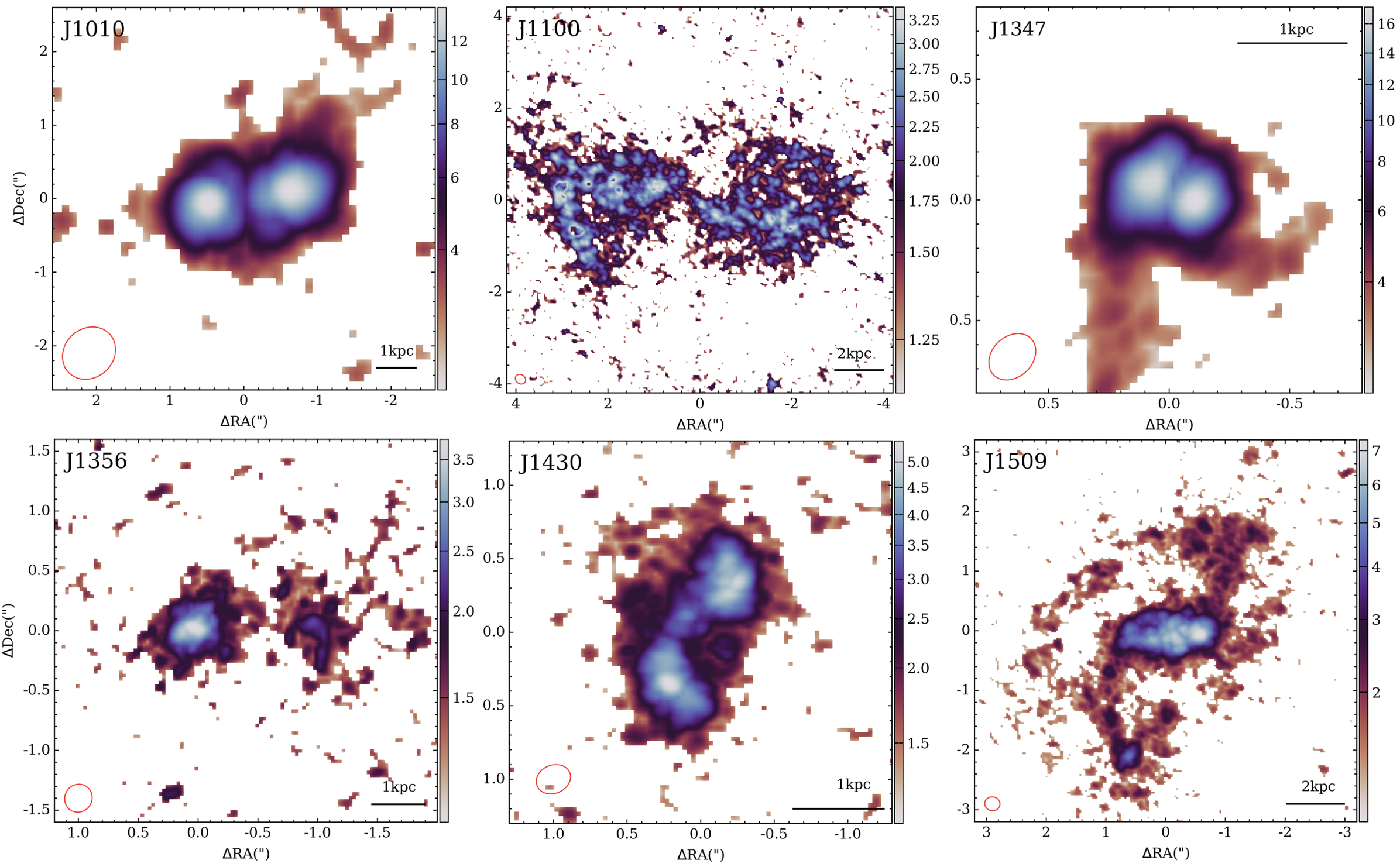}
    \caption{ALMA CO(2-1) peak intensity maps of the six QSO2s. 
    The colour bars indicate the scales of flux density in mJy. The black horizontal lines indicate the physical scales for each galaxy and the beam sizes are shown in the bottom-left corner of each panel as red ellipses. The CO(2-1) distribution present a variety of morphologies, such as spiral arms, double-peaks, and merger signatures. East is to the left and north to the top.}
    \label{fig:tpeaks}
\end{figure*}

The properties of the molecular gas traced by the CO(2-1) emission are presented in \cra~for five of the QSO2s and in \citet{lamperti22} for J1347. In order to show the distribution of the CO(2-1) gas from a slightly different perspective and to allow a comparison with the CO(3-2) emission, we computed the peak intensity maps of CO(2-1), which are shown in Figure~\ref{fig:tpeaks}. 
As discussed in \cra, the CO(2-1) emission of the QSO2s reveals a variety of morphologies: from spiral arms in the case of the J1100 and J1509 to double-peaked structures in J1010, J1347, and J1430, and a central concentration in J1356 together with a west arc that is part of an on-going merger. One of the advantages of the peak maps is that clumpy and narrow structures can be better seen, since they are smoothed in the integrated intensity (moment 0) maps. The CO(2-1) peak maps clearly exhibit the double-peaked features in J1010, J1347, and J1430 (the merging/interacting systems) but also enhance conspicuous regions in the spiral systems, that appear diluted in the integrated intensity maps. The CO(2-1) moment 0 map of J1100, shown in \cra, revealed only some of the clumps mainly along the bar, while the peak map shows clumps spread along the two spiral arms. In the case of J1509, the CO(2-1) moment 0 map shows centrally concentrated gas, and the CO(2-1) peak map reveals several clumpy structures within the central region. 

\begin{figure*}
    \centering
    \includegraphics[width=16cm]{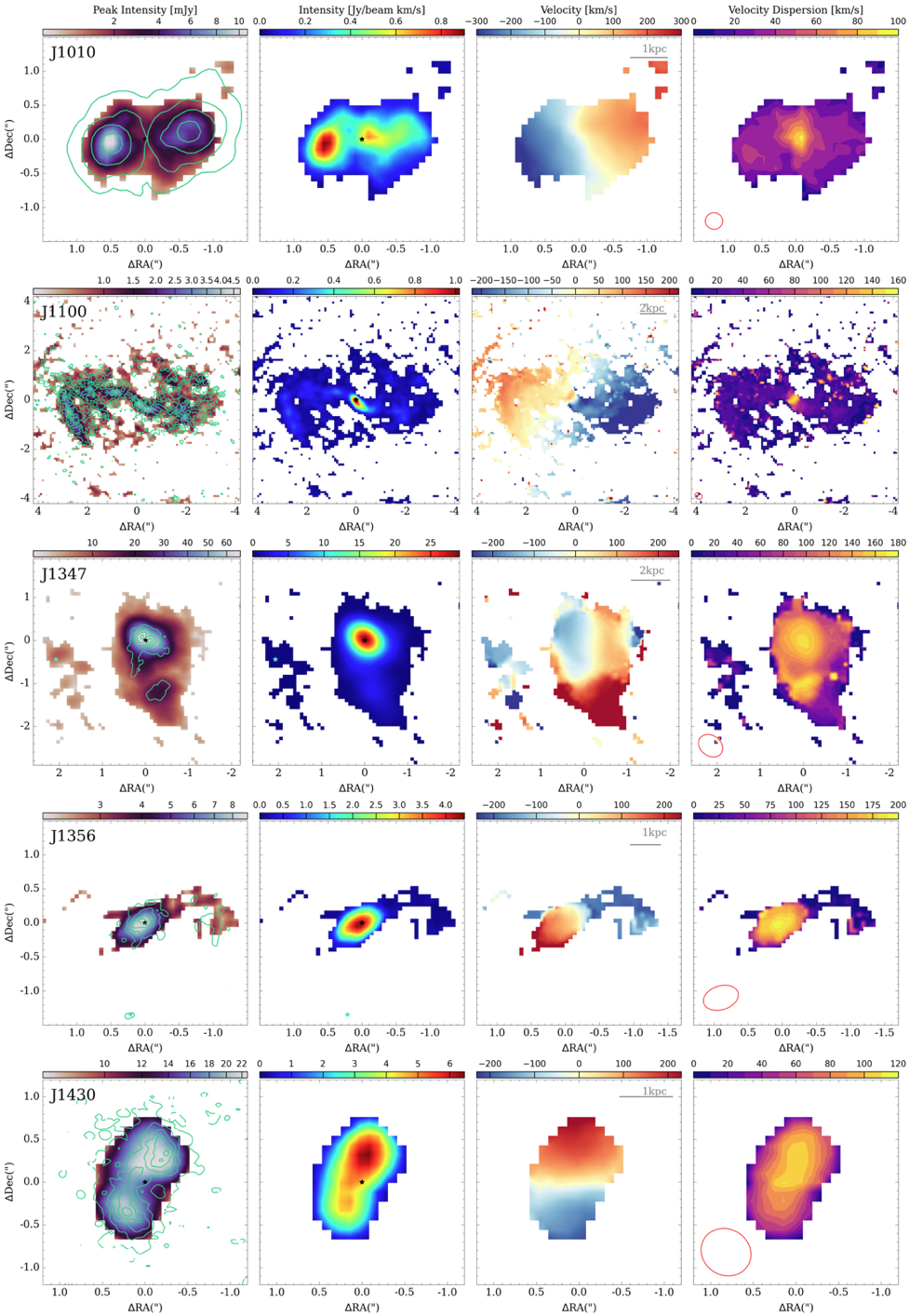}
    \vspace{-0.2cm}\caption{Moment maps of the CO(3-2) emission for the QSO2s. From left to right panels correspond to the peak intensities (in mJy units), integrated intensity (moment 0, in Jy beam$^{\rm {-1}}$\kms units), intensity weighted velocity field (moment 1, in \kms), and velocity dispersion (moment 2, in \kms), respectively. East is to the left and north to the top. The red ellipses in the bottom left corner of the velocity dispersion maps indicate the beam sizes. The black stars show the nuclei, defined as the peak of the 220GHz continuum as in \cra. For the cases of J1347 and J1356, a secondary nucleus (blue star) is identified in the optical and/or with ALMA continuum to the East and South directions, respectively. Green contours on the left panels indicate the $\sim$0\farcs2 (0\farcs8 in the case of J1010) resolution CO(2-1) peak distribution shown in Figure~\ref{fig:tpeaks} for comparison. The gray horizontal lines indicate the physical sizes.}
    \label{fig:co32moms}
\end{figure*}

Using the CO(3-2) data cubes, we computed the peak intensity, integrated intensity (moment 0), intensity-weighted velocity field (moment 1), and velocity dispersion (moment 2) maps, which are shown in Figure~\ref{fig:co32moms}. The moment maps were created using the CASA task \texttt{immoments} and a 3$\sigma_{\rm rms}$ clipping. The CO(3-2) maps of J1430 (Teacup) were presented in Appendix A of \ane, but they are included here for comparison with those of the other QSO2s. Double-peaked morphologies\footnote{We note that the double-peaked morphologies seen in CO do not mean the galaxy necessarily host two nuclei. For the cases in which there are two nuclei identified in the optical images and in ALMA continuum at 200\,GHz, namely J1347 and J1356, we show the position of the second nucleus in Figure~\ref{fig:co32moms} as blue stars.} are also seen in the CO(3-2) peak intensity maps (left panels of Figure~\ref{fig:co32moms}) for the interacting quasars J1010, J1347, and J1430. The CO(3-2) distribution shown in the moment 0 maps (central-left panels) is more compact than the CO(2-1) in the cases of J1010 and J1356, although the western arc part of the merger in the latter is less noticeable in CO(3-2) than in CO(2-1). For J1347, the coarser resolution of the CO(3-2) data makes the southern part of the merger clearly detected. In the case of J1100, the bar and spiral arms are prominent in CO(3-2), similar to the morphology seen in CO(2-1).

The integrated velocity field maps (moment 1) in the middle-right panels of Figure~\ref{fig:co32moms} show a rotation pattern for the main disk of the five quasars displayed, with projected velocities of $v\approx \pm$220\kms, except for J1010 that shows a higher amplitude up to $v\approx \pm$300\kms. We note that the interacting western arc in J1356 has blueshifted velocity components, while the southern region part the merging system in J1347 shows redshifted velocities. The fact that the velocity pattern of the merging regions coincides with the blue(red)shifted sides of the main disk indicates that the gas there is starting to settle into them.

\begin{figure*}
    \centering
    \includegraphics[width=17cm]{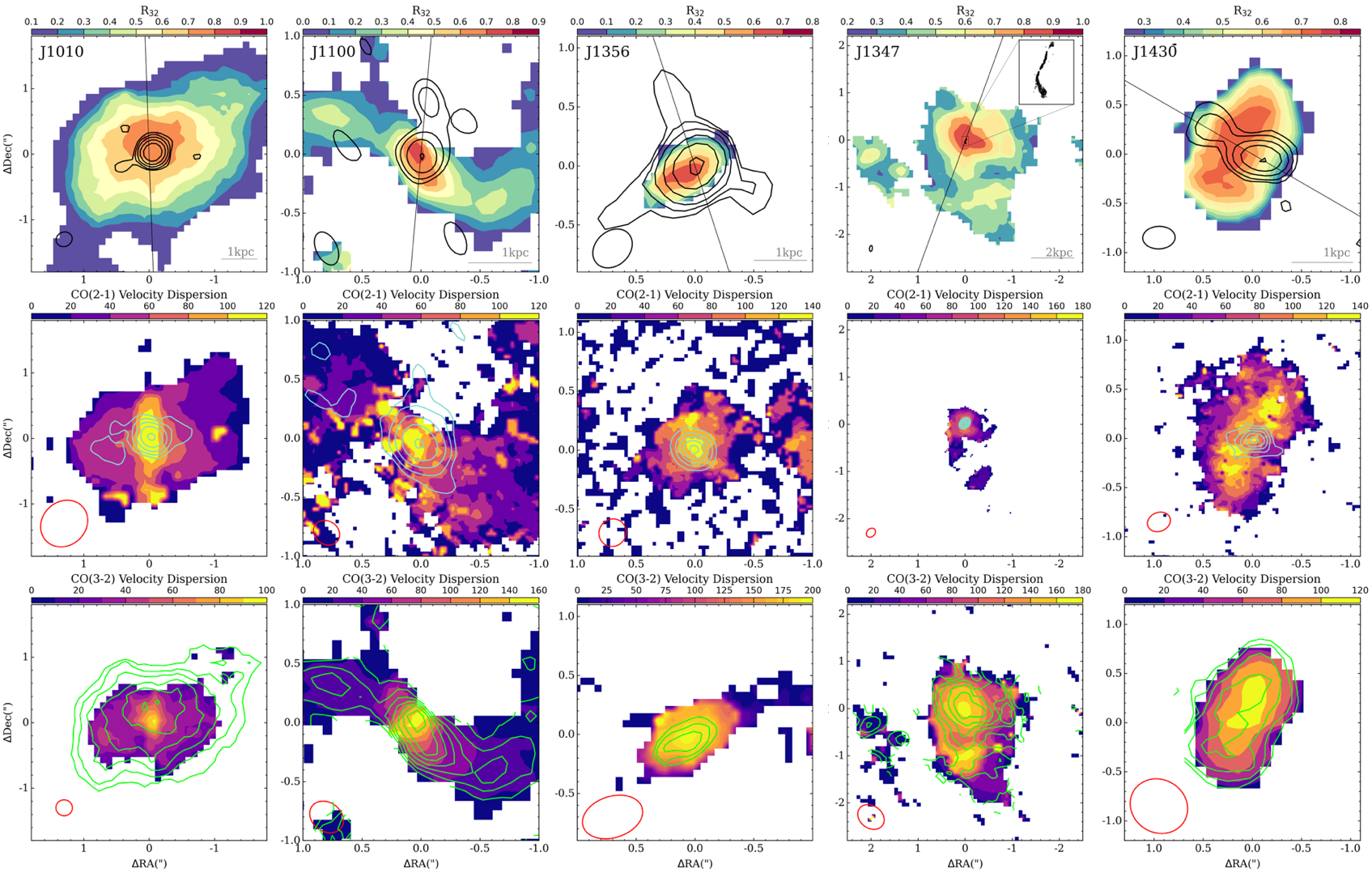}
    \caption{Brightness temperature ratio \ratio~ and CO(2-1) and CO(3-2) velocity dispersion maps. The top panels show the \ratio~ map at the coarser resolution of either CO(2-1) or CO(3-2), with the VLA 6\,GHz HR contours overlaid in black (at 4, 8, 16, 32, and 64$\sigma_{rms}$, with $\sigma_{rms}$ listed in Table~\ref{tab:radio} in the Appendix, also in Table~3 of \citealt{jarvis19}). For J1100 we present the e-MERLIN observations at 1.5\,GHz at a 0\farcs3 resolution presented in \citet{jarvis19} and for J1347 we show in the inset box the 100\,pc scale jet seen with VLBA data at 15\,GHz \citep{kellermann04}. The beam size of the radio observations are indicated as black ellipses on the bottom left corner of the top panels. For J1347, the beam size of the VLBA data is multiplied by a factor of 100 for display purposes. \pajet~from \citet{jarvis19} are indicated as black solid lines (values are listed in Table~\ref{pas}). In J1010 and J1356, \pajet~was determined by these authors from the radio emission detected at larger scales in the low-resolution radio images. The middle and bottom panels correspond to the velocity dispersion maps of CO(2-1) and CO(3-2), respectively. The deepest ALMA continuum images are overlaid as cyan contours in the middle panels: at 300\,GHz for J1010 and J1100 and at 200\,Ghz for J1356, J1347, and J1430. The red ellipses in the bottom left corner of the velocity dispersion maps indicate the corresponding beam sizes. Green contours on the bottom panels are \ratio~from the top panels.}
    \label{fig:ratiosdisp}
\end{figure*}

In the right panels of Figure~\ref{fig:co32moms} we present the CO(3-2) velocity dispersion (moment 2) maps. The central velocity dispersions are typically $v_{\rm disp}\gtrsim$100\,\kms~in the nucleus of the galaxies, with decreasing values outwards. 
J1356 shows high values of $v_{\rm disp}$, of $\sim$200\,\kms, which are larger than the maximum value of $\sim$160\kms~measured from CO(2-1) by \cra. The same is observed for the barred spiral J1100, where the CO(3-2) and CO(2-1) maximum dispersion values are $\sim$160\,\kms~and $\sim$120\,\kms, respectively. The opposite is found for J1010, since the CO(3-2) $v_{\rm disp}$ in the nucleus peaks at 100\,\kms, smaller then the 120\,\kms~measured for the CO(2-1). The velocity dispersion values of CO(3-2) for J1347 are centrally peaked at $\sim$140\kms, same as those reported in CO(2-1) by \citet{lamperti22}. There is a difference in the southern merging region, where CO(2-1) exhibits low values ($<$40\kms), indicative of gas settling in the merger (see the CO(2-1) velocity dispersion map in Figure~\ref{fig:ratiosdisp}), while in CO(3-2) the same region shows values of up to 160\kms. We note that higher velocity dispersions measured in CO(3-2) can be due to beam smearing, as these observations have coarser resolution then the CO(2-1) data. Indeed, when convolving to a common resolution the values measured for CO(3-2) and CO(2-1) are more similar.


\subsection{Integrated fluxes and line ratios}\label{sec:integrated}

\begin{table*}
\caption{Integrated properties of the QSO2s and radii of the corresponding regions.}
\centering
\begin{tabular}{llcccccc}
\hline
\hline
\multirow{ 2}{*}{Name} 	 &    CO     & S$_{\rm CO}$       &  $L^\prime_{\rm CO(J\rightarrow {J-1})}$  &   M$_{\rm H_2}$
&    \multicolumn{2}{c}{Radius}   & \multirow{ 2}{*}{R$_{\rm 32}$}   \\
         & (J$\rightarrow$J-1) &  (Jy\kms) & 10$^9$($\rm K~km~s^{-1}~pc^{2}$)     & 10$^9$(\msol)  
         & (\arcsec) & (kpc) &   \\
\hline \noalign{\smallskip}
\multirow{ 2}{*}{J1010}   &   (2-1)    & 7.92 $\pm$ 0.96   &  0.89 $\pm$ 0.11  &  3.87 $\pm$ 1.59   
& \multirow{ 2}{*}{1.33} & \multirow{ 2}{*}{2.4} & \multirow{ 2}{*}{0.75}    \\
                          &   (3-2)    & 13.35 $\pm$ 1.36  & 0.66 $\pm$ 0.07  &  2.89 $\pm$ 1.16   
                          &  &   \\
\noalign{\smallskip}
\multirow{ 2}{*}{J1100}   &   (2-1)    & 30.6 $\pm$ 3.4    &  3.63 $\pm$ 0.40  &  15.8 $\pm$ 6.3   &  \multirow{ 2}{*}{2.70} & \multirow{ 2}{*}{5.0} & \multirow{ 2}{*}{0.22}   \\
                          &   (3-2)    & 15.34 $\pm$ 1.54  &  0.81 $\pm$ 0.08  &  3.53 $\pm$ 1.41   &  &  \\
\noalign{\smallskip}
\multirow{ 2}{*}{J1347}   &   (2-1)   & 41.5 $\pm$ 0.6   &  7.30 $\pm$ 0.11   &  5.8 $\pm$ 3.7   &  \multirow{ 2}{*}{1.92} & \multirow{ 2}{*}{4.2} & \multirow{ 2}{*}{0.69}  \\
                          &   (3-2)    & 64.1 $\pm$ 8.1   &  5.01 $\pm$ 0.63   &  3.0 $\pm$ 1.3   &  &   \\
\noalign{\smallskip}        
\multirow{ 2}{*}{J1356}   &   (2-1)    & 14.7 $\pm$ 2.3   &  2.65 $\pm$ 0.42  &  11.5 $\pm$ 5.2   &  \multirow{ 2}{*}{2.28} & \multirow{ 2}{*}{5.0}  & \multirow{ 2}{*}{0.30}  \\
                          &   (3-2)    & 9.9 $\pm$ 1.1    &  0.79 $\pm$ 0.09  &  3.46 $\pm$ 1.42   &  &    \\
\noalign{\smallskip}
\multirow{ 2}{*}{J1430}   &   (2-1)    & 16.9 $\pm$ 1.8   &  1.43 $\pm$ 0.15  &  6.24 $\pm$ 2.48   &  \multirow{ 2}{*}{0.81} & \multirow{ 2}{*}{1.3}  & \multirow{ 2}{*}{0.54}   \\
        &   (3-2)    & 20.6 $\pm$ 3.6   &  0.78 $\pm$ 0.14 &  3.38 $\pm$ 1.60   &   &  \\
\noalign{\smallskip}
\hline	   					 			    					 			      
\end{tabular}	
\tablefoot{The CO(2-1) measurements of J1347 are from \citet{lamperti22}, and since this QSO2 is a ULIRG, the mass was calculated using an $\alpha_{\rm CO}$=0.8\,\msol(K\,\kms\,pc$^2$)$^{-1}$. For the other five QSO2s the values are from \cra~and the masses were computed using an $\alpha_{\rm CO}$=4.35\,\msol(K\,\kms\,pc$^2$)$^{-1}$.} 
\label{tab:co}
\end{table*}

We report the integrated fluxes, $\rm S_{CO}\Delta V$, detected in CO(2-1) and CO(3-2) in Table~\ref{tab:co}. To compute the total CO masses, we first convert the integrated flux
measurements to CO luminosities $L^\prime_{\rm CO}$ , in units of $\rm K~km~s^{-1}~pc^{2}$, using the following equation of \citet{solomon05}:
\begin{equation}
{L^\prime_{\rm CO}}(K km s^{-1} pc^2) = 3.25\times10^7 \frac{S_{\rm CO}\Delta V}{(1+z)^3}  \left( \frac{D_{\rm L}}{\nu_{\rm obs}} \right)^2, 
\end{equation}
where $\nu_{\rm obs}$ is the observed frequency of the CO line, in GHz, $D_{\rm L}$ is the luminosity distance of the galaxy, in Mpc, $z$ is the redshift and $\rm S_{\rm CO}\Delta V$ is the integrated flux in Jy\kms. The CO transitions can be converted to $L^\prime_{\rm CO(1-0)}$ adopting a line ratio, called R$_{\rm J-1}=\frac{L^\prime_{\rm CO(J\rightarrow {J-1} )}}{L^\prime_{\rm CO(1\rightarrow0)}}$. Under the assumption that the gas is thermalised and optically thick, $L^\prime_{\rm CO(3-2)} = L^\prime_{\rm CO(2-1)} = L^\prime_{\rm CO(1-0)}$, and therefore the brightness temperature ratio R$_{\rm J-1}$=1.

The molecular masses can be calculated from ${L^\prime_{\rm CO(1-0)}}$ using the CO-to-H$_2$ conversion factor $M_{H_2}=\alpha_{\rm CO} {L^\prime_{\rm CO(1-0)}}$.
We adopted the  standard Galactic CO-to-H$_2$, conversion factor of $\alpha_{\rm CO}=4.36M_\odot (K kms^{-1}pc^{2})^{-1}$  \citep{tacconi13,bolatto13}. 
The molecular masses inferred from the CO(3-2) emission are of the order of (3-4)$\times 10^9$\msol~(see Table \ref{tab:co}), which are systematically lower than the ones derived using the CO(2-1) observations in \cra~and \citet{lamperti22}. Both mass estimates were computed under the same assumptions of $\alpha_{\rm CO}$ and thermalised gas $R_{\rm 31}$=$R_{\rm 21}$=1, therefore indicating that the molecular gas in the QSO2s is less excited and has temperature and density conditions below the thermalised limit. Indeed, we derived the integrated $R_{\rm 32}$ line rations, also listed in Table~\ref{tab:co}, and we find values of $R_{\rm 32}<$1. The values derived for the integrated line ratios range 0.22$<$\ratio$<$0.75, with an average value of $<$\ratio$>$=0.5$\pm$0.2, similar to the line ratios typically measured in spiral and disk galaxies \citep{leroy22}. However, in the next section we will present spatially resolved line ratios to examine possible localised changes in the gas conditions of the QSO2s. A discussion of the physical condition of the gas is addressed in Section~\ref{sec:gasphysics}.

\subsection{Spatially resolved line ratios}\label{sec:spaline}

The spatial resolution of our ALMA CO(2-1) and CO(3-2) observations allows us to produce spatially resolved line ratio maps, which can provide insights into the changes in gas excitation due to the AGN ionization, jets, and star formation. Line ratio maps allow us to unveil signatures that differentiate outflows from ambient ISM. In Figure~\ref{fig:ratiosdisp}  we present the \ratio~maps (top panels) together with the velocity dispersion (bottom panels) of the five quasars, similar to the analysis reported for the Teacup in \ane. The radio continuum emission from the VLA HR data at 6\,GHz is indicated as black contours (except for J1100 and J1347, for which we report e-MERLIN data at 1.5\,GHz and VLBA data at 15\,GHz, respectively). The PA orientation of the radio emission 
is indicated with black solid lines. The ALMA continuum at either 200 or 300\,GHz is shown in blue contours overlaid on the velocity dispersion maps.

The \ratio~maps shown in the top panels of Figure~\ref{fig:ratiosdisp} show higher values in the nuclear region, of \ratio$\sim$0.7-0.9, compared to the smaller values found across the CO disks, of \ratio$\sim$0.2-0.5. The regions of enhanced \ratio~are usually co-spatial with those showing high central velocity dispersion in the bottom panels. 

No clear trend regarding the orientation of high \ratio~and velocity dispersion with the PA of the radio jets is found for the QSO2s. Apparent enhanced turbulence and gas excitation perpendicular to the radio jet is only found for the Teacup, as reported in \ane. Tentative perpendicular orientation of the \ratio~and velocity dispersion is found for J1100, where the \pajet=175\degree~derived from the e-MERLIN contours is misaligned by $\sim$135\degree~with the axis of high \ratio~and dispersion (PA$\sim$40\degree). 
The region showing high values of \ratio~in J1347 appears perpendicular to the orientation of the 100\,pc jet, whilst the region of high velocity dispersion is aligned with the jet. 
For J1010 and J1356, the VLA HR data is unresolved, and therefore the orientation of possible jets (as suggested by the spectral index analysis in \citealt{jarvis19}) is merely indicative and derived from the larger scale radio structures seen in the low-resolution radio images.

Therefore, although some targets exhibit a degree of (anti-)alignment between \ratio, velocity dispersion, and radio emission, the results are not consistent across the sample. One of the goals of the QSOFEED project is to understand how different outflow, jet, and ISM properties — and their interactions, including orientation and energy exchange among others — shape the observed gas kinematics and excitation. Given the diverse and complex nature of the individual targets, a larger sample is required to explore the broad parameter space and derive more general conclusions.

\subsection{Modelling the kinematics of the molecular gas}\label{sec:model}

We analysed the CO(2-1) and CO(3-2) kinematics using the “3D-Based Analysis of Rotating Objects from Line Observations'' (\barolo) software by \citet{barolo15}. \barolo~ performs a 3D tilted-ring modelling of the emission line datacubes to derive the parameters that better describe the kinematics of the data. We ran \barolo~ on both CO(3-2) and CO(2-1) datacube in order to investigate non circular motions, since the code allows us to infer radial velocities in the fit of the rotation curves. 

\begin{figure}[h]
\resizebox{\hsize}{!}{\includegraphics{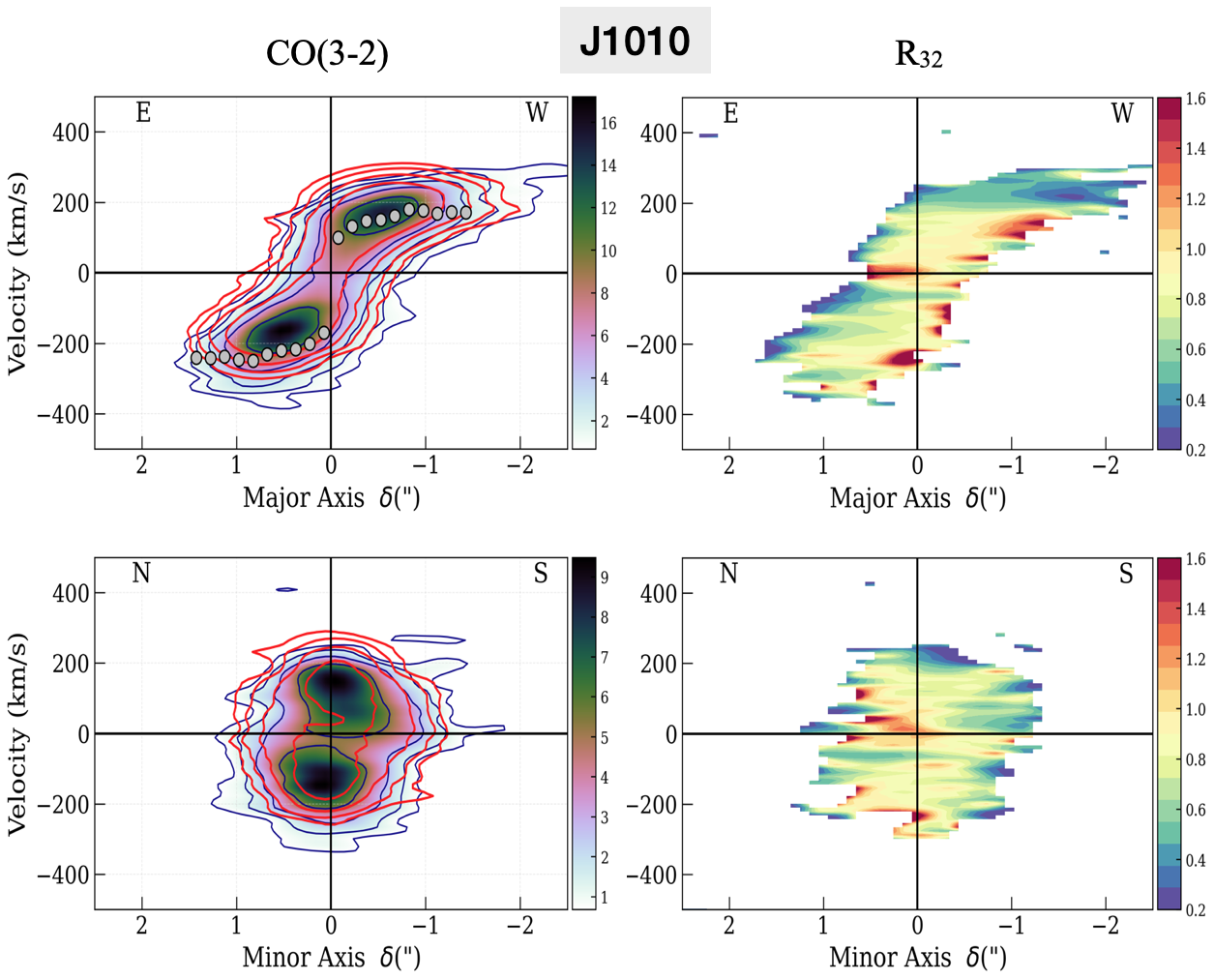}}
\caption{PVDs along the major (top panels) and minor (bottom panels) axis of the of the CO(3-2) emission (left) and the \ratio~ line ratios (right) for J1010. PVDs were extracted in a slit of 0\farcs6 width  along the major (PA=288\degree) and minor (PA=198\degree) axis of the CO disk and are displayed above 2$\sigma_{\rm rms}$. The contours are 2, 4, 8, and 16$\sigma_{\rm rms}$, with $\sigma_{\rm rms}$=0.3\,mJy~beam$^{-1}$ for CO(3-2) in the left panels. The red contours shown in the left and middle panels correspond to the corresponding \barolo~ model at 2, 4, and 6$\sigma_{\rm rms}$, and the grey dots are the derived rotation velocities. Right panels are the PVDs of the \hbox{CO(3-2)/CO(2-1)} line ratios. The colour bars correspond to flux density in mJy for the \hbox{CO(3-2)} PVDs and brightness temperature ratio \ratio, which is dimensionless.}
\label{fig:pv1010}
\end{figure}

\begin{figure}[b]
\resizebox{\hsize}{!}{\includegraphics{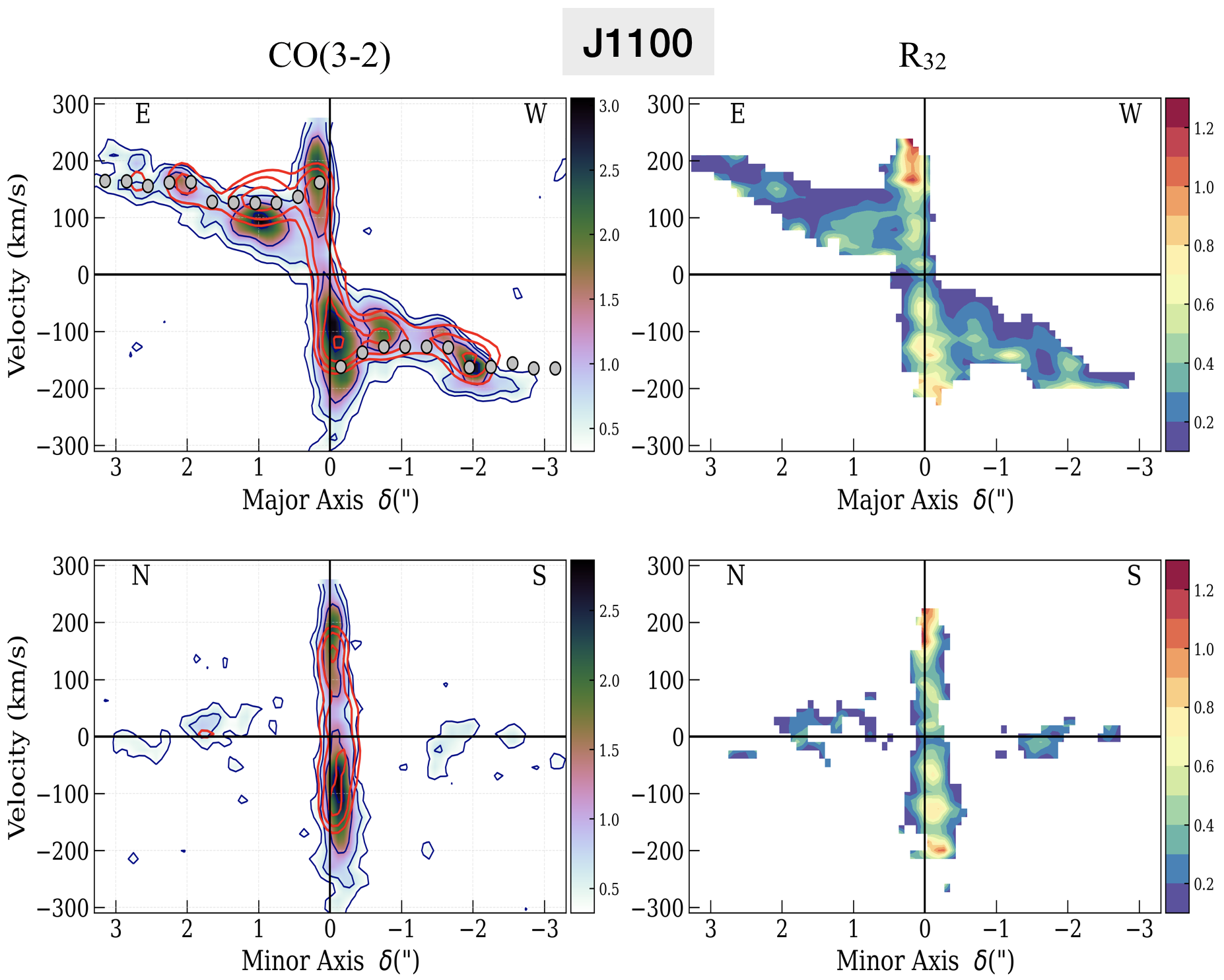}}
\caption{Same as Fig~\ref{fig:pv1010} for J1100. The PVDs were extracted in a slit of 0\farcs3 width along the major (PA=69\degree) and minor (PA=159\degree) axis of the CO disk. The contours are 2, 4, 8, and 16$\sigma_{\rm rms}$, with $\sigma_{\rm rms}$=0.2\,mJy~beam$^{-1}$ for CO(3-2) in the left panels.}
\label{fig:pv1100}
\end{figure}

\begin{figure}[h]
\resizebox{\hsize}{!}{\includegraphics{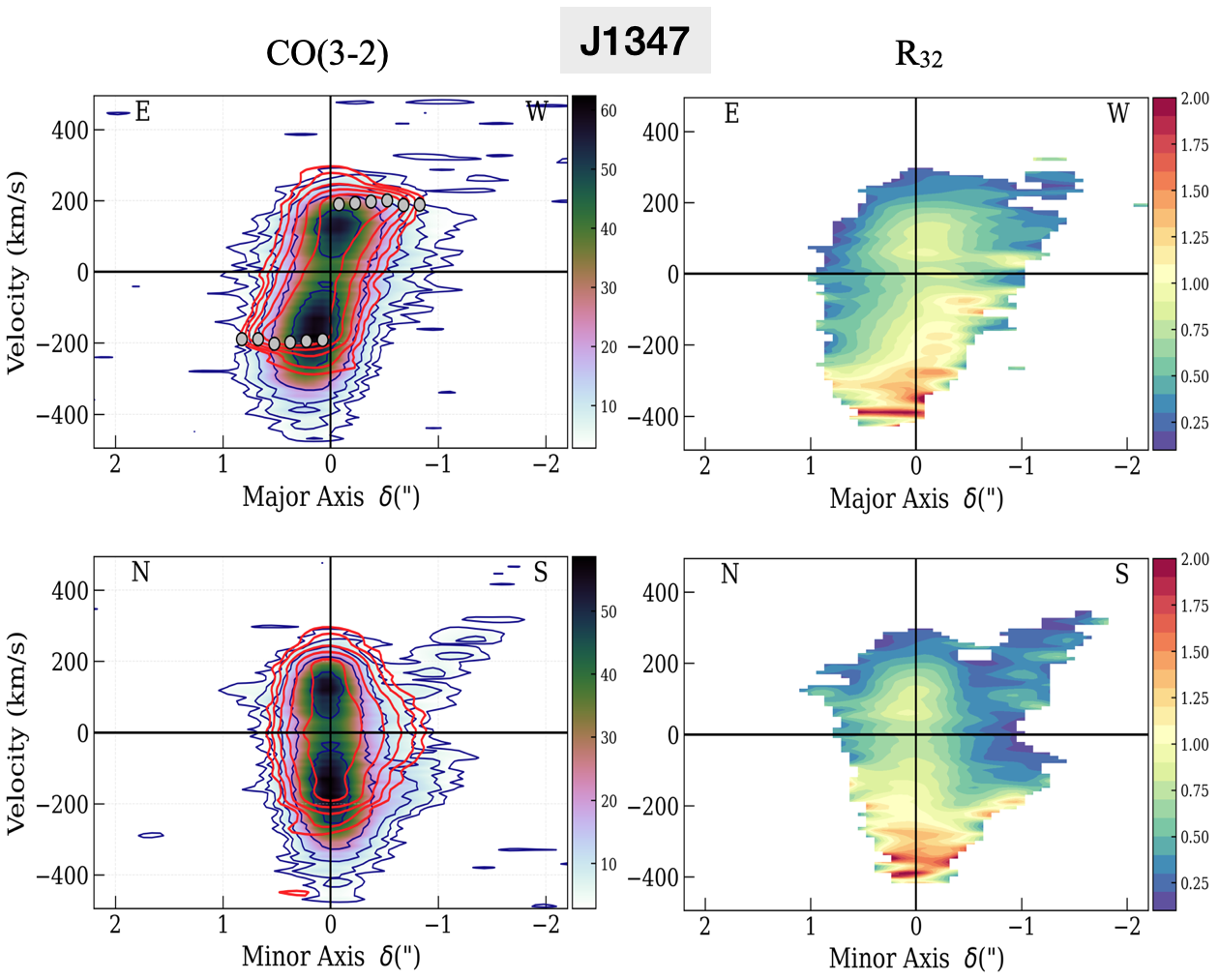}}
\caption{Same as Fig~\ref{fig:pv1010} for J1347. The PVDs were extracted in a slit of 0\farcs3 width along the major (PA=248\degree) and minor (PA=-22\degree) axis of the CO disk. The contours are 2, 4, 8, and 16$\sigma_{\rm rms}$, with $\sigma_{\rm rms}$=1.4\,mJy~beam$^{-1}$ for CO(3-2) in the left panels.}
\label{fig:pv1347}
\end{figure}

\begin{figure}[h]
\resizebox{\hsize}{!}{\includegraphics{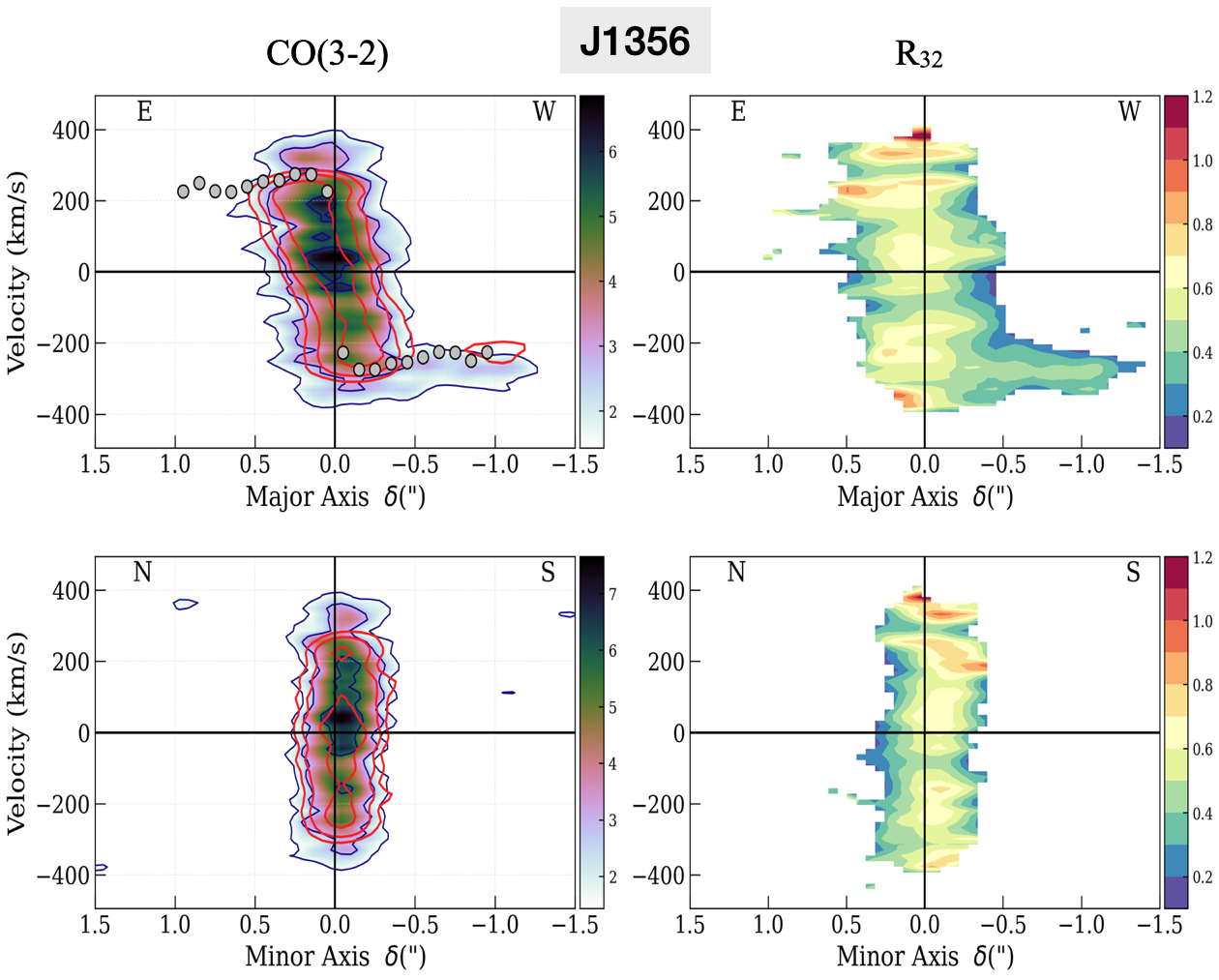}}
\caption{Same as Fig~\ref{fig:pv1010} for J1356. The PVDs were extracted in a slit of 0\farcs3 width along the major (PA=110\degree) and minor (PA=200\degree) axis of the CO disk. The contours are 2, 4, 8, and 16$\sigma_{\rm rms}$, with $\sigma_{\rm rms}$=0.7\,mJy~beam$^{-1}$ for CO(3-2) in the left panels.}
\label{fig:pv1356}
\end{figure}

The center was fixed to the position of the continuum peak at 200\,GHz and we allowed \barolo~to fit 
the inclination, PA, systemic velocity, rotation velocity, and dispersion as free parameters. For the initial guess on the inclination and PA, we used the parameters derived in \cra~and for the case of J1347 in \citet{luke24}. We then fixed the inclination, PA, and systemic velocity to the average values derived from the first run\footnote{Overall the average values derived for the CO(3-2) inclination and PA are similar or the same as those derived using the CO(2-1) line, see Table~\ref{tab:barolo} in Appendix~\ref{sec:baloro}.} and allow the code to fit only rotation and dispersion. We present the results of the analysis with \barolo~in form of position-velocity diagrams (PVDs) along the major and minor axis of the CO(3-2) disk in Figures~\ref{fig:pv1010}-\ref{fig:pv1356}. Since the CO(2-1) \barolo~fits are already shown in detail in \cra, here we just show the results for CO(3-2), although in the right panels of Figures~\ref{fig:pv1010}-\ref{fig:pv1356} we also present the PVDs of the \ratio~to identify regions having different gas excitation conditions. The \ratio~PVDs were smoothed using a Gaussian kernel with a $\sigma$=1.5 to reveal intrinsic variations and reduce the noise due to a pixel-by-pixel ratio.

In Figure~\ref{fig:pv1010} the CO(3-2) PVDs for the quasar J1010 show a clear rotation pattern along the major axis (PA=288\degree, or PA=-72\degree). The \barolo~ model is displayed as red contours and it reproduces well rotation up to velocities of $\sim$300\kms, although it cannot fit the high-velocity gas within 0 and 1\farcs5 to the east, which shows velocities of up to -400\kms~(see top left panel in Figure~\ref{fig:pv1010}). This non-circular velocity feature is more clearly visualised in the high velocity map shown in Figure~\ref{fig:highvel} and discussed in Section~\ref{energetics}. The line ratio PVDs shown on the right panels of Figure~\ref{fig:pv1010} have typical values of \ratio$\sim$0.8, in agreement with the global ratio derived from the integrated fluxes listed in Table~\ref{tab:co}. However, there are several regions that show values \ratio$>1$, indicating the presence of optically thin gas and/or gas with higher excitation temperatures. However, these regions with \ratio$>1$ are distributed all over the minor and major axis slits, for both low and high velocity components.

The CO(3-2) PVDs for the spiral J1100 are presented in the left panels of Figure~\ref{fig:pv1100} and show clear rotation along the major axis (PA=69\degree, same as the PA derived for the CO(2-1) line) that is well reproduced by the \barolo~fit shown as red contours, and the typical 0th velocities along the minor axis. In the central region, concentrated in a radius of less than 0\farcs4 (740\,pc), the PVDs along the major and minor axes show a large velocity gradient with velocities reaching $\pm$300\kms. The \barolo~ model reproduces well rotation up to velocities of $\sim$180\kms, but it cannot fit the high velocities in the nuclear region. These high velocity regions coincide with the highest line ratios seen in the \ratio~PVDs on the right panels of Figure~\ref{fig:pv1100}, with \ratio~values between 0.6-1.2, higher than the typical values of \ratio$<$0.5 found along the spiral arms at larger radii. Compared to the integrated value of $<$\ratio$>$=0.22 in Table~\ref{tab:co}, this indicates that the global ratio traces mostly the contribution from the galaxy disk.

For the merging galaxy J1347, we show the PVDs in Figure~\ref{fig:pv1347}. The CO(3-2) along the major axis (top left panel) shows a rotation pattern oriented along the west-east direction (PA=248\degree) that can be fit with the \barolo~model up to velocities of $\sim$300\kms. There is a clear excess of blueshifted velocities up to 420\kms~in the central to east direction that cannot be fit by rotation. Along the minor axis (PA=-22\degree, bottom left panel), the high blueshifted velocities are also present, and we can see the contribution from the southern part of the merger mostly in redshifted velocities. The line ratio PVDs (right panels of Figure~\ref{fig:pv1347}) of the main disk (r$<$1\arcsec) have values of \ratio$\sim$0.8, slightly larger than the global ratio derived from the integrated fluxes of $<$\ratio$>$=0.69, as listed in Table~\ref{tab:co}, and values \ratio$<$0.5 across the southern part of the merger. However, the regions having the high blueshifted velocities both along the major and minor axes show the highest line ratios, of \ratio$\sim$1-2, indicating the presence of optically thin/warmer fast gas.

In Figure~\ref{fig:pv1356} the CO(3-2) PVDs of J1356 show a less prominent contribution from rotation along the major axis (PA=110\degree) for the main galaxy disk (r$<$0\farcs5 $\simeq$1.1\,kpc) and the system present an overall large velocity dispersion (see Figure \ref{fig:ratiosdisp}). The blueshifted velocities seen between 0 and 1\arcsec~to the west correspond to the western arc of the merging system. The \barolo~model, displayed as red contours, can reproduce rotation up to velocities of $\sim$300\kms, but it fails to reproduce the high velocity gas at velocities up to {\bf $\pm$}400\kms, mostly detected at redshifted velocities. The line ratio PVDs on the right panels of Figure~\ref{fig:pv1356} have typical values of \ratio$\sim$0.6 along the main disk (or north nucleus in \cra) and \ratio$<$0.5 along the western arc, higher than the value derived from the integrated intensities of $<$\ratio$>$0.3. One possible reason is that the sensitivity of the CO(2-1) is much better than the CO(3-2), allowing for a clearer detection of the western arm and the southern nucleus of the system (see Appendix D.1 in \cra), while in the shallower CO(3-2) observations only part of the western arm is detected and the southern nucleus is undetected, making the integrated \ratio~lower. On the other hand, we find \ratio~values of up to 1.2 in regions that are co-spatial with the redshifted high velocity gas, indicating different excitation conditions there.

The PVDs of CO(2-1), CO(3-2), and \ratio~of the Teacup (J1430) can be seen in Figure 2 of \ane~and are described in full there. We find values of \ratio$\sim$0.4-0.8 across the disc, and the highest values, of up to \ratio$\sim$1, in the region occupied by high-velocity blueshifted gas ($\sim$200-350\kms) within 0 and 1\arcsec~to the south.

The regions of the QSO2s where we detected high velocity gas components and gas excitation differing from ambient gas conditions in the main disks are interpreted as outflowing gas.  This interpretation is based on the fact that their kinematics cannot be explained by regular rotation, based on our analysis using \barolo, and they show higher excitation temperatures than those  typically found in the galaxy disks of non-active star-forming galaxies.
We note that this is the same outflow definition employed in \citet{cra22}, but adding gas excitation. We note that in the case of the merging galaxies, J1347 and J1356, the observed non-rotational motions could be attributable to merger-induced flows. However, numerical simulations suggest that molecular gas velocity dispersions in mergers typically reach only tens of \kms~\citep{bournaud08, bournaud11}, which is significantly lower than the values that we measure in the QSO2s. Observationally, ALMA CO(3-2) observations of the overlapping region between the merging Antennae galaxies (NGC 4038/39), a prototypical merger, revealed knots of low-velocity dispersion ($\sim$10\kms) and maximum values of up to $\sim$80\kms~in supergiant molecular clouds \citep{whitmore14}. Considering this, we favor the outflow interpretation to explain the observed kinematics and gas excitation.

\subsection{Molecular outflow properties}\label{sec:outprop}

\begin{figure*}
    \centering
    \includegraphics[width=17cm]{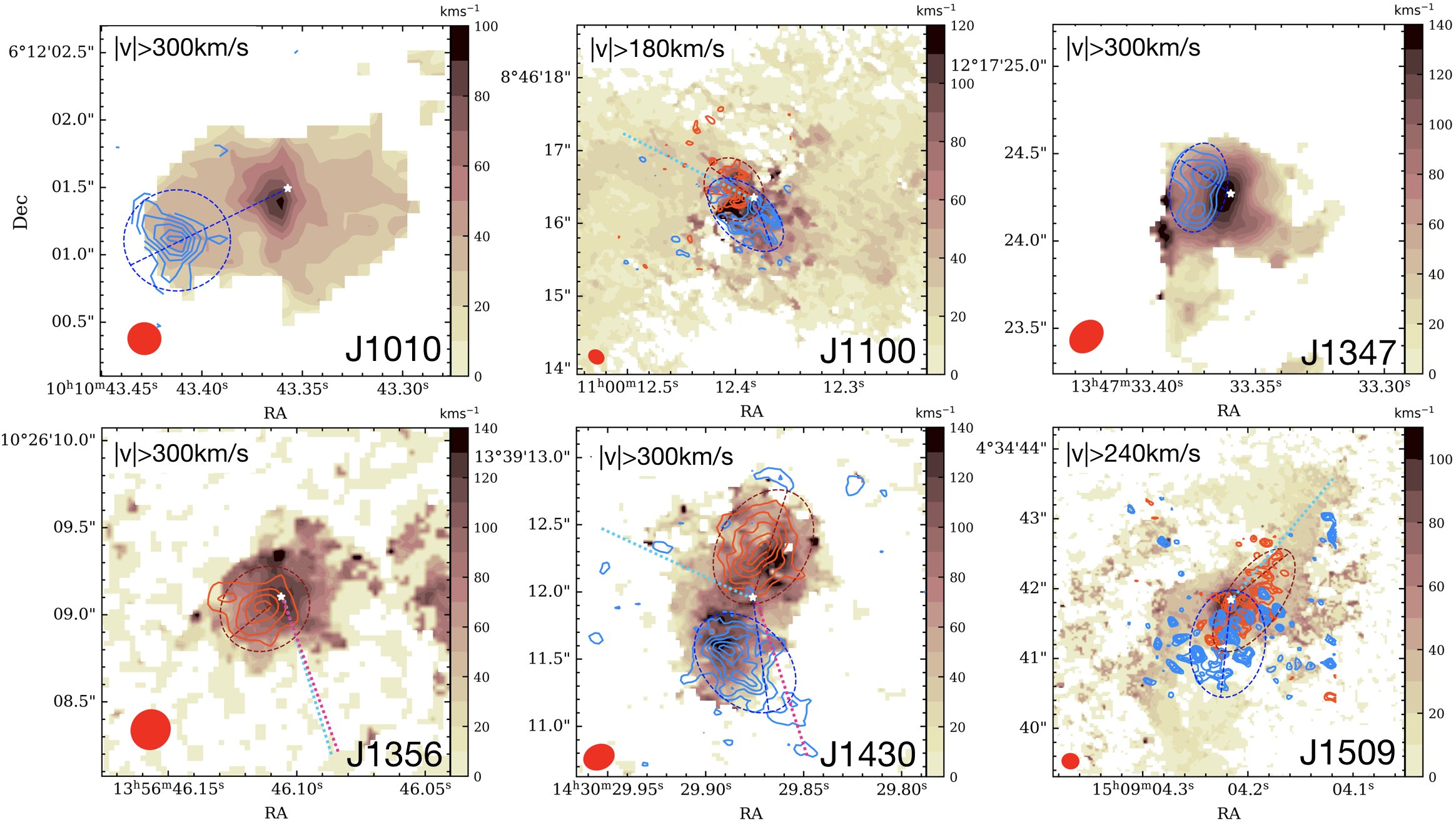}
    \caption{High-velocity flux maps overlaid on the CO velocity dispersion maps for the 6 QSO2s. 
    The colour maps corresponds to the moment 2 of the total CO(2-1) emission at 0\farcs2 resolution for all the QSO2s except for J1010, for which CO(3-2) at the same angular resolution is shown instead. The colour bars are in \kms~units. The contributions from the high velocity gas above (below) the 
    velocity value indicated on the top left corner of each panel are shown as red (blue) contours at (0.35,0.5,0.65,0.8,0.95)$\times\sigma_{\rm max}$. 
    The beam size is indicated with a red ellipse in the bottom left corner of each panel. Ellipses were fitted to the high-velocity gas to determine the orientation of the blueshifted and redshifted outflows, and they are shown with corresponding colors. The dashed lines connecting the galaxy nucleus, defined by the 200\,GHz continuum peak (white star), and the most distant point within the ellipse were used to estimate the PA of the outflows. For comparison, the PA of the redshifted and blueshifted sides of the ionized outflows detected in [OIII] by \citet{speranza24} are indicated with magenta and cyan dotted lines.}
    \label{fig:highvel}
\end{figure*}

The determination of the outflow mass rates in the literature is performed using heterogeneous methods. From the observational point of view, it is challenging to obtain a precise determination of the outflow mass, radius, and velocities, all affected by uncertainties. Different methodologies and assumptions can lead to estimates that diverge in more than an order of magnitude \citep{kiara23,harrison24}. For these reasons, here we consider three scenarios for estimating the outflow masses and mass rates, from least to most conservative, following \ane. To measure the fluxes and corresponding masses, we used the CO(2-1) data at the original 0\farcs2 resolution to avoid artefacts or wrong interpretations of the results due to possible beam smearing effects because of the larger CO(3-2) beam size. The only exception is for J1010, in which CO(3-2) was used because it has better spatial resolution (0\farcs25) than the CO(2-1) data (0\farcs75).

\textsc{i}) In the first scenario, we assume that all the molecular gas that cannot be reproduced with our rotating disc model is outflowing. Thus, we just integrated the emission from the \barolo~model and subtracted it from the CO(2-1) datacube. We find that for the spirals J1100 and J1509 and the interacting early-type galaxy J1010, most of the CO(2-1) emission can be reproduced with rotation ($\sim$75\%), while the mergers J1347 and J1356, and the post-merger system J1430 present more complex kinematics, with J1356 being dispersion-dominated (see Figure~\ref{fig:vsig} in Appendix~\ref{sec:vsig}) and therefore rotation accounts for $\sim$50\% of the total CO emission. In this scenario, we adopted an outflow velocity, $v_{\rm out}$, 
calculated as the average of the mean redshifted and the mean blueshifted velocity residuals 
from the \barolo~fit shown in \cra~and in Appendix~\ref{sec:baloro}, using absolute values. These velocities are in the range 50$<v_{\rm out}<$100\kms~(see Table \ref{tab:prop}).

\begin{table*}
\caption{Outflow measurements derived from the three scenarios proposed for the QSO2s.}             
\label{tab:prop}      
\centering   
\begin{tabular}{c c c c c c c c c c c c}         
\hline\hline  

Name & Scenario & $S_{\rm CO}$ & $v_{\rm out}$ & \multicolumn{2}{c}{$r_{\rm out}$} &  $M_{\rm out}$ & $\dot{M}_{\rm out}$  & $M_{\rm out}/M_{\rm H_2}$ & $\eta$ & $P_{\rm jet}$ & $\dot{E}_{\rm out}/P_{\rm jet}$\\  
 &         & (Jy\kms) & (\kms) & (\arcsec) & (kpc) & (10$^7$ M$_\odot$) & (\mout)  & (\%) &  & (10$^{43}$\ergs) &   \\
\hline  

      & \textsc{i}       & 3.10$\pm$0.3  & 50  &    &     & 27.9$\pm$20   &  7.9$\pm$5.7    &  7.2  &   &  & 0.0001 \\
J1010 & \textsc{ii}      & 0.34$\pm$0.03 & 300 &  1.0 & 1.8 & 3.1$\pm$2.1   &  5.2$\pm$3.6    &  0.8  & 0.2  & 8.9 & 0.0017  \\
      & \textsc{iii}     & 0.30$\pm$0.05 & 300 &    &     & 2.7$\pm$2.1   &  4.6$\pm$3.6    &  0.7  &   &  & 0.0015 \\

\hline      

      & \textsc{i}       & 13$\pm$0.9   & 70  &    &     & 123$\pm$85.7 & 36.7$\pm$25.5 & 7.8 & & & 0.001\\
J1100 & \textsc{ii}      & 2.7$\pm$0.1  & 180 &1.3 & 2.4 & 26$\pm$17.5  & 14.4$\pm$9.7 & 1.6 & 0.4 &7.2&0.001\\
      & \textsc{iii}     & 1.0$\pm$0.1  & 180 &    &     & 9.5$\pm$6.9  & 5.2$\pm$3.8 & 0.6 &  & &0.0004  \\

\hline

      & \textsc{i}$^a$   &   & 514 & 0.05 & 0.12 & 3.4 &  148 & 0.6 &  & & 0.023 \\
J1347 & \textsc{ii}      & 0.9$\pm$0.2 & 300 & 0.29 & 0.65 & 13.2$\pm$10.5 & 62.9$\pm$50 & 2.3 & 1.1 & 53 & 0.003 \\
      & \textsc{iii}$^b$ &   & 390 & 0.23 & 0.52 & 4.2 & 31.6 & 0.72 &  & & 0.003  \\

\hline  

      & \textsc{i}$^c$   & 0.99$\pm$0.1  & 400 & 0.14 & 0.3 & 7.0  &  92.8  & 1.1 & &  & 0.053 \\
J1356 & \textsc{ii}      & 0.89$\pm$0.07  & 300 & 0.58 & 1.3 & 12.9$\pm$9   &  26.2$\pm$18   & 2.0 & 0.4 & 8.9 & 0.008 \\
      & \textsc{iii}$^d$ & 0.1$\pm$0.02  & 310 &  0.2 & 0.4   & 1.4$\pm$1.2  & 10$\pm$8.8 & 0.22 &  & &0.003  \\

\hline

      & \textsc{i}       & 7.6$\pm$0.9  & 100 & & & 51.5$\pm$48.2   &  44.0 $\pm$ 32.7   & 8.3 & & &0.004\\
J1430 & \textsc{ii}      & 2.4$\pm$0.1  & 300 & 0.75 & 1.2 & 16.0$\pm$10.9   &  41.0$\pm$27.9   & 2.6 & 3.4 &
 3.3 &0.035 \\
      & \textsc{iii}     & 0.9$\pm$0.1  & 300 & & & 5.6$\pm$4.3  &  15.1$\pm$11.0  &  0.9 & & & 0.013\\

\hline

      & \textsc{i}       & 10.1$\pm$1.1  & 50 & & & 119$\pm$87   &  26.6 $\pm$ 19.6   & 6.7 &  & &  0.0003\\
J1509 & \textsc{ii}      & 1.52$\pm$0.15  & 240 & 1.1 & 2.1 & 17.9$\pm$12.9 & 19.2$\pm$13.9 & 1.0 & 0.6  &6.1 & 0.0058 \\
      & \textsc{iii}     & 1.29$\pm$0.14  & 240 & & & 15.2$\pm$11.1  &  16.3$\pm$12.0  & 0.9 &  & & 0.0049\\
\hline
\end{tabular}
\\
\tablefoot{The reported outflow velocities, $v_{\rm out}$, correspond to the projected velocities. The uncertainty in $\alpha_{\rm CO}=$0.8$\pm$0.5\,\msol(K\,\kms\,pc$^2$)$^{-1}$ \citep{downes98} is included in the mass error estimates. For J1347, the values reported for scenario \textsc{i}$^a$ are the nuclear outflow properties derived from CO(1-0) data at 0\farcs05 resolution by \citet{luke24} and for scenario \textsc{iii}$^b$, the upper limits reported by \citet{lamperti22} for the unresolved outflow detected in CO(2-1), using the average outflow velocity and radius of 390\kms~and 0.52\,kpc that they adopted for non-detections. For J1356, the values reported for scenario \textsc{i}$^c$ correspond to the CO(3-2) outflow measurements reported by \citet{sun14}, but adopting an outflow velocity of 400\kms~instead of the maximum outflow velocity used by these authors (500\kms), and scenario \textsc{iii}$^d$ to the measurements from \cra, which correspond to outflowing gas along the kinematic minor axis. 
The outflow mass fractions were computed using the total masses of $M_{\rm H_2}$ listed in Table~\ref{tab:co}, 
except for J1356, for which we used the mass of J1356N reported by \cra, of $M_{\rm H_2}$=6.39$\times$10$^9$ \msol. For J1509, $M_{\rm H_2}$=17.6$\pm$7.4 \msol~(\cra).}
\end{table*}

\textsc{ii}) In the second scenario, we consider that only the gas with the highest velocities is outflowing. The determination of the 
velocity cut comes from the comparison between the PVDs shown in Figures \ref{fig:pv1010}-\ref{fig:pv1356} and the \barolo~rotation model, shown in red contours in the same figures. 
The outflow velocity, $v_{\rm out}$, is defined as the velocity cut above the maximum rotation velocity that is modeled with \barolo. 
To calculate the flux of the high velocity gas, we then created moment 0 maps by selecting only channels above the rotation velocities modelled with \barolo, shown in Figure~\ref{fig:highvel}. In the case of J1010 we only see the blueshifted side of the outflow, reaching velocities up to 300\,\kms~in the corresponding PVD (see Figure~\ref{fig:pv1010}). The same happens in the case of J1347, 
based on the detection of high-velocity gas and excitation seen in the PVDs in Figure~\ref{fig:pv1347}. As for J1356, we only considered the more prominent redshifted velocities seen in the PVDs in Figure~\ref{fig:pv1356} that also correspond to the region with high values of \ratio.

\textsc{iii}) The third scenario is a mix of the first two. We assume that all the high-velocity gas is outflowing, but the contribution from rotation is subtracted following a similar approach as in \citet{santi19}. We first subtracted the contribution of circular motions by de-projecting the one dimensional rotation curve derived with \barolo~ to the corresponding velocity field on a pixel by pixel basis. We re-shuffled the channels in order to remove the rotation component from the velocity axis of the CO(2-1) datacube. Then, we created the residual integrated spectrum, resulting in a narrower residual CO profile (i.e., without the rotation component), as the one we presented for the Teacup in Appendix C of \ane. Since this method is suited to optimize the signal-to-noise of the emission associated with non-circular motions, it can reveal high-velocity wings that otherwise are too faint and/or immersed in the total CO profiles to be detected. 
Using the residual CO profiles, we estimate the flux contribution of the outflowing gas by integrating the flux of the high-velocity gas in the blue and red wings, using the same velocity cut as in Scenario \textsc{ii}.

In Table~\ref{tab:prop} we list the derived outflow integrated intensities ($S_{\rm CO}$), velocities ($v_{\rm out}$), and radii ($r_{\rm out}$) of the QSO2s in each of the three scenarios considered. We note that for J1347 and J1356 we did not measure the outflow fluxes as described in scenarios \textsc{i} and \textsc{iii} because they are ongoing mergers with very complex kinematics showing a small contribution from rotation (as discussed in Section~\ref{sec:model} and Appendix~\ref{sec:vsig}), and since scenarios \textsc{i} and \textsc{iii} depend on the subtraction of the rotation model, respectively, they cannot be performed accurately. Instead, in the case of J1347 we used as the least conservative scenario (Scenario \textsc{i} in Table \ref{tab:prop}) the recent  molecular outflow properties reported by \cite{luke24} based on CO(1-0) observations at 0\farcs05 resolution. The CO(1-0) outflow is compact (120\,pc) and likely driven by the small scale radio jet detected with VLBI and shown in Figure \ref{fig:ratiosdisp} \citep{morganti13,luke24}. As for the most conservative scenario (Scenario \textsc{iii} in Table \ref{tab:prop}), we adopted the upper limit derived by \citet{lamperti22} from the CO(2-1) observations at 0\farcs22 resolution. We note that the CO(3-2) data, first presented in \citet{fotopoulou19}, show blueshifted velocities of up to $\sim$400\kms~that are not seen in CO(2-1). However, for this QSO2 the high angular resolution CO(1-0) data reported by \citet{luke24} were crucial to detect the nuclear outflow of r$_{\rm out}\sim$120 pc, which shows velocities of up to $\sim$600\kms. In the case of J1356, we adopt as the least conservative scenario (Scenario \textsc{i} in Table \ref{tab:prop}) the outflow properties reported by \citet{sun14}, based on the CO(3-2) observations, but using an average outflow velocity of 400\kms~instead of the maximum outflow velocity adopted by these authors, of $v_{\rm out}$=500\kms. The most conservative scenario (Scenario \textsc{iii} in Table \ref{tab:prop}) in the case of J1356 corresponds to the outflow properties reported by \cra, which were calculated considering just outflowing gas along the kinematic minor axis of the galaxy. 

We computed the outflow masses from the integrated flux measurements of the outflows. First, the integrated fluxes were converted to CO luminosities, L$_{\rm CO(2-1)}^\prime$, in units of ${\rm K\,km\,s^{-1}\,pc^2}$ using Equation 3 of \citet{solomon05} and then translated into masses using the CO-to-H$_2$ ($\alpha_{\rm CO}$) conversion factor M$_{\rm H_2}$=$\alpha_{\rm CO}$R$_{\rm 12}$L$^\prime_{\rm CO(2-1)}$. Under the assumption that the gas is thermalised and optically thick, the brightness temperature ratio $R_{\rm 12}={L_{\rm CO(1-0)}^\prime}/{L_{\rm CO(2-1)}^\prime}$=1. We adopted the conservative value of $\alpha_{\rm CO}$ = 0.8, commonly used to derive outflow masses (see \cra~and references therein). The values of the outflow masses derived from each scenario are listed in Table~\ref{tab:prop}.


To compute outflow mass rates, we use the values for the outflow radii, $r_{\rm out}$, listed in Table~\ref{tab:prop}. These radii correspond to the maximum distances from the nucleus measured for the high-velocity gas shown with contours in Figure~\ref{fig:highvel}. The cold molecular outflow radii range between 0.65$<r_{\rm out}<$2.4\,kpc. For J1347 and J1356 we also include the outflow radii reported by the corresponding authors in the case of scenarios \textsc{i} and \textsc{iii}. Considering them, the range of outflow radii includes smaller values, of 0.12 and 0.52\,kpc in the case of J1347, and 0.3 and 0.4\,kpc in J1356. The outflow velocities that we measure range from 50 to 100\kms~in the case of scenario \textsc{i}, while in scenarios \textsc{ii} and \textsc{iii} the high velocity maps were created considering emission above/below $v_{\rm out}$=180-300\kms~and integrating the flux in the high-velocity residual wings, respectively (see Table \ref{tab:prop}). In the case of J1347 and J1356, faster outflow velocities were reported in the corresponding works that we used here as more and less conservative scenarios (see Table~\ref{tab:prop}).

For a time-averaged thin expelled shell geometry \citep{rupke05}, $\dot{M}_{\rm out}$=$M_{\rm out} \left(v_{\rm out}/r_{\rm out}\right)$, which corresponds to the outflow mass averaged over the flow timescale, $t_{\rm out}={r_{\rm out}}/{v_{\rm out}}$. This geometry has been previously adopted in other studies of molecular outflows in the local Universe \citep{fluetsch19,lutz20,cra22}. The outflow velocities and radii we reported for the three scenarios correspond to the projected values, since the determination of the outflow angle is uncertain. For a given outflow angle, the mass outflow rates will be then corrected to \mdot= $M_{\rm out} \times \frac{v_{\rm out}}{r_{\rm out}} \times tan(\alpha)$. For the case of outflows co-planar to the CO disks, then $tan(\alpha) = tan(90-i)=1/tan(i)$, where $i$ is the inclination angle of the CO disk (\cra).

Considering scenario \textsc{i}, we measure outflow masses that correspond to $\sim$0.6-8\% of the total molecular gas mass of the QSO2s, and outflow rates in the range 8$<$\mdot$<$148\mout. From scenario \textsc{ii} we derive outflow mass fractions of $M_{\rm out}/M_{\rm H_2}$=0.8-2.6\%, and 5$<$\mdot$<$63\mout. Finally, in the case of the most conservative scenario \textsc{iii}, the outflowing mass fractions are $\lesssim$1\% of the total molecular gas and 5$<$\mdot$<$32\mout, similar to the measurements reported by \cra~for the same QSO2s. The largest values of the outflow 
mass rate in the three scenarios considered here correspond to those of J1347, the only radio-loud QSO2 in our sample. It is noteworthy that this is not the case for the outflow masses, for which we measure the highest values in J1100 and J1509. We note that the outflow mass fractions depend on the choice of the $\alpha_{\rm CO}$ conversion factors used to compute the total and outflow masses. We adopted the standard Galactic value ($\alpha_{\rm CO}=4.36M_\odot (K kms^{-1}pc^{2})^{-1}$) for the total molecular mass and $\alpha_{\rm CO}=0.8 M_\odot (K kms^{-1}pc^{2})^{-1}$ for the outflow masses, resulting in a factor of $\sim$5 larger fractions if same $\alpha_{\rm CO}$ was adopted.

The mass loading factors, defined as $\eta = \dot{M}_{\rm out}$/SFR, for scenario \textsc{ii} are listed in Table~\ref{tab:prop}, with measured values ranging from $\eta \sim$ 0.2 to 3.4. We find $\eta > 1$, which is indicative of outflows efficiently removing molecular gas, for the Teacup and J1347. 
For the remaining QSO2s, the relatively low values of $\eta$ (0.2-0.6) might be indicating that star formation is the primary mechanism consuming molecular gas (see \cra~and \citealt{speranza24} for further discussion).
In Section~\ref{energetics} we discuss the energetics of these outflows in the context of radio and quasar mode AGN feedback.

\section{Discussion}\label{sec:discussion}

\subsection{Energetics of the molecular outflows}\label{energetics}

Empirical scaling relations 
using molecular gas observations of AGN and ULIRGs of different luminosities showed that the mass outflow rate increases with AGN luminosity \citep{cicone14,fiore17,fluetsch19}, 
supporting the scenario of 
accretion disc winds pushing away the surrounding gas and launching galaxy-scale outflows (i.e., quasar feedback). However, it is noteworthy that these scaling relations were inferred using galaxies that were known to host powerful molecular outflows detected using CO and/or OH observations. Indeed, more recent studies of less biased AGN and ULIRG samples have revealed important deviations from the classical scaling relations, and many luminous AGN including the QSO2s studied here lie well below them \citep{cra22,lamperti22,speranza24}. This indicates that AGN luminosity itself is not a sufficient condition for driving powerful outflows, and other factors might be relevant for an AGN to drive more or less massive molecular outflows. 

\begin{figure*}
    \centering
    \includegraphics[width=17cm]{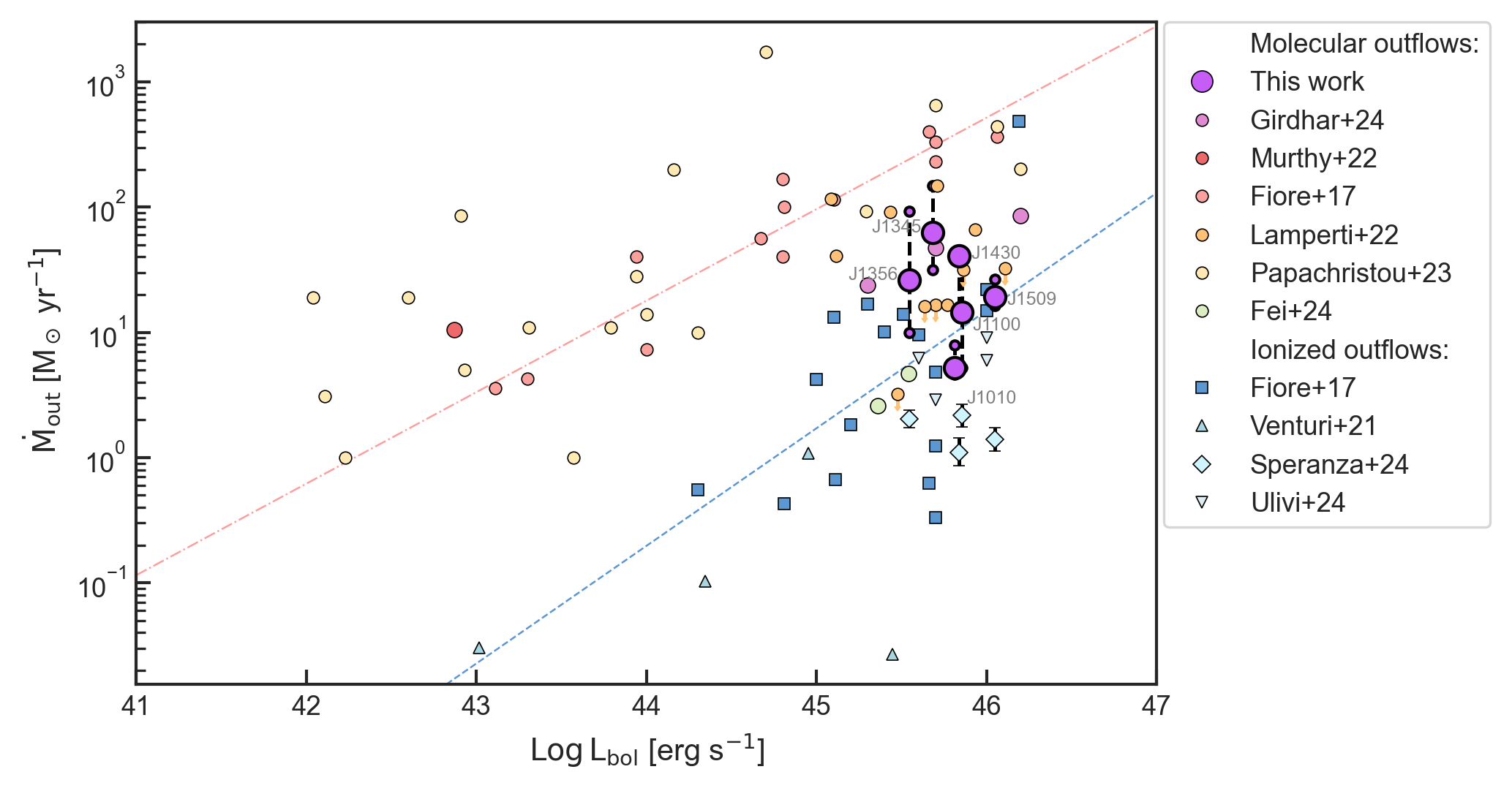}
    \caption{Outflow rate versus bolometric luminosity. The molecular outflow mass rates derived for the QSO2s using the three scenarios are displayed as large purple circles for the intermediate value, and dashed black lines connecting the most and least conservative values (small purple circles). For comparison, we included other molecular mass outflow rates measurements from the literature as circles of different colors: AGN and ULIRGs from \citet{fiore17}, QSO2s from \citet{girdhar24}, two type-1 QSOs from \citet{fei24}, AGN-dominated ULIRGs from \citet{lamperti22}, AGN with possibly jet-driven outflows from \citet{michalis23}, and the low-power jet-driven outflow from \citet{murthy22}. Mass outflow rates reported for ionized outflows in the literature are shown with different symbols and shades of blue: AGN and ULIRGs from \citet{fiore17} as squares, local jetted-Seyfert galaxies from \citet{venturi21} as upward triangles, QSO2s at z<0.4 from \citet{ulivi24} as downward triangles, and four of the QSO2s studied here from \citet{speranza24} as diamonds. The corresponding linear fits from \citet{fiore17} are shown as a salmon dot-dashed line for molecular outflows and as a blue dashed line for the ionized ones. The mass outflow rates from the literature have been converted to the thin shell geometry adopted in this work when necessary.}
    \label{fig:moutrates}
\end{figure*}

In this work we used three scenarios to compute the mass outflow rates, as described in Section~\ref{sec:outprop}, and found values of \mdot=5-150\mout. The mass outflow rates that we derive from the most conservative scenario \textsc{iii} if we exclude the radio-loud QSO2 J1347, are \mdot=5-16\mout. These values are similar to those reported by \cra~using a different method, consisting of considering as outflowing gas that along the kinematic minor axis of the galaxies, and assuming that the outflows are coplanar with the CO disc. In Figure~\ref{fig:moutrates} we show the \mdot~versus L$_{\rm bol}$ plot including our QSO2s, 
with the intermediate values (big purple circles) representing scenario \textsc{ii} and the black dashed lines connecting the least and most conservative values derived from scenarios \textsc{i} and \textsc{iii} (small purple circles). In agreement with the results presented in \cra, our \mdot~values lie below the \citet{fiore17} empirical relation, even when we consider the least conservative scenario \textsc{i}. This is also the case for the AGN-dominated ULIRGs presented in \citet{lamperti22}, 
also shown in Figure~\ref{fig:moutrates}, whose outflow mass rates range from 3 to 145\mout. Based on their bolometric luminosities (L$_{\rm bol}\sim$10$^{45-46}$\ergs), both our QSO2s and the AGN-dominated ULIRGs were expected to have molecular outflow rates of $\gtrsim$100\mout~based on the \citet{fiore17} scaling relation. The four $z<$0.2 QSO2s with nuclear molecular outflows reported in \citet{girdhar24} also occupy the same region in the \mdot - L$_{\rm bol}$ plane as the QSOFEED QSO2s.


This and other recent works are starting to reveal a population of AGN with less massive molecular outflows that those expected from empirical scaling relations. 
This is likely due to the advent of sub-mm facilities with high sensitivity such as ALMA and NOEMA, which enable the observation of more diverse, and hence less biased, AGN samples. Based on all the outflow measurements included in Figure~\ref{fig:moutrates}, from this and other works, we claim that the scaling relations found by e.g. \citet{cicone14}, \citet{fiore17}, and \citet{fluetsch19} 
most likely represent the upper boundary of the \mdot - L$_{\rm AGN}$ relation, as first suggested by \cra.

We note that part of the discrepancy between the empirical \mdot - L$_{\rm AGN}$ relation and the observations lying below it in Figure~\ref{fig:moutrates} could be associated with the different outflow velocity definitions (e.g., $v_{\rm max}$ versus $v_{\rm out}$). For example, \citet{venturi23} reported that the use of $v_{\rm max}$, defined as $v_{\rm out}$+FWHM/2, instead of $v_{\rm out}$ results in an increase of the ionized mass outflow rates by a factor of 3-20. 
In the case of molecular outflows, which have lower velocities than the ionized ones, $v_{\rm out}$ cannot be measured from the line wings in most cases, but rather from the residuals obtained after kinematical modeling, such as the analysis performed here. This is also the case of some of the molecular outflows used by \citet{fiore17} to derive the empirical relation shown in Figure \ref{fig:moutrates}, such as NGC\,1068 \citep{santi14} and NGC\,1433 \citep{combes13}. We refer the reader to \citet{kiara23} and \citet{speranza24} for further details on how the use of different methodologies and outflow parameter definitions affect the derived outflow energetics.

\begin{figure*}
    \centering
    \includegraphics[width=18cm]{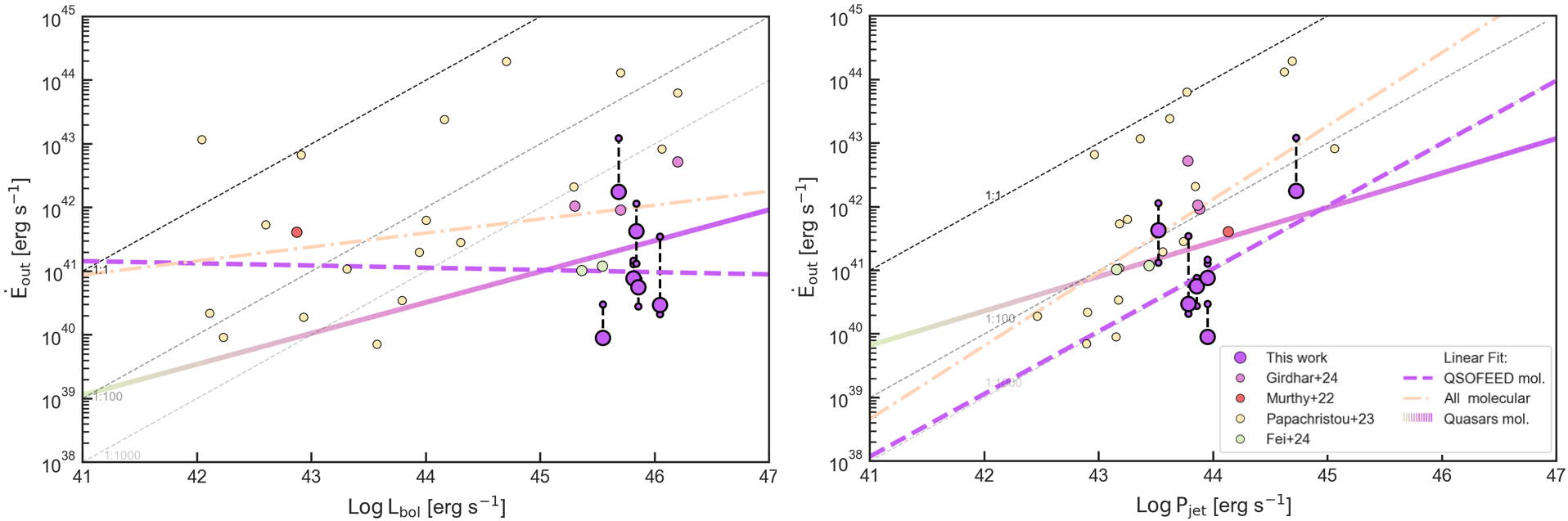}
    \caption{Kinetic power of the outflow versus bolometric luminosity (left panel) and jet power (right panel). The 1:1, 1:100, and 1:1000 relations are shown as dashed lines in both panels. The dashed purple line in both panels corresponds to the linear fit of the purple circles, which correspond to the molecular outflows of the QSO2s studied here (``QSOFEED mol.'').  The diverging color line (``Quasars mol.'') represents the linear fit of the type-1 PG quasars in \citet[][green circles]{fei24}, the QSO2s from \citet[][magenta circles]{girdhar24}, and the QSOFEED QSO2s (purple circles). The light orange dot-dashed line (``All molecular'') shows the fit of all the molecular outflows mentioned before 
    and those reported by \citet[][yellow circles]{michalis23} and \citet[][red circles]{murthy22}. 
    Symbols and colors are the same as in Figure~\ref{fig:moutrates}.}
    \label{fig:ekin}
\end{figure*}

To explain these ``mild'' molecular outflows detected in luminous AGN, \cra~suggested that other factors including jet power, coupling between winds, jets, and/or ionized outflows and the CO disks, and the ISM distribution might also be relevant (see also \citealt{harrison24} for a recent review). To investigate the possible role of compact jets as drivers of the molecular outflows, we computed the kinetic power of the outflows ($\dot{E}_{\rm out}$=0.5\mdot$v_{\rm out}^2$) and compared them with the values of $L_{\rm bol}$ and the jet-power ($P_{\rm jet}$) of the QSO2s and other AGN in Figure~\ref{fig:ekin}. The values of $P_{\rm jet}$ were computed using the 1.4 GHz VLA fluxes and spectral indices corresponding to the HR data from \citet{jarvis19} and applying the $L_{\rm 1.4GHz}\times P_{\rm jet}$ relation of \citet{birzan08}, as in \ane. They are listed in Table~\ref{tab:prop}.

The observed kinetic powers of the QSO2 outflows are in the range of 6.4$\times 10^{39}<\dot{E}_{\rm out}<1.2 \times 10^{43}$\ergs, and in the left panel of Figure~\ref{fig:ekin} we can see that the radiative coupling efficiencies, defined as $\epsilon_{\rm AGN}\equiv \dot{E}_{\rm out}/L_{\rm bol}$, lie below the 1:1000 relation, except for the least conservative value measured for J1347. This is showing that the kinetic power of the outflows correspond to $\ll$0.1\% of the AGN bolometric luminosities shown in Table \ref{tab:main}. There is a large scatter, of more than three orders of magnitude, in the $\epsilon_{\rm AGN}$ derived when we considered the measurements from the three different scenarios ($10^{-6}<\epsilon_{\rm AGN}<10^{-4}$ if we only consider scenario \textsc{ii}; 0.0001-0.01\%). This scatter is similar to the one found for observationally-derived outflow kinetic
powers compiled by \citet{harrison18}. Although the majority of the cosmological simulations require $\epsilon_{\rm AGN}\gtrsim$1\% for AGN feedback to efficiently suppress star formation \citep[e.g.][]{dubois14,schaye15, costa18}, these fiducial efficiencies adopted in the simulations should not be directly compared to the observed coupling efficiencies \citep{costa18,harrison18,harrison24}. This is because theoretical efficiencies are calibrated quantities, and not a prediction from the simulations, and also because only a fraction of the injected energy is converted into outflow kinetic energy, depending on the properties of ISM, the gravitational potential, and other factors, so the observed coupling efficiencies are often small \citep{harrison24}. 

Low radiative coupling efficiencies can also be due to the fact that radiation pressure–driven outflows depend strongly on the distribution and geometry of gas and dust around the AGN \citep{bieri17,costa18}. Radiative coupling efficiency depends on the degree to which radiation is trapped and reprocessed within an optically thick, dusty medium. In clumpy or porous environments, radiation can escape through low-density channels, significantly reducing the momentum imparted to the gas \citep[e.g.][]{ishibashi15, costa18}. This strong dependence on the small-scale structure of the ISM can naturally explain the lower radiative coupling efficiencies observed in the quasars in this work. The purpose of showing Figure~\ref{fig:ekin} including the 1:100 relation is to highlight that for the six QSO2s and for the majority of the observational data shown in that Figure, $\epsilon_{\rm AGN}\ll$1\%.

For comparison, in the right panel of Figure~\ref{fig:ekin} we present $\dot{E}_{\rm out}$ versus $P_{\rm jet}$. The jet coupling efficiencies considering scenario \textsc{ii}, defined as $\epsilon_{\rm jet}\equiv \dot{E}_{\rm out}/P_{\rm jet}$, are in the range 0.001$<\epsilon_{\rm jet}<$0.035 (0.1-3.5\%). These are similar to the jet coupling efficiencies of the four QSO2s at $z<$0.2 reported by \citet{girdhar24}, of 0.05-4\% (one of them is Teacup, so in the following we will refer to the other three), and are significantly higher than the corresponding radiative coupling efficiencies.

We investigated possible correlations exiting between the outflow kinetic power, $\dot{E}_{\rm out}$, and $L_{\rm bol}$ and $P_{\rm jet}$. The first test was to apply a logarithmic linear regression to the six QSOFEED sources studied in this work. The second test was to group the six QSOFEED QSO2s and other nearby quasars with molecular outflows reported in the literature: three QSO2s at $z\lesssim$0.2 reported in \citet{girdhar24} 
and the two type-1 PG quasars at $z\sim$0.06 recently reported by \citet{fei24}. Finally, the third test includes all the quasars included in the second test and the low-power jet-driven outflow in a low-luminosity AGN from \citet{murthy22} and the compilation of possibly jet-driven outflows in AGN of different bolometric luminosities presented in \citet{michalis23}.

In Table~\ref{tab:correlations} we list the derived parameters for the logarithmic linear regression in the format $\log(Y)= a*\log(X)+ b$, with the corresponding Pearson correlation coefficients, with possible values in the range of -1$<r_{\rm Pearson}<$1 and the $p$-values. $r_{\rm Pearson}$ values close to zero indicate no correlation, close to 1 a strong correlation and close to -1 an anti-correlation.

\begin{table}[h]
 \caption{Correlation parameters of the outflow energetics}
    \centering
    \begin{tabular}{lcccc}
          \hline
       \hline
       Log correlation  &  $a$ & $b$ & $r_{\rm Pearson}$ & $p$-value  \\
       \hline
       \noalign{\smallskip}
       \multicolumn{5}{c}{QSOFEED molecular}  \\
   
    $\dot{E}_{\rm out} \times L_{\rm bol}$ & -0.19 & 54.0 & -0.44 & 0.38 \\
     $\dot{E}_{\rm out} \times P_{\rm jet}$ & 0.98 & -2.3 & 0.48 & 0.33 \\
    \hline
    \noalign{\smallskip}
    \multicolumn{5}{c}{Quasars molecular}  \\
  
    $\dot{E}_{\rm out} \times L_{\rm bol}$ & 0.12 & 40.6 & 0.17 & 0.62 \\  
      $\dot{E}_{\rm out} \times P_{\rm jet}$ & 0.54 & 17.6 & 0.25 & 0.45 \\
        \hline
        \noalign{\smallskip}
      \multicolumn{5}{c}{All molecular}  \\
    
    $\dot{E}_{\rm out} \times L_{\rm bol}$ & 1.43 & -18.0 & 0.61 & 0.0003 \\
    $\dot{E}_{\rm out} \times P_{\rm jet}$ & 1.15 & -8.6 & 0.55 & 0.0017 \\
    \noalign{\smallskip}
    \hline
    \end{tabular}  
    \label{tab:correlations}
\end{table}

The $p$-values shown in Table~\ref{tab:correlations} show that the only statistically significant correlations are found when considering all the measurements in the literature, i.e.,  the ``All molecular'' case, with $p$-value$<$ 0.2\% and $r_{\rm Pearson}\sim$0.6, indicating a positive correlation with both $L_{\rm bol}$ and $P_{\rm jet}$. On the other hand, when using 
the ``QSOFEED molecular'' (6 galaxies) and ``Quasars molecular'' (11 galaxies) samples, the $p$-values are high and indicate no correlations, possibly due to the small sample sizes. 
However, we note that the linear regression using only the values of $\dot{E}_{\rm out}$ and and $P_{\rm jet}$ derived for ``QSOFEED molecular'' (indicated with a purple dashed line in the right panel of Figure \ref{fig:ekin}) follows the 1:1000 relation, indicating jet coupling efficiencies of $\epsilon_{\rm jet}\sim$0.1\%. This might indicate that the compact, low-power jets often detected in radio-quiet AGN, are potentially capable of launching the molecular outflows that we observe. To confirm or discard any of these trends, more observations covering larger parameter space in $P_{\rm jet}$ and $L_{\rm bol}$ are required.

Therefore, despite the fact that our QSO2s have high AGN luminosities, 
low-power jets might be driving some of the molecular outflows that we see,  based on the higher jet coupling efficiencies reported here. This result is
predicted by simulations of jet-ISM interactions \citep{mukherjee18sim,meenakshi22} and 
in agreement with previous studies using similar samples of obscured quasars 
\citep[e.g.][]{mullaney13,zakamska14,jarvis19,moly19,girdhar22,girdhar24}. In fact, low-to-moderate power kiloparsec scale jets are starting to be recognized as potential drivers of multi-phase outflows and as relevant mechanisms for AGN feedback \citep{morganti15,venturi21}. In ``radio-quiet'' quasars, spatially resolved radio structures are often associated with morphological and kinematic distinct features in the ionized and molecular gas, such as increased turbulence and outflowing bubbles \citep[e.g.][]{vm17,jarvis19,jarvis21,girdhar22,cra22,ane23,ulivi24}. 
This suggest that jets can be an important feedback mechanism in highly accreting radio-quiet quasars, where radiative feedback would be expected to prevail. Alternatively, it has been proposed that the radio structures detected in e.g. VLA data of radio-quiet AGN could be produced by shocks induced by the outflows themselves as they made progress through the ambient ISM \citep{fischer19,fischer23}.

We note that even if a jet-like morphology is resolved only for J1430 (Teacup) in the HR VLA radio images of our QSO2s, the unresolved radio emission of the others might still be attributed to small scale radio jets. \citet{jarvis19} analysed the VLA data of J1010, J1100, J1356\footnote{J1356 shows a jet-like morphology extending up to 2\farcs6 (5.7\,kpc) in the LR data (1\arcsec~resolution), but not in the HR data (0\farcs25 resolution).}, and J1430. They all show a radio excess and lie above the FIR–radio correlation of star-forming galaxies \citep{bell03}, indicating that only star formation processes are not enough to produce their radio emission. In addition, \citet{jarvis19} classified the unresolved nuclear radio emission of these QSO2s as ``jet/lobe/wind'', based on their steep spectral indices \citep[$\alpha \sim -0.8$, see Table 4 in][]{jarvis19}. This hypothesis is confirmed for the case of J1347, which shows diffuse emission on large scales \citep[$r\sim$80\,kpc;][]{stanghe05} and an unresolved core 
using low angular resolution radio observations, but high angular resolution VLBI data revealed a 100\,pc scale jet \citep{kellermann04,morganti13}, shown as black contours in the inset panel of J1347 in Figure~\ref{fig:ratiosdisp}.

The sensitivity and resolution of next generation radio facilities such as the LOw Frequency ARray (LOFAR), the Square Kilometre Array (SKA), and the new-generation Very Large Array (ngVLA) will help us to understand the origin of the radio emission in ``radio-quiet'' AGN \citep{panessa19} and they likely will reveal a wealth of galactic scale jets \citep{nyland18,morabito22,ye23}. This will be particularly important to advance our understanding of the role of galactic scale low-power jets for AGN feedback, since ``radio-quiet'' sources dominate the AGN population \citep{padovani16}.

\subsection{Relative orientation of the multi-phase outflows}\label{pas}

In order to provide a more comprehensive view of the multi-phase outflows detected in the quasars studied here, we compare our findings for the cold molecular gas (T$\sim$10--100\,K) with observations in the optical and NIR, allowing us to probe a broad range of gas temperatures and densities \citep{cicone18,harrison24}. In the optical, the QSOFEED outflows have been studied in the warm ionized phase using the [O\,III]$\lambda$5007\,$\AA$ emission line (T$\sim$10$^4$\,K), and in the NIR using the ro-vibrational H$_2$ lines to trace the warm molecular gas (T$\gtrsim$10$^3$\,K).

Using GTC/MEGARA 
integral field spectroscopic observations of five QSO2s (same as \cra~and this work, except for J1347), \citet{speranza24} characterized the morphology and kinematics of 
the [O\,III] emitting gas. According to the outflow properties there reported, the ionized outflows in these five QSO2s carry less mass than their molecular counterparts and are more extended (3.2$<r_{\rm out}<$12.6\,kpc) and faster (500$<|v_{\rm out}|<$1300\kms) than the molecular outflows reported here (0.65$<r_{\rm out}<$2.4\,kpc and 180$<|v_{\rm out}|<$300\kms~from scenario \textsc{ii}). The orientations (PAs) of the ionized outflows are indicated in Figure~\ref{fig:highvel} and Table~\ref{tab:pa} for comparison with the PAs measured here for the cold molecular outflows. The PA of J1010 could not be accurately measured from the GTC/MEGARA data, since this was the only QSO2 analyzed in \citet{speranza24} in which the [O\,III] outflow is not spatially resolved. However, from inspection of the kinematic maps shown in \citet{speranza24}, the blueshifted side of the ionized outflow would have a PA of $\sim$-45\degree, i.e. towards the north-west. \citet{ulivi24} analyzed a VLT/MUSE cube of this QSO2 and also detected broad, blueshifted [OIII] emission, but without a clear outflow PA. 

 
We followed a similar procedure to that adopted in \citet{speranza24} to determine the PA of the blueshifted and/or redshifted sides of the molecular outflows. \citet{speranza24} fitted an ellipse to the 3$\sigma$ contour of the O[III] outflow flux maps using a Python script \citep{hill16}. Here we used the same script to fit ellipses to the global shape of the high-velocity contours in Figure~\ref{fig:highvel}. We then determined the outflow PA using the line connecting the QSO2 nucleus, defined by the 200\,GHz continuum peak, and the most distant point within the ellipse. The cold molecular outflow PAs are listed in Table~\ref{tab:pa} and shown as dashed blue and red lines in Figure~\ref{fig:highvel} for the blueshifted and redshifted sides of the outflows, respectively.

\begin{table}[h]
\caption{PAs of the cold, warm molecular and ionized outflows and radio jets.}
\centering
\begin{tabular}{lccccccc}
\hline
\hline
Name 	 &   \multicolumn{2}{c}{PA$\rm_{CO}$}  &  \multicolumn{2}{c}{PA$\rm_{H_2}$} & \multicolumn{2}{c}{PA$\rm_{[OIII]}$} & \pajet   \\
         & blue & red  &  blue & red & blue & red &  \\
\hline \noalign{\smallskip}
J1010  &  116  & \dots & \dots & \dots & \dots  & \dots & 182 \\
\noalign{\smallskip}
J1100  &  200 & 55  & \dots& \dots&  63 & \dots & 175\\
\noalign{\smallskip}
J1347  & 56 &\dots& & \dots& \dots\dots    & \dots& 160\\
\noalign{\smallskip}        
J1356  & \dots & 129 & \dots & 135 &  197 &  200 & 18\\
\noalign{\smallskip}
J1430  & 188 & -18 & 175 & -5 &  65  &  198 & 60\\
\noalign{\smallskip}
J1509 & 174 & -51 & \dots & \dots &   -40 & \dots & \dots\\
\noalign{\smallskip}
\hline	   					 			    					 	\end{tabular}	
\tablefoot{The values of PA$\rm_{H_2}$ are from \citet{mavi25}, PA$\rm_{[OIII]}$ from \citet{speranza24}, shown in Figure \ref{fig:highvel} together with the the values of PA$_{\rm CO}$, and \pajet~from \citet{jarvis19}, shown in Figure \ref{fig:ratiosdisp}.}
\label{tab:pa}
\end{table}

The projected PAs of the ionized and molecular outflows do not have similar orientations. In fact, in the cases of J1100, J1430, and J1509, their orientation appears to be inverted (see Figure \ref{fig:highvel}). For instance, in J1430, the redshifted [O\,III] outflow (PA$\rm_{[OIII],red}$=198\degree) is almost aligned with the blueshifted CO outflow (PA$\rm_{CO,blue}$=188\degree). The same is found for J1100 and J1509, where the redshifted sides of the CO outflows (PA$\rm_{CO,red}$ of 55\degree and -51\degree, respectively) are closely aligned with the blueshifted ionized outflow sides (PA$\rm_{[OIII],blue}$ of 63\degree and -40\degree). A caveat to consider is that the MEGARA observations are seeing-limited, with a resolution of approximately 1\farcs1, which is significantly coarser than the resolution of the ALMA data ($\sim$0\farcs2), although in the case of well-resolved ionized outflows as it is the case of J1100, J1356, J1430, and J1509 should not be an issue when it comes to determining the outflow orientation.

On the other hand, the warm molecular gas kinematics in J1356 and J1430, traced by the H$_2$1-0\,S(1) and S(2) emission lines observed with VLT/SINFONI (seeing of $\sim$0\farcs8), are consistent with that of the cold molecular gas observed by ALMA \citep{mavi25}. Both the blue- and redshifted sides of the warm molecular outflow are detected in J1430, while in J1356 only the redshifted outflow side is detected, having the same orientation observed for CO in this work (see Table \ref{tab:pa}), and similar radii and velocities (1.8$<r_{\rm out}<$1.9\,kpc and 370<|v$_{\rm out}|<$470\kms). The main difference is that the warm molecular outflows represent just a small fraction of the total molecular gas mass: \citet{mavi25} measured warm-to-cold gas ratios of $\sim$1$\times$10$^{-5}$ in the two QSO2s \citep[see][for further details on the comparison between cold and warm molecular gas in J1356 and J1430]{mavi25}.

In general, the warm and cold molecular phases seem to be tracing the same outflow, with the main distinction being the mass fraction carried by them. On the contrary, the warm ionized outflows do not seem to be a different face of the same outflow, showing different orientation, velocity, and radius. This is likely related to the different temperatures and densities of the molecular and ionized gas, which respond differently to the action of AGN-driven winds and jets. This should be considered when comparing with simulations, which often consider ``cold'' gas that at temperatures of $\sim$10$^3$-10$^4$ K (e.g. \citealt{mukherjee16,Ward24}).


\subsection{Global versus localised physical conditions of the molecular gas}\label{sec:gasphysics}

\begin{figure}
\centering
\resizebox{\hsize}{!}{\includegraphics{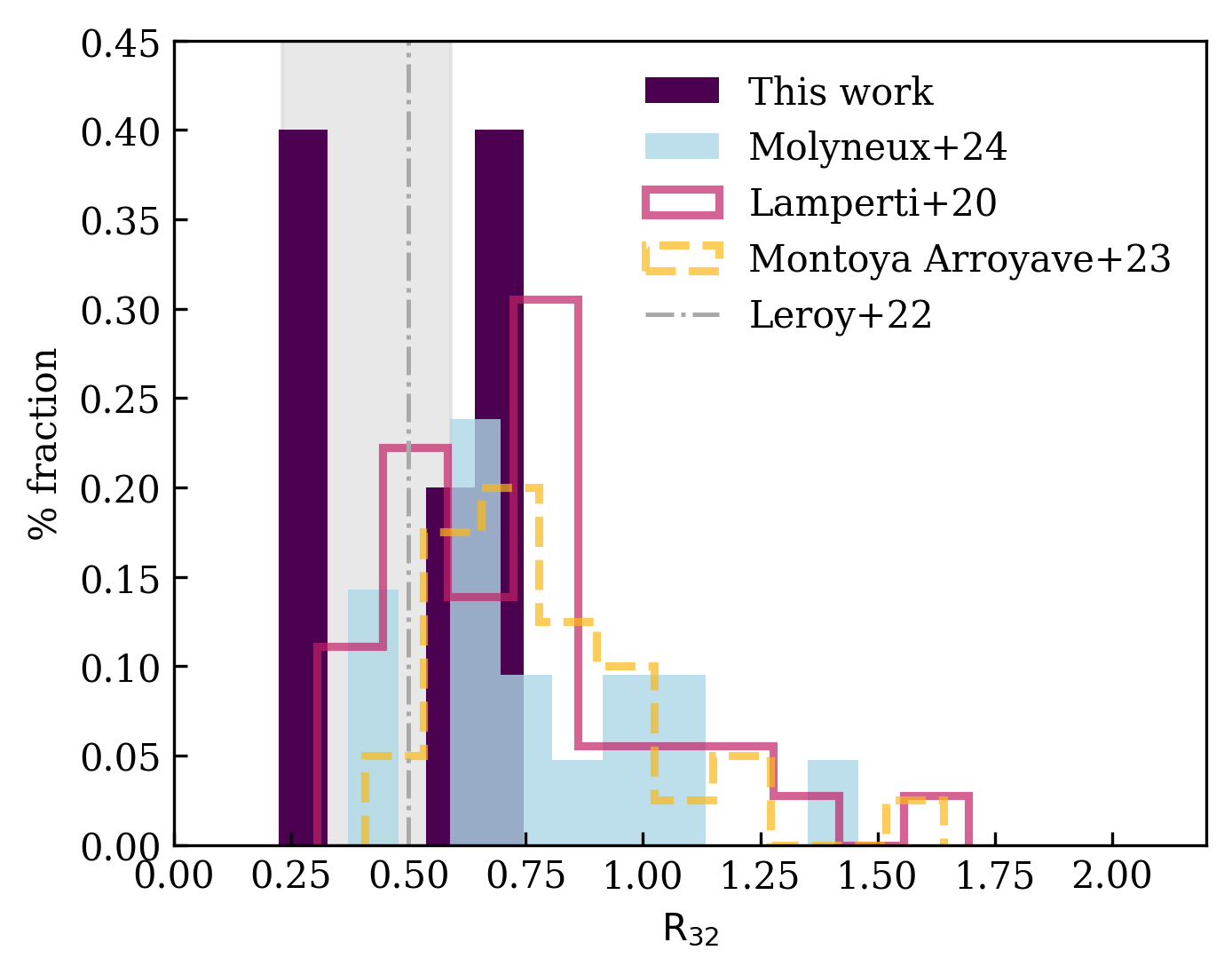}}
\caption{Histogram of the integrated \ratio~for the 5 QSO2s with both CO(2-1) and CO(3-2) transitions available, shown as filled purple bars. 
The values measured for a sample of QSO2s at $z<$0.2 from \citet{moly24} are also shown in light blue, for the X-ray selected AGN reported by \citet{lamperti20} in pink, of the local ULIRGs presented in \citet{montoya23} in yellow, and the mean value measured for star-forming disk galaxies from \citet{leroy22} is the gray dot-dashed line, with the corresponding interval shown as a shaded gray area.}
\label{fig:r32hist}
\end{figure}

In Sections~\ref{sec:integrated} and \ref{sec:spaline} we presented the \ratio~line ratios measured from the galaxy-integrated CO fluxes and the spatially resolved \ratio~maps of the QSO2s. The global values derived from the integrated CO(3-2) and CO(2-1) intensities listed in Table~\ref{tab:co} are shown in the histogram of Figure~\ref{fig:r32hist}.
We compare our global \ratio~measurements with those derived for other samples of galaxies: a similar sample of 17 QSO2s at $z<$0.2 from \citet{moly24} using either ACA or APEX CO integrated fluxes from J $\rightarrow$ 1 up to 7 transitions; 36 hard X-ray-selected AGN host galaxies from the BAT AGN Spectroscopic Survey (BASS) analyzed in \citet{lamperti20}, observed with the 15\,m single-dish James Clerk Maxwell Telescope; a sample of 36 local ULIRGs observed with APEX by \citet{montoya23}; and the low-J transitions study of a large sample of nearby star-forming disk galaxies using single-dish CO mapping surveys and PHANGS-ALMA by \citet{leroy22}.

We derived integrated values of 0.22$<$\ratio$<$0.75 and an average value of $<$\ratio$>$0.5$\pm$0.2 for the QSO2s, which is similar to the average values derived for normal disk galaxies in \citet{leroy22}, of 0.23$<$\ratio$<$0.59. 
\citet{moly24} reported a median value of <\ratio>=0.6 for their sample of QSO2s, similar to ours, but their values range from 0.37-1.46, a broader range than the one we found for our smaller sample. A slightly higher average value of $<$\ratio$>$0.74$\pm$0.3 was measured for the AGN BASS sample by \citet{lamperti20}, with ratios ranging from 0.3-1.7. A similarly broad interval of \ratio~was found by \citet{montoya23} for a sample of ULIRGs including active and non-active galaxies, with a median value of <\ratio>=0.76 and a range of 0.4$<$\ratio$<$1.7. 


The CO SLEDs derived for QSO2s at $z<$0.2 in \citet{moly24} do not show evidence for high gas excitation for high J$\rightarrow$6,7 transitions, as it is observed for more luminous high-redshift quasars \citep{carilli13}. We note however that only one of the four QSO2s observed in CO(6-5) by \citet{moly24} was detected and none of the three QSO2s observed in CO(7-6) were detected. These authors concluded that AGN feedback is not exerting any influence in the overall gas excitation on global scales, and instead, they proposed that feedback from winds/jets might by occurring on localized regions that they cannot resolve with the ACA/APEX observations. 
For lower J transitions, the CO SLEDs of a sample of 35 local AGN ($z\leq$0.15) reported by \citet{esposito22} indicate no evidence for AGN influence on the cold and low-density gas on more localised kiloparsec-scales.

The high angular resolution ALMA observations employed here allow us to probe physical scales of a few hundreds of parsecs of the QSO2s. The spatially resolved \ratio~maps shown in Figure~\ref{fig:ratiosdisp} reveal a gradient of excitation, with higher values of \ratio$\sim$0.7-0.9 within the central kiloparsec of the galaxies and of $\sim$0.2-0.5 across the galaxy disks. Spatially resolved CO line ratio emission analyses have been reported for a few nearby Seyferts and radio galaxies \citep{viti14,morganti15,dasyra16,oosterloo17,oosterloo19,ruffa22}, but not for quasars, with the exception of the Teacup (\ane). 

Using a plethora of molecular lines (i.e., CO, HCN, HCO$^+$, and CS lines) and the RADEX \citep{radex} radiative transfer code, \citet{viti14} modelled the excitation conditions of the molecular gas in the circumnuclear disc of NGC\,1068. The authors claimed that elevated values of \hbox{$R_{\rm 31} \equiv L^\prime_{\rm CO(3-2)}/L^\prime_{\rm CO(1-0)}\sim$5} 
and \hbox{$R_{\rm 21} \equiv L^\prime_{\rm CO(2-1)}/L^\prime_{\rm CO(1-0)}\gtrsim$4} correspond to hot dense gas (kinetic temperatures $T>$150\,K and gas density $n_{\rm H}>10^5$\,cm$^{-3}$) excited by the AGN.

Higher values of different emission line ratios have been reported for the molecular gas interacting with jets in the radio galaxy PKS\,1549-79 \citep{oosterloo19}, and also in the Seyfert galaxy IC\,5063 \citep{dasyra16,oosterloo17}. IC\,5063 is a clear-cut example of jet coplanar with the galaxy disc, which is driving a multi-phase outflow \citep{tadhunter14,morganti15}. The jet power estimated for IC\,5063 is $P_{\rm jet}\sim$10$^{\rm 44}$\ergs~and the PVD along the jet axis revealed \ratio~values ranging from 1.0$-$1.5 in the outflowing regions, clearly different from the gas following regular rotation ($R_{\rm 32}<0.7$). Under local thermodynamic equilibrium (LTE) conditions, $R_{\rm 32}>$1 corresponds to excitation temperatures ($T_{\rm ex}$) of $\sim$50\,K \citep{oosterloo17}. A similar temperature was found for the fast outflow gas component of IC\,5063 using \hbox{$R_{\rm 42}\equiv L^\prime_{\rm CO(4-3)}/L^\prime_{\rm CO(2-1)}>>$1}, suggesting that the gas in the outflow is optically thin \citep{dasyra16}. 

\begin{figure*}
    \centering
    \includegraphics[width=17cm]{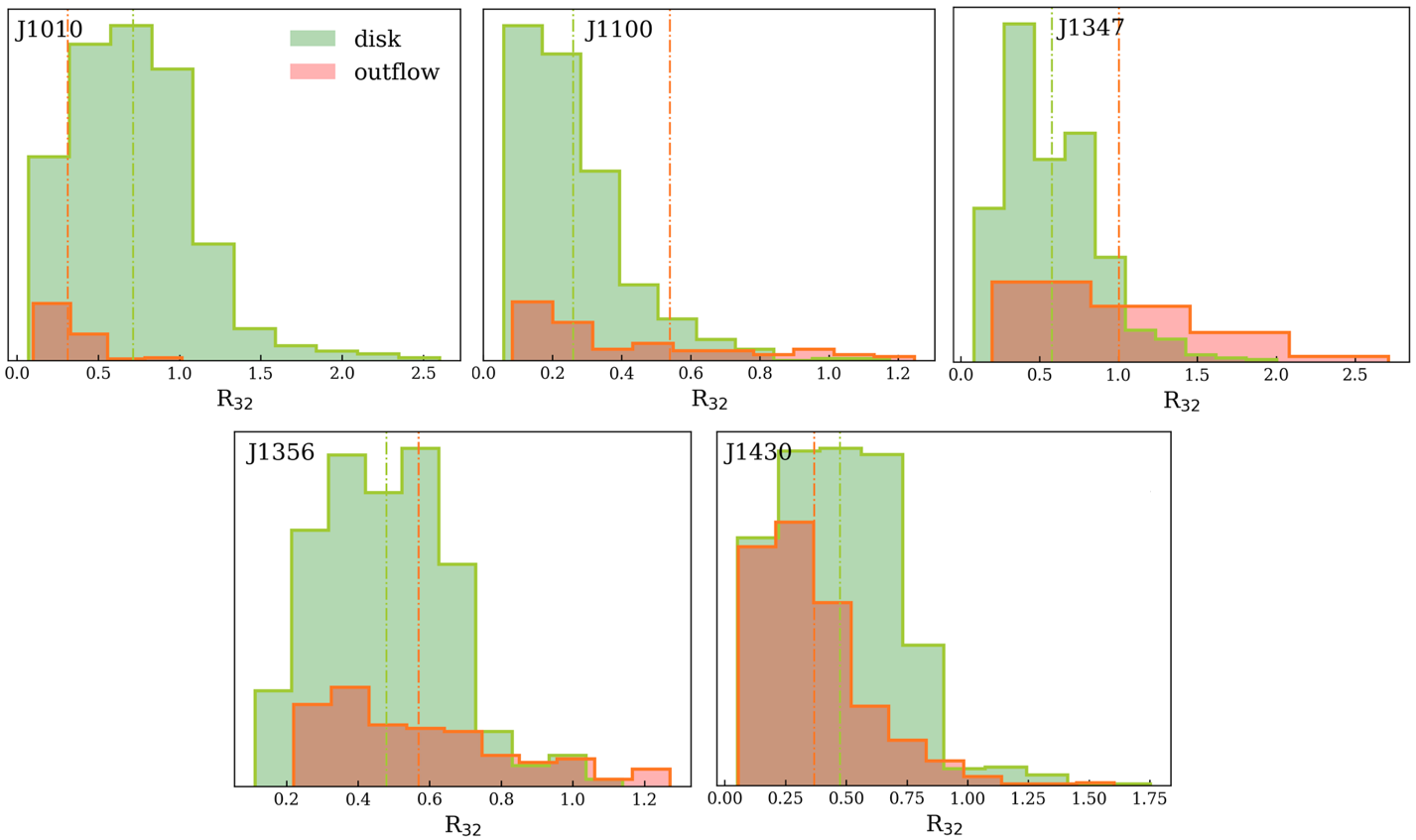}
    \caption{Distribution of the \ratio~values in the PVD along the minor-axis for the high-velocity (red) and disk regions (green) shown in Figures~\ref{fig:pv1010}-\ref{fig:pv1356}.}
    \label{fig:pv-r32hist}
\end{figure*}

Another example of jet-induced molecular gas excitation is the radio galaxy NGC\,3100. The cold molecular gas is distributed in a ring-like morphology, with the central region ($\sim$500 pc) depleted of CO. Most of the CO emission presents \hbox{$R_{\rm 21}$} and \hbox{$R_{\rm 31}$} ratios of $<$1, implying sub-thermal excitation conditions \citep{ruffa22}, but ratios of $R_{\rm 21}\sim$4 and $R_{\rm 31}\sim$3 are observed at the inner part of the disc, where there is spatial correlation with the radio lobes. All these cases indicate that jet-impacted regions have different gas excitation conditions due to shocks and/or reduced optical thickness \citep{morganti21}.

In an attempt to quantify the differences in \ratio~found between the disk and outflow regions, Figure~\ref{fig:pv-r32hist} shows the histograms of \ratio~values included in the PVDs along the minor axis (see  Figures~\ref{fig:pv1010}-\ref{fig:pv1356}). 
The green histograms correspond to molecular gas with velocities v$<v_{\rm out}$ from scenario \textsc{ii}, which we attributed to the disk, and the red histograms to gas with v$>v_{\rm out}$, which we associate with the outflow regions. We note that this method only includes gas along the minor axis, within a simulated slit of a given size, and therefore does not include all the gas in the disk or outflow. The outflow histograms show higher values of \ratio~than the disk ones in J1100, J1347, and J1356. In the case of J1010, as we described in Section~\ref{sec:model}, the regions with \ratio$>$1 are distributed all over the minor and major axis PVDs, for both low and high velocity gas. This is likely related to coarser resolution of the CO(2-1) data in the case of J1010, of $\sim$0\farcs8, which prevents us from accurately constraining the \ratio~distribution in this QSO2. For J1430, studied in detail in \ane, the jet is almost coplanar with the CO disk, producing elevated \ratio~values across the whole disk, and particularly along the direction perpendicular to the jet.


Therefore, based on the analysis presented in this work, for our QSO2s we find that the regions with high-velocity gas that cannot be attributed to rotation, as modelled with \barolo, have values of \ratio~of up to $\approx$1-2, with J1347 showing the highest values. 
These values are significantly higher than those measured in areas of the PVDs dominated by rotation (\ratio$\approx$0.3-0.7) and than the values reported by \citet{leroy22} for normal disk galaxies (mean \ratio~of 0.5),  
indicating that the molecular gas participating in the outflows exhibits enhanced temperature and is optically thin. Shocks could potentially induce a different temperature profile and thus affect the observed line ratios in these outflows \citep{meijerink13}. This could also be due to higher X-ray emission in the vicinity of the AGN, since \ratio~ tends to increase with the X-ray luminosity in high density gas ($n_{\rm H}>10^4$\,cm$^{-3}$), as shown in \citet{lamperti20}.
However, we are unable to infer the physical conditions ($T_{\rm ex}$ and density) of the gas using only one molecular line ratio, since more transitions, specially from higher J, would be necessary for that. 

\section{Conclusions}\label{sec:sum}

We presented the analysis of ALMA CO(2-1) and CO(3-2) observations at 0\farcs2-0\farcs6 resolution (370-960\,pc) of a sample of six obscured quasars at $z\sim$0.1, which belong to the QSOFEED sample. Our goal was to investigate the relevance of quasar-driven winds and/or jets on their cold molecular ISM and kinematics. To do so, we performed kinematic modelling together with spatially resolved line ratios, \ratio, that indicate changes in the gas excitation conditions. Our main findings can be summarized as follows.

\noindent\hspace{0.1cm}-The CO(3-2) emission of the QSO2s, even if measured from slightly lower angular resolution data (0\farcs6), shows similar morphology and kinematics than those of the CO(2-1) emission first presented in \cra. 
The peak intensity maps better reveal clumpy structures and double-peaked morphologies than the corresponding moment 0 maps.

\noindent\hspace{0.1cm}-The line ratios using the integrated CO(2-1) and CO(3-2) fluxes measured for our sample range from 0.22$<$\ratio$<$0.75, with an average value of <\ratio>=0.5$\pm$0.2, similar to the line ratios typically found in the galaxy disks of nearby star-forming galaxies. In principle, this implies no impact of the AGN on the global molecular gas excitation. However, our spatially resolved \ratio~maps reveal a gradient of excitation, with higher values of \ratio$\sim$0.7-0.9 within the central kiloparsec of the galaxies that decrease to values of $\sim$0.2-0.5 across the galaxy disks. The regions of higher excitation coincide with areas of high velocity dispersion, but we do not find a clear trend between the orientation of this enhancement of gas excitation and dispersion either along or perpendicular to the radio emission, with the exception of the Teacup (J1430), previously reported by \ane.

\noindent\hspace{0.1cm}-The combination of kinematic modelling with \barolo~and the \ratio~PVDs reveals that the high velocity regions associated with outflows have the highest line ratios seen in the PVD, exceeding \ratio$>$1. This indicates that either the high velocity gas presents higher gas excitation (higher temperatures) or it is optically thin.

\noindent\hspace{0.1cm}-Using three scenarios to compute the mass outflow rates, we found values ranging from \mdot=5-150\mout. 
These values lie below the \mdot-L$_{\rm bol}$ empirical relations reported in literature (e.g. \citealt{fiore17}), even when we consider the least conservative scenario \textsc{i}. 

\noindent\hspace{0.1cm}-We measured outflow kinetic powers in the range of 6.4$\times 10^{39}<\dot{E}_{\rm out}<1.2 \times 10^{43}$\ergs~for our QSO2s. The radiative coupling efficiencies ($\epsilon_{\rm AGN}\equiv \dot{E}_{\rm out}/L_{\rm bol}$) 
range from $10^{-6}<\epsilon_{\rm AGN}<10^{-4}$ when we use the kinetic powers derived from scenario \textsc{ii}, 
corresponding to coupling efficiencies of $\ll$0.1\%. 
On the other hand, the jet coupling efficiencies ($\epsilon_{\rm jet}\equiv \dot{E}_{\rm out}/P_{\rm jet}$) range from 0.001 < $\epsilon_{\rm jet}$ < 0.035 ($\epsilon_{\rm jet}\sim$0.1\%). 
These findings might indicate that, despite 
the high AGN luminosities of the QSO2s (L$_{\rm bol}\sim$10$^{45.5-46}$ erg~s$^{-1}$), 
compact low-power jets might be playing a key role in driving the molecular outflows. 

\noindent\hspace{0.1cm}-By comparing the cold molecular outflow properties with those derived for some of the QSO2s in the ionized \citep{speranza24} and warm molecular gas phases \citep{mavi25}, we find that the warm and cold molecular phases appear to be tracing the same outflow, while  
the warm ionized one does not, as it exhibits distinct orientation and kinematics.

To better understand the role of jets and winds on the distribution, kinematics, and gas excitation of molecular gas 
in ``radio quiet'' quasars, high-angular resolution CO and radio observations of larger samples are required. 
In particular, high angular resolution data of multiple CO transitions will be crucial to disentangle the outflow from regular kinematics, characterize the impact of winds and jets on gas with different molecular gas densities, and constrain the excitation mechanisms.



\begin{acknowledgements}

The authors acknowledge the anonymous referee for their constructive report. AA and CRA acknowledge support from the Agencia Estatal de Investigaci\'on of the Ministerio de Ciencia, Innovaci\'on y Universidades (MCIU/AEI) under the grant ``Tracking active galactic nuclei feedback from parsec to kiloparsec scales'', with reference PID2022$-$141105NB$-$I00 and the European Regional Development Fund (ERDF). They also acknowledge the projects ``Quantifying the impact of quasar feedback on galaxy evolution'', with reference EUR2020-112266, funded by MICINN-AEI/10.13039/501100011033 and the European Union NextGenerationEU/PRTR, and the Consejer\' ia de Econom\' ia, Conocimiento y Empleo del Gobierno de Canarias and the European Regional Development Fund (ERDF) under grant ``Quasar feedback and molecular gas reservoirs'', with reference ProID2020010105, ACCISI/FEDER, UE. AA acknowledges funding from the MICINN (Spain) through the Juan de la Cierva-Formación program under contract FJC2020-046224-I and from the European Union grant WIDERA ExGal-Twin, GA 101158446. SGB acknowledges support from the Spanish grant PID2022-138560NB-I00, funded by MCIN/AEI/10.13039/501100011033/FEDER, EU. MPS acknowledges support under grants RYC2021-033094-I, CNS2023-145506, and PID2023-146667NB-I00 funded by MCIN/AEI/10.13039/501100011033 and the European Union NextGenerationEU/PRTR. IL acknowledges support from PRIN-MUR project "PROMETEUS" (202223XPZM). This paper makes use of the following ALMA data: ADS/JAO.ALMA\#2018.1.00870.S, and ADS/JAO.ALMA\#2016.1.01535.S. ALMA is a partnership of ESO (representing its member states), NSF (USA) and NINS (Japan), together with NRC(Canada) and NSC and ASIAA (Taiwan), in cooperation with the Republic of Chile. The Joint ALMA Observatory is operated by ESO, AUI/NRAO and NAOJ. The National Radio Astronomy Observatory is a facility of the National Science Foundation operated under cooperative agreement by Associated Universities, Inc. 
\end{acknowledgements}

\bibliographystyle{aa} 
\bibliography{REF.bib}

\begin{thebibliography}{125}
\expandafter\ifx\csname natexlab\endcsname\relax\def\natexlab#1{#1}\fi

\bibitem[{{Audibert} {et~al.}(2019){Audibert}, {Combes}, {Garc{\'\i}a-Burillo},
  {Hunt}, {Eckart}, {Aalto}, {Casasola}, {Boone}, {Krips}, {Viti}, {Muller},
  {Dasyra}, {van der Werf}, \& {Mart{\'\i}n}}]{ane19}
{Audibert}, A., {Combes}, F., {Garc{\'\i}a-Burillo}, S., {et~al.} 2019, \aap,
  632, A33

\bibitem[{{Audibert} {et~al.}(2021){Audibert}, {Combes}, {Garc{\'\i}a-Burillo},
  {Hunt}, {Eckart}, {Aalto}, {Casasola}, {Boone}, {Krips}, {Viti}, {Muller},
  {Dasyra}, {van der Werf}, \& {Mart{\'\i}n}}]{ane21}
{Audibert}, A., {Combes}, F., {Garc{\'\i}a-Burillo}, S., {et~al.} 2021, \aap,
  656, A60

\bibitem[{{Audibert} {et~al.}(2023){Audibert}, {Ramos Almeida},
  {Garc{\'\i}a-Burillo}, {Combes}, {Bischetti}, {Meenakshi}, {Mukherjee},
  {Bicknell}, \& {Wagner}}]{ane23}
{Audibert}, A., {Ramos Almeida}, C., {Garc{\'\i}a-Burillo}, S., {et~al.} 2023,
  \aap, 671, L12

\bibitem[{{Becker} {et~al.}(1995){Becker}, {White}, \& {Helfand}}]{becker95}
{Becker}, R.~H., {White}, R.~L., \& {Helfand}, D.~J. 1995, \apj, 450, 559

\bibitem[{{Bell}(2003)}]{bell03}
{Bell}, E.~F. 2003, \apj, 586, 794

\bibitem[{{Bessiere} {et~al.}(2024){Bessiere}, {Ramos Almeida}, {Holden},
  {Tadhunter}, \& {Canalizo}}]{bessiere24}
{Bessiere}, P.~S., {Ramos Almeida}, C., {Holden}, L.~R., {Tadhunter}, C.~N., \&
  {Canalizo}, G. 2024, arXiv e-prints, arXiv:2405.06421

\bibitem[{{Bieri} {et~al.}(2017){Bieri}, {Dubois}, {Rosdahl}, {Wagner}, {Silk},
  \& {Mamon}}]{bieri17}
{Bieri}, R., {Dubois}, Y., {Rosdahl}, J., {et~al.} 2017, \mnras, 464, 1854

\bibitem[{{B{\^i}rzan} {et~al.}(2008){B{\^i}rzan}, {McNamara}, {Nulsen},
  {Carilli}, \& {Wise}}]{birzan08}
{B{\^i}rzan}, L., {McNamara}, B.~R., {Nulsen}, P.~E.~J., {Carilli}, C.~L., \&
  {Wise}, M.~W. 2008, \apj, 686, 859

\bibitem[{{Bolatto} {et~al.}(2013){Bolatto}, {Wolfire}, \& {Leroy}}]{bolatto13}
{Bolatto}, A.~D., {Wolfire}, M., \& {Leroy}, A.~K. 2013, \araa, 51, 207

\bibitem[{{Bournaud} {et~al.}(2008){Bournaud}, {Duc}, \&
  {Emsellem}}]{bournaud08}
{Bournaud}, F., {Duc}, P.~A., \& {Emsellem}, E. 2008, \mnras, 389, L8

\bibitem[{{Bournaud} {et~al.}(2011){Bournaud}, {Powell}, {Chapon}, \&
  {Teyssier}}]{bournaud11}
{Bournaud}, F., {Powell}, L.~C., {Chapon}, D., \& {Teyssier}, R. 2011, in
  Astrophysical Dynamics: From Stars to Galaxies, ed. N.~H. {Brummell}, A.~S.
  {Brun}, M.~S. {Miesch}, \& Y.~{Ponty}, Vol. 271, 160--169

\bibitem[{{Bourne} \& {Sijacki}(2017)}]{bourne17}
{Bourne}, M.~A. \& {Sijacki}, D. 2017, \mnras, 472, 4707

\bibitem[{{Bourne} \& {Yang}(2023)}]{bourne23}
{Bourne}, M.~A. \& {Yang}, H.-Y.~K. 2023, Galaxies, 11, 73

\bibitem[{{Carilli} \& {Walter}(2013)}]{carilli13}
{Carilli}, C.~L. \& {Walter}, F. 2013, \araa, 51, 105

\bibitem[{{Cicone} {et~al.}(2018){Cicone}, {Brusa}, {Ramos Almeida}, {Cresci},
  {Husemann}, \& {Mainieri}}]{cicone18}
{Cicone}, C., {Brusa}, M., {Ramos Almeida}, C., {et~al.} 2018, Nature
  Astronomy, 2, 176

\bibitem[{{Cicone} {et~al.}(2014){Cicone}, {Maiolino}, {Sturm},
  {Graci{\'a}-Carpio}, {Feruglio}, {Neri}, {Aalto}, {Davies}, {Fiore},
  {Fischer}, {Garc{\'\i}a-Burillo}, {Gonz{\'a}lez-Alfonso}, {Hailey-Dunsheath},
  {Piconcelli}, \& {Veilleux}}]{cicone14}
{Cicone}, C., {Maiolino}, R., {Sturm}, E., {et~al.} 2014, \aap, 562, A21

\bibitem[{{Combes} {et~al.}(2013){Combes}, {Garc{\'\i}a-Burillo}, {Casasola},
  {Hunt}, {Krips}, {Baker}, {Boone}, {Eckart}, {Marquez}, {Neri}, {Schinnerer},
  \& {Tacconi}}]{combes13}
{Combes}, F., {Garc{\'\i}a-Burillo}, S., {Casasola}, V., {et~al.} 2013, \aap,
  558, A124

\bibitem[{{Costa} {et~al.}(2018){Costa}, {Rosdahl}, {Sijacki}, \&
  {Haehnelt}}]{costa18}
{Costa}, T., {Rosdahl}, J., {Sijacki}, D., \& {Haehnelt}, M.~G. 2018, \mnras,
  473, 4197

\bibitem[{{Croton} {et~al.}(2006){Croton}, {Springel}, {White}, {De Lucia},
  {Frenk}, {Gao}, {Jenkins}, {Kauffmann}, {Navarro}, \& {Yoshida}}]{croton06}
{Croton}, D.~J., {Springel}, V., {White}, S. D.~M., {et~al.} 2006, \mnras, 365,
  11

\bibitem[{{Dasyra} {et~al.}(2016){Dasyra}, {Combes}, {Oosterloo}, {Oonk},
  {Morganti}, {Salom{\'e}}, \& {Vlahakis}}]{dasyra16}
{Dasyra}, K.~M., {Combes}, F., {Oosterloo}, T., {et~al.} 2016, \aap, 595, L7

\bibitem[{{Di Matteo} {et~al.}(2008){Di Matteo}, {Colberg}, {Springel},
  {Hernquist}, \& {Sijacki}}]{dimatteo08}
{Di Matteo}, T., {Colberg}, J., {Springel}, V., {Hernquist}, L., \& {Sijacki},
  D. 2008, \apj, 676, 33

\bibitem[{{Di Teodoro} \& {Fraternali}(2015)}]{barolo15}
{Di Teodoro}, E.~M. \& {Fraternali}, F. 2015, \mnras, 451, 3021

\bibitem[{{Downes} \& {Solomon}(1998)}]{downes98}
{Downes}, D. \& {Solomon}, P.~M. 1998, \apj, 507, 615

\bibitem[{{Dubois} {et~al.}(2016){Dubois}, {Peirani}, {Pichon}, {Devriendt},
  {Gavazzi}, {Welker}, \& {Volonteri}}]{dubois16}
{Dubois}, Y., {Peirani}, S., {Pichon}, C., {et~al.} 2016, \mnras, 463, 3948

\bibitem[{{Dubois} {et~al.}(2014){Dubois}, {Pichon}, {Welker}, {Le Borgne},
  {Devriendt}, {Laigle}, {Codis}, {Pogosyan}, {Arnouts}, {Benabed}, {Bertin},
  {Blaizot}, {Bouchet}, {Cardoso}, {Colombi}, {de Lapparent}, {Desjacques},
  {Gavazzi}, {Kassin}, {Kimm}, {McCracken}, {Milliard}, {Peirani}, {Prunet},
  {Rouberol}, {Silk}, {Slyz}, {Sousbie}, {Teyssier}, {Tresse}, {Treyer},
  {Vibert}, \& {Volonteri}}]{dubois14}
{Dubois}, Y., {Pichon}, C., {Welker}, C., {et~al.} 2014, \mnras, 444, 1453

\bibitem[{{Esposito} {et~al.}(2024){Esposito}, {Vallini}, {Pozzi}, {Casasola},
  {Alonso-Herrero}, {Garc{\'\i}a-Burillo}, {Decarli}, {Calura}, {Vignali},
  {Mingozzi}, {Gruppioni}, \& {Sengupta}}]{esposito24}
{Esposito}, F., {Vallini}, L., {Pozzi}, F., {et~al.} 2024, \mnras, 527, 8727

\bibitem[{{Esposito} {et~al.}(2022){Esposito}, {Vallini}, {Pozzi}, {Casasola},
  {Mingozzi}, {Vignali}, {Gruppioni}, \& {Salvestrini}}]{esposito22}
{Esposito}, F., {Vallini}, L., {Pozzi}, F., {et~al.} 2022, \mnras, 512, 686

\bibitem[{{Fabian}(2012)}]{fabian12}
{Fabian}, A.~C. 2012, \araa, 50, 455

\bibitem[{{Fei} {et~al.}(2024){Fei}, {Wang}, {Molina}, {Ho}, {Shangguan},
  {Bauer}, \& {Treister}}]{fei24}
{Fei}, Q., {Wang}, R., {Molina}, J., {et~al.} 2024, arXiv e-prints,
  arXiv:2409.15611

\bibitem[{{Feruglio} {et~al.}(2020){Feruglio}, {Fabbiano}, {Bischetti},
  {Elvis}, {Travascio}, \& {Fiore}}]{feruglio20}
{Feruglio}, C., {Fabbiano}, G., {Bischetti}, M., {et~al.} 2020, \apj, 890, 29

\bibitem[{{Feruglio} {et~al.}(2010){Feruglio}, {Maiolino}, {Piconcelli},
  {Menci}, {Aussel}, {Lamastra}, \& {Fiore}}]{feruglio10}
{Feruglio}, C., {Maiolino}, R., {Piconcelli}, E., {et~al.} 2010, \aap, 518,
  L155

\bibitem[{{Fiore} {et~al.}(2017){Fiore}, {Feruglio}, {Shankar}, {Bischetti},
  {Bongiorno}, {Brusa}, {Carniani}, {Cicone}, {Duras}, {Lamastra}, {Mainieri},
  {Marconi}, {Menci}, {Maiolino}, {Piconcelli}, {Vietri}, \&
  {Zappacosta}}]{fiore17}
{Fiore}, F., {Feruglio}, C., {Shankar}, F., {et~al.} 2017, \aap, 601, A143

\bibitem[{{Fischer} {et~al.}(2019){Fischer}, {Smith}, {Kraemer}, {Schmitt},
  {Crenshaw}, {Koss}, {Mushotzky}, {Larson}, {U}, \& {Rigby}}]{fischer19}
{Fischer}, T., {Smith}, K.~L., {Kraemer}, S., {et~al.} 2019, \apj, 887, 200

\bibitem[{{Fischer} {et~al.}(2023){Fischer}, {Johnson}, {Secrest}, {Crenshaw},
  \& {Kraemer}}]{fischer23}
{Fischer}, T.~C., {Johnson}, M.~C., {Secrest}, N.~J., {Crenshaw}, D.~M., \&
  {Kraemer}, S.~B. 2023, \apj, 953, 87

\bibitem[{{Fluetsch} {et~al.}(2019){Fluetsch}, {Maiolino}, {Carniani},
  {Marconi}, {Cicone}, {Bourne}, {Costa}, {Fabian}, {Ishibashi}, \&
  {Venturi}}]{fluetsch19}
{Fluetsch}, A., {Maiolino}, R., {Carniani}, S., {et~al.} 2019, \mnras, 483,
  4586

\bibitem[{{Fotopoulou} {et~al.}(2019){Fotopoulou}, {Dasyra}, {Combes},
  {Salom{\'e}}, \& {Papachristou}}]{fotopoulou19}
{Fotopoulou}, C.~M., {Dasyra}, K.~M., {Combes}, F., {Salom{\'e}}, P., \&
  {Papachristou}, M. 2019, \aap, 629, A30

\bibitem[{{Garc{\'\i}a-Burillo} {et~al.}(2021){Garc{\'\i}a-Burillo},
  {Alonso-Herrero}, {Ramos Almeida}, {Gonz{\'a}lez-Mart{\'\i}n}, {Combes},
  {Usero}, {H{\"o}nig}, {Querejeta}, {Hicks}, {Hunt}, {Rosario}, {Davies},
  {Boorman}, {Bunker}, {Burtscher}, {Colina}, {D{\'\i}az-Santos}, {Gandhi},
  {Garc{\'\i}a-Bernete}, {Garc{\'\i}a-Lorenzo}, {Ichikawa}, {Imanishi},
  {Izumi}, {Labiano}, {Levenson}, {L{\'o}pez-Rodr{\'\i}guez}, {Packham},
  {Pereira-Santaella}, {Ricci}, {Rigopoulou}, {Rouan}, {Shimizu}, {Stalevski},
  {Wada}, \& {Williamson}}]{santi21}
{Garc{\'\i}a-Burillo}, S., {Alonso-Herrero}, A., {Ramos Almeida}, C., {et~al.}
  2021, \aap, 652, A98

\bibitem[{{Garc{\'\i}a-Burillo} {et~al.}(2019){Garc{\'\i}a-Burillo}, {Combes},
  {Ramos Almeida}, {Usero}, {Alonso-Herrero}, {Hunt}, {Rouan}, {Aalto},
  {Querejeta}, {Viti}, {van der Werf}, {Vives-Arias}, {Fuente}, {Colina},
  {Mart{\'\i}n-Pintado}, {Henkel}, {Mart{\'\i}n}, {Krips}, {Gratadour}, {Neri},
  \& {Tacconi}}]{santi19}
{Garc{\'\i}a-Burillo}, S., {Combes}, F., {Ramos Almeida}, C., {et~al.} 2019,
  \aap, 632, A61

\bibitem[{{Garc{\'\i}a-Burillo} {et~al.}(2015){Garc{\'\i}a-Burillo}, {Combes},
  {Usero}, {Aalto}, {Colina}, {Alonso-Herrero}, {Hunt}, {Arribas},
  {Costagliola}, {Labiano}, {Neri}, {Pereira-Santaella}, {Tacconi}, \& {van der
  Werf}}]{santi15}
{Garc{\'\i}a-Burillo}, S., {Combes}, F., {Usero}, A., {et~al.} 2015, \aap, 580,
  A35

\bibitem[{{Garc{\'\i}a-Burillo} {et~al.}(2014){Garc{\'\i}a-Burillo}, {Combes},
  {Usero}, {Aalto}, {Krips}, {Viti}, {Alonso-Herrero}, {Hunt}, {Schinnerer},
  {Baker}, {Boone}, {Casasola}, {Colina}, {Costagliola}, {Eckart}, {Fuente},
  {Henkel}, {Labiano}, {Mart{\'\i}n}, {M{\'a}rquez}, {Muller}, {Planesas},
  {Ramos Almeida}, {Spaans}, {Tacconi}, \& {van der Werf}}]{santi14}
{Garc{\'\i}a-Burillo}, S., {Combes}, F., {Usero}, A., {et~al.} 2014, \aap, 567,
  A125

\bibitem[{{Garc{\'\i}a-Burillo} {et~al.}(2024){Garc{\'\i}a-Burillo}, {Hicks},
  {Alonso-Herrero}, {Pereira-Santaella}, {Usero}, {Querejeta},
  {Gonz{\'a}lez-Mart{\'\i}n}, {Delaney}, {Ramos Almeida}, {Combes},
  {Angl{\'e}s-Alc{\'a}zar}, {Audibert}, {Bellocchi}, {Davies}, {Davis},
  {Elford}, {Garc{\'\i}a-Bernete}, {H{\"o}nig}, {Labiano}, {Leist}, {Levenson},
  {L{\'o}pez-Rodr{\'\i}guez}, {Mercedes-Feliz}, {Packham}, {Ricci}, {Rosario},
  {Shimizu}, {Stalevski}, \& {Zhang}}]{santi24}
{Garc{\'\i}a-Burillo}, S., {Hicks}, E.~K.~S., {Alonso-Herrero}, A., {et~al.}
  2024, \aap, 689, A347

\bibitem[{{Girdhar} {et~al.}(2022){Girdhar}, {Harrison}, {Mainieri}, {Bittner},
  {Costa}, {Kharb}, {Mukherjee}, {Arrigoni Battaia}, {Alexander}, {Calistro
  Rivera}, {Circosta}, {De Breuck}, {Edge}, {Farina}, {Kakkad}, {Lansbury},
  {Molyneux}, {Mullaney}, {S}, {Thomson}, \& {Ward}}]{girdhar22}
{Girdhar}, A., {Harrison}, C.~M., {Mainieri}, V., {et~al.} 2022, \mnras, 512,
  1608

\bibitem[{{Girdhar} {et~al.}(2024){Girdhar}, {Harrison}, {Mainieri},
  {Fern{\'a}ndez Aranda}, {Alexander}, {Arrigoni Battaia}, {Bianchin},
  {Calistro Rivera}, {Circosta}, {Costa}, {Edge}, {Farina}, {Kakkad}, {Kharb},
  {Molyneux}, {Mukherjee}, {Njeri}, {Silpa}, {Venturi}, \& {Ward}}]{girdhar24}
{Girdhar}, A., {Harrison}, C.~M., {Mainieri}, V., {et~al.} 2024, \mnras, 527,
  9322

\bibitem[{{Harrison}(2017)}]{harrison17}
{Harrison}, C.~M. 2017, Nature Astronomy, 1, 0165

\bibitem[{{Harrison} {et~al.}(2018){Harrison}, {Costa}, {Tadhunter},
  {Fl{\"u}tsch}, {Kakkad}, {Perna}, \& {Vietri}}]{harrison18}
{Harrison}, C.~M., {Costa}, T., {Tadhunter}, C.~N., {et~al.} 2018, Nature
  Astronomy, 2, 198

\bibitem[{{Harrison} \& {Ramos Almeida}(2024)}]{harrison24}
{Harrison}, C.~M. \& {Ramos Almeida}, C. 2024, Galaxies, 12, 17

\bibitem[{{Hervella Seoane} {et~al.}(2023){Hervella Seoane}, {Ramos Almeida},
  {Acosta-Pulido}, {Speranza}, {Tadhunter}, \& {Bessiere}}]{kiara23}
{Hervella Seoane}, K., {Ramos Almeida}, C., {Acosta-Pulido}, J.~A., {et~al.}
  2023, \aap, 680, A71

\bibitem[{{Hickox} {et~al.}(2014){Hickox}, {Mullaney}, {Alexander}, {Chen},
  {Civano}, {Goulding}, \& {Hainline}}]{hickox14}
{Hickox}, R.~C., {Mullaney}, J.~R., {Alexander}, D.~M., {et~al.} 2014, \apj,
  782, 9

\bibitem[{Hill(2016)}]{hill16}
Hill, C. 2016, Learning Scientific Programming with Python (Cambridge
  University Press)

\bibitem[{{Holden} {et~al.}(2024){Holden}, {Tadhunter}, {Audibert},
  {Oosterloo}, {Ramos Almeida}, {Morganti}, {Pereira-Santaella}, \&
  {Lamperti}}]{luke24}
{Holden}, L.~R., {Tadhunter}, C., {Audibert}, A., {et~al.} 2024, \mnras, 530,
  446

\bibitem[{{Hollenbach} \& {Tielens}(1999)}]{hollenbach99}
{Hollenbach}, D.~J. \& {Tielens}, A.~G.~G.~M. 1999, Reviews of Modern Physics,
  71, 173

\bibitem[{{Ishibashi} \& {Fabian}(2015)}]{ishibashi15}
{Ishibashi}, W. \& {Fabian}, A.~C. 2015, \mnras, 451, 93

\bibitem[{{Jarvis} {et~al.}(2021){Jarvis}, {Harrison}, {Mainieri}, {Alexander},
  {Arrigoni Battaia}, {Calistro Rivera}, {Circosta}, {Costa}, {De Breuck},
  {Edge}, {Girdhar}, {Kakkad}, {Kharb}, {Lansbury}, {Molyneux}, {Mukherjee},
  {Mullaney}, {Farina}, {Silpa}, {Thomson}, \& {Ward}}]{jarvis21}
{Jarvis}, M.~E., {Harrison}, C.~M., {Mainieri}, V., {et~al.} 2021, \mnras, 503,
  1780

\bibitem[{{Jarvis} {et~al.}(2019){Jarvis}, {Harrison}, {Thomson}, {Circosta},
  {Mainieri}, {Alexander}, {Edge}, {Lansbury}, {Molyneux}, \&
  {Mullaney}}]{jarvis19}
{Jarvis}, M.~E., {Harrison}, C.~M., {Thomson}, A.~P., {et~al.} 2019, \mnras,
  485, 2710

\bibitem[{{Kellermann} {et~al.}(2004){Kellermann}, {Lister}, {Homan},
  {Vermeulen}, {Cohen}, {Ros}, {Kadler}, {Zensus}, \& {Kovalev}}]{kellermann04}
{Kellermann}, K.~I., {Lister}, M.~L., {Homan}, D.~C., {et~al.} 2004, \apj, 609,
  539

\bibitem[{{King}(2003)}]{king03}
{King}, A. 2003, \apjl, 596, L27

\bibitem[{{King} \& {Nixon}(2015)}]{king15}
{King}, A. \& {Nixon}, C. 2015, \mnras, 453, L46

\bibitem[{{Kong} \& {Ho}(2018)}]{kong18}
{Kong}, M. \& {Ho}, L.~C. 2018, \apj, 859, 116

\bibitem[{{Lamastra} {et~al.}(2009){Lamastra}, {Bianchi}, {Matt}, {Perola},
  {Barcons}, \& {Carrera}}]{lamastra09}
{Lamastra}, A., {Bianchi}, S., {Matt}, G., {et~al.} 2009, \aap, 504, 73

\bibitem[{{Lamperti} {et~al.}(2022){Lamperti}, {Pereira-Santaella}, {Perna},
  {Colina}, {Arribas}, {Garc{\'\i}a-Burillo}, {Gonz{\'a}lez-Alfonso}, {Aalto},
  {Alonso-Herrero}, {Combes}, {Labiano}, {Piqueras-L{\'o}pez}, {Rigopoulou}, \&
  {van der Werf}}]{lamperti22}
{Lamperti}, I., {Pereira-Santaella}, M., {Perna}, M., {et~al.} 2022, \aap, 668,
  A45

\bibitem[{{Lamperti} {et~al.}(2020){Lamperti}, {Saintonge}, {Koss}, {Viti},
  {Wilson}, {He}, {Shimizu}, {Greve}, {Mushotzky}, {Treister}, {Kramer},
  {Sanders}, {Schawinski}, \& {Tacconi}}]{lamperti20}
{Lamperti}, I., {Saintonge}, A., {Koss}, M., {et~al.} 2020, \apj, 889, 103

\bibitem[{{Leroy} {et~al.}(2022){Leroy}, {Rosolowsky}, {Usero}, {Sandstrom},
  {Schinnerer}, {Schruba}, {Bolatto}, {Sun}, {Barnes}, {Belfiore}, {Bigiel},
  {den Brok}, {Cao}, {Chiang}, {Chevance}, {Dale}, {Eibensteiner}, {Faesi},
  {Glover}, {Hughes}, {Jim{\'e}nez Donaire}, {Klessen}, {Koch}, {Kruijssen},
  {Liu}, {Meidt}, {Pan}, {Pety}, {Puschnig}, {Querejeta}, {Saito}, {Sardone},
  {Watkins}, {Weiss}, \& {Williams}}]{leroy22}
{Leroy}, A.~K., {Rosolowsky}, E., {Usero}, A., {et~al.} 2022, \apj, 927, 149

\bibitem[{{Lutz} {et~al.}(2020){Lutz}, {Sturm}, {Janssen}, {Veilleux}, {Aalto},
  {Cicone}, {Contursi}, {Davies}, {Feruglio}, {Fischer}, {Fluetsch},
  {Garcia-Burillo}, {Genzel}, {Gonz{\'a}lez-Alfonso}, {Graci{\'a}-Carpio},
  {Herrera-Camus}, {Maiolino}, {Schruba}, {Shimizu}, {Sternberg}, {Tacconi}, \&
  {Wei{\ss}}}]{lutz20}
{Lutz}, D., {Sturm}, E., {Janssen}, A., {et~al.} 2020, \aap, 633, A134

\bibitem[{{Mandal} {et~al.}(2021){Mandal}, {Mukherjee}, {Federrath},
  {Nesvadba}, {Bicknell}, {Wagner}, \& {Meenakshi}}]{mandal21}
{Mandal}, A., {Mukherjee}, D., {Federrath}, C., {et~al.} 2021, \mnras, 508,
  4738

\bibitem[{{Martini}(2004)}]{martini04}
{Martini}, P. 2004, in Coevolution of Black Holes and Galaxies, ed. L.~C. {Ho},
  169

\bibitem[{{McMullin} {et~al.}(2007){McMullin}, {Waters}, {Schiebel}, {Young},
  \& {Golap}}]{casa}
{McMullin}, J.~P., {Waters}, B., {Schiebel}, D., {Young}, W., \& {Golap}, K.
  2007, in Astronomical Society of the Pacific Conference Series, Vol. 376,
  Astronomical Data Analysis Software and Systems XVI, ed. R.~A. {Shaw},
  F.~{Hill}, \& D.~J. {Bell}, 127

\bibitem[{{McNamara} \& {Nulsen}(2012)}]{mcnamara12}
{McNamara}, B.~R. \& {Nulsen}, P.~E.~J. 2012, New Journal of Physics, 14,
  055023

\bibitem[{{Meenakshi} {et~al.}(2022){Meenakshi}, {Mukherjee}, {Wagner},
  {Nesvadba}, {Bicknell}, {Morganti}, {Janssen}, {Sutherland}, \&
  {Mandal}}]{meenakshi22}
{Meenakshi}, M., {Mukherjee}, D., {Wagner}, A.~Y., {et~al.} 2022, \mnras
  [\eprint[arXiv]{2203.10251}]

\bibitem[{{Meijerink} {et~al.}(2013){Meijerink}, {Kristensen}, {Wei{\ss}}, {van
  der Werf}, {Walter}, {Spaans}, {Loenen}, {Fischer}, {Israel}, {Isaak},
  {Papadopoulos}, {Aalto}, {Armus}, {Charmandaris}, {Dasyra}, {Diaz-Santos},
  {Evans}, {Gao}, {Gonz{\'a}lez-Alfonso}, {G{\"u}sten}, {Henkel}, {Kramer},
  {Lord}, {Mart{\'\i}n-Pintado}, {Naylor}, {Sanders}, {Smith}, {Spinoglio},
  {Stacey}, {Veilleux}, \& {Wiedner}}]{meijerink13}
{Meijerink}, R., {Kristensen}, L.~E., {Wei{\ss}}, A., {et~al.} 2013, \apjl,
  762, L16

\bibitem[{{Meijerink} {et~al.}(2007){Meijerink}, {Spaans}, \&
  {Israel}}]{meijerink07}
{Meijerink}, R., {Spaans}, M., \& {Israel}, F.~P. 2007, \aap, 461, 793

\bibitem[{{Mingozzi} {et~al.}(2018){Mingozzi}, {Vallini}, {Pozzi}, {Vignali},
  {Mignano}, {Gruppioni}, {Talia}, {Cimatti}, {Cresci}, \&
  {Massardi}}]{mingozi18}
{Mingozzi}, M., {Vallini}, L., {Pozzi}, F., {et~al.} 2018, \mnras, 474, 3640

\bibitem[{{Molyneux} {et~al.}(2024){Molyneux}, {Calistro Rivera}, {De Breuck},
  {Harrison}, {Mainieri}, {Lundgren}, {Kakkad}, {Circosta}, {Girdhar}, {Costa},
  {Mullaney}, {Kharb}, {Arrigoni Battaia}, {Farina}, {Alexander}, {Ward},
  {Silpa}, \& {Smit}}]{moly24}
{Molyneux}, S.~J., {Calistro Rivera}, G., {De Breuck}, C., {et~al.} 2024,
  \mnras, 527, 4420

\bibitem[{{Molyneux} {et~al.}(2019){Molyneux}, {Harrison}, \&
  {Jarvis}}]{moly19}
{Molyneux}, S.~J., {Harrison}, C.~M., \& {Jarvis}, M.~E. 2019, \aap, 631, A132

\bibitem[{{Montoya Arroyave} {et~al.}(2023){Montoya Arroyave}, {Cicone},
  {Makroleivaditi}, {Weiss}, {Lundgren}, {Severgnini}, {De Breuck},
  {Baumschlager}, {Schimek}, {Shen}, \& {Aravena}}]{montoya23}
{Montoya Arroyave}, I., {Cicone}, C., {Makroleivaditi}, E., {et~al.} 2023,
  \aap, 673, A13

\bibitem[{{Morabito} {et~al.}(2022){Morabito}, {Jackson}, {Mooney}, {Sweijen},
  {Badole}, {Kukreti}, {Venkattu}, {Groeneveld}, {Kappes}, {Bonnassieux},
  {Drabent}, {Iacobelli}, {Croston}, {Best}, {Bondi}, {Callingham}, {Conway},
  {Deller}, {Hardcastle}, {McKean}, {Miley}, {Moldon}, {R{\"o}ttgering},
  {Tasse}, {Shimwell}, {van Weeren}, {Anderson}, {Asgekar}, {Avruch}, {van
  Bemmel}, {Bentum}, {Bonafede}, {Brouw}, {Butcher}, {Ciardi}, {Corstanje},
  {Coolen}, {Damstra}, {de Gasperin}, {Duscha}, {Eisl{\"o}ffel}, {Engels},
  {Falcke}, {Garrett}, {Griessmeier}, {Gunst}, {van Haarlem}, {Hoeft}, {van der
  Horst}, {J{\"u}tte}, {Kadler}, {Koopmans}, {Krankowski}, {Mann}, {Nelles},
  {Oonk}, {Orru}, {Paas}, {Pandey}, {Pizzo}, {Pandey-Pommier}, {Reich},
  {Rothkaehl}, {Ruiter}, {Schwarz}, {Shulevski}, {Soida}, {Tagger}, {Vocks},
  {Wijers}, {Wijnholds}, {Wucknitz}, {Zarka}, \& {Zucca}}]{morabito22}
{Morabito}, L.~K., {Jackson}, N.~J., {Mooney}, S., {et~al.} 2022, \aap, 658, A1

\bibitem[{{Morganti} {et~al.}(2013){Morganti}, {Fogasy}, {Paragi}, {Oosterloo},
  \& {Orienti}}]{morganti13}
{Morganti}, R., {Fogasy}, J., {Paragi}, Z., {Oosterloo}, T., \& {Orienti}, M.
  2013, Science, 341, 1082

\bibitem[{{Morganti} {et~al.}(2021){Morganti}, {Oosterloo}, {Murthy}, \&
  {Tadhunter}}]{morganti21}
{Morganti}, R., {Oosterloo}, T., {Murthy}, S., \& {Tadhunter}, C. 2021,
  Astronomische Nachrichten, 342, 1135

\bibitem[{{Morganti} {et~al.}(2015){Morganti}, {Oosterloo}, {Oonk},
  {Frieswijk}, \& {Tadhunter}}]{morganti15}
{Morganti}, R., {Oosterloo}, T., {Oonk}, J.~B.~R., {Frieswijk}, W., \&
  {Tadhunter}, C. 2015, \aap, 580, A1

\bibitem[{{Mukherjee} {et~al.}(2016){Mukherjee}, {Bicknell}, {Sutherland}, \&
  {Wagner}}]{mukherjee16}
{Mukherjee}, D., {Bicknell}, G.~V., {Sutherland}, R., \& {Wagner}, A. 2016,
  \mnras, 461, 967

\bibitem[{{Mukherjee} {et~al.}(2018){Mukherjee}, {Bicknell}, {Wagner},
  {Sutherland}, \& {Silk}}]{mukherjee18sim}
{Mukherjee}, D., {Bicknell}, G.~V., {Wagner}, A.~Y., {Sutherland}, R.~S., \&
  {Silk}, J. 2018, \mnras, 479, 5544

\bibitem[{{Mullaney} {et~al.}(2013){Mullaney}, {Alexander}, {Fine}, {Goulding},
  {Harrison}, \& {Hickox}}]{mullaney13}
{Mullaney}, J.~R., {Alexander}, D.~M., {Fine}, S., {et~al.} 2013, \mnras, 433,
  622

\bibitem[{{Murthy} {et~al.}(2022){Murthy}, {Morganti}, {Wagner}, {Oosterloo},
  {Guillard}, {Mukherjee}, \& {Bicknell}}]{murthy22}
{Murthy}, S., {Morganti}, R., {Wagner}, A.~Y., {et~al.} 2022, Nature Astronomy,
  6, 488

\bibitem[{{Nelson} {et~al.}(2019){Nelson}, {Pillepich}, {Springel}, {Pakmor},
  {Weinberger}, {Genel}, {Torrey}, {Vogelsberger}, {Marinacci}, \&
  {Hernquist}}]{nelson19}
{Nelson}, D., {Pillepich}, A., {Springel}, V., {et~al.} 2019, \mnras, 490, 3234

\bibitem[{{Nyland} {et~al.}(2018){Nyland}, {Harwood}, {Mukherjee},
  {Jagannathan}, {Rujopakarn}, {Emonts}, {Alatalo}, {Bicknell}, {Davis},
  {Greene}, {Kimball}, {Lacy}, {Lonsdale}, {Lonsdale}, {Maksym}, {Moln{\'a}r},
  {Morabito}, {Murphy}, {Patil}, {Prandoni}, {Sargent}, \&
  {Vlahakis}}]{nyland18}
{Nyland}, K., {Harwood}, J.~J., {Mukherjee}, D., {et~al.} 2018, \apj, 859, 23

\bibitem[{{Oosterloo} {et~al.}(2019){Oosterloo}, {Morganti}, {Tadhunter},
  {Raymond Oonk}, {Bignall}, {Tzioumis}, \& {Reynolds}}]{oosterloo19}
{Oosterloo}, T., {Morganti}, R., {Tadhunter}, C., {et~al.} 2019, \aap, 632, A66

\bibitem[{{Oosterloo} {et~al.}(2017){Oosterloo}, {Raymond Oonk}, {Morganti},
  {Combes}, {Dasyra}, {Salom{\'e}}, {Vlahakis}, \& {Tadhunter}}]{oosterloo17}
{Oosterloo}, T., {Raymond Oonk}, J.~B., {Morganti}, R., {et~al.} 2017, \aap,
  608, A38

\bibitem[{{Padovani}(2016)}]{padovani16}
{Padovani}, P. 2016, \aapr, 24, 13

\bibitem[{{Panessa} {et~al.}(2019){Panessa}, {Baldi}, {Laor}, {Padovani},
  {Behar}, \& {McHardy}}]{panessa19}
{Panessa}, F., {Baldi}, R.~D., {Laor}, A., {et~al.} 2019, Nature Astronomy, 3,
  387

\bibitem[{{Papachristou} {et~al.}(2023){Papachristou}, {Dasyra},
  {Fern{\'a}ndez-Ontiveros}, {Audibert}, {Ruffa}, {Combes}, {Polkas}, \&
  {Gkogkou}}]{michalis23}
{Papachristou}, M., {Dasyra}, K.~M., {Fern{\'a}ndez-Ontiveros}, J.~A., {et~al.}
  2023, \aap, 679, A115

\bibitem[{{Pereira-Santaella} {et~al.}(2018){Pereira-Santaella}, {Colina},
  {Garc{\'\i}a-Burillo}, {Combes}, {Emonts}, {Aalto}, {Alonso-Herrero},
  {Arribas}, {Henkel}, {Labiano}, {Muller}, {Piqueras L{\'o}pez}, {Rigopoulou},
  \& {van der Werf}}]{pereira18}
{Pereira-Santaella}, M., {Colina}, L., {Garc{\'\i}a-Burillo}, S., {et~al.}
  2018, \aap, 616, A171

\bibitem[{{Pierce} {et~al.}(2023){Pierce}, {Tadhunter}, {Ramos Almeida},
  {Bessiere}, {Heaton}, {Ellison}, {Speranza}, {Gordon}, {O'Dea}, {Grimmett},
  \& {Makrygianni}}]{pierce23}
{Pierce}, J.~C.~S., {Tadhunter}, C., {Ramos Almeida}, C., {et~al.} 2023,
  \mnras, 522, 1736

\bibitem[{{Ramos Almeida} {et~al.}(2019){Ramos Almeida}, {Acosta-Pulido},
  {Tadhunter}, {Gonz{\'a}lez-Fern{\'a}ndez}, {Cicone}, \&
  {Fern{\'a}ndez-Torreiro}}]{cra19}
{Ramos Almeida}, C., {Acosta-Pulido}, J.~A., {Tadhunter}, C.~N., {et~al.} 2019,
  \mnras, 487, L18

\bibitem[{{Ramos Almeida} {et~al.}(2022){Ramos Almeida}, {Bischetti},
  {Garc{\'\i}a-Burillo}, {Alonso-Herrero}, {Audibert}, {Cicone}, {Feruglio},
  {Tadhunter}, {Pierce}, {Pereira-Santaella}, \& {Bessiere}}]{cra22}
{Ramos Almeida}, C., {Bischetti}, M., {Garc{\'\i}a-Burillo}, S., {et~al.} 2022,
  \aap, 658, A155

\bibitem[{{Ramos Almeida} {et~al.}(2025){Ramos Almeida}, {Garcia-Bernete},
  {Pereira-Santaella}, {Speranza}, {Maiolino}, {Ji}, {Audibert}, {Cezar},
  {Acosta-Pulido}, {Alonso-Herrero}, {Garcia-Burillo}, {Gonzalez-Martin},
  {Rigopoulou}, {Tadhunter}, {Labiano}, {Levenson}, \& {Donnan}}]{cra25}
{Ramos Almeida}, C., {Garcia-Bernete}, I., {Pereira-Santaella}, M., {et~al.}
  2025, arXiv e-prints, arXiv:2504.01595

\bibitem[{{Ramos Almeida} {et~al.}(2017){Ramos Almeida}, {Piqueras L{\'o}pez},
  {Villar-Mart{\'\i}n}, \& {Bessiere}}]{cra17}
{Ramos Almeida}, C., {Piqueras L{\'o}pez}, J., {Villar-Mart{\'\i}n}, M., \&
  {Bessiere}, P.~S. 2017, \mnras, 470, 964

\bibitem[{{Reyes} {et~al.}(2008){Reyes}, {Zakamska}, {Strauss}, {Green},
  {Krolik}, {Shen}, {Richards}, {Anderson}, \& {Schneider}}]{reyes08}
{Reyes}, R., {Zakamska}, N.~L., {Strauss}, M.~A., {et~al.} 2008, \aj, 136, 2373

\bibitem[{{Ruffa} {et~al.}(2022){Ruffa}, {Prandoni}, {Davis}, {Laing},
  {Paladino}, {Casasola}, {Parma}, \& {Bureau}}]{ruffa22}
{Ruffa}, I., {Prandoni}, I., {Davis}, T.~A., {et~al.} 2022, \mnras, 510, 4485

\bibitem[{{Rupke} {et~al.}(2005){Rupke}, {Veilleux}, \& {Sanders}}]{rupke05}
{Rupke}, D.~S., {Veilleux}, S., \& {Sanders}, D.~B. 2005, \apjs, 160, 115

\bibitem[{{Schawinski} {et~al.}(2015){Schawinski}, {Koss}, {Berney}, \&
  {Sartori}}]{schawinski15}
{Schawinski}, K., {Koss}, M., {Berney}, S., \& {Sartori}, L.~F. 2015, \mnras,
  451, 2517

\bibitem[{{Schaye} {et~al.}(2015){Schaye}, {Crain}, {Bower}, {Furlong},
  {Schaller}, {Theuns}, {Dalla Vecchia}, {Frenk}, {McCarthy}, {Helly},
  {Jenkins}, {Rosas-Guevara}, {White}, {Baes}, {Booth}, {Camps}, {Navarro},
  {Qu}, {Rahmati}, {Sawala}, {Thomas}, \& {Trayford}}]{schaye15}
{Schaye}, J., {Crain}, R.~A., {Bower}, R.~G., {et~al.} 2015, \mnras, 446, 521

\bibitem[{{Solomon} \& {Vanden Bout}(2005)}]{solomon05}
{Solomon}, P.~M. \& {Vanden Bout}, P.~A. 2005, \araa, 43, 677

\bibitem[{{Speranza} {et~al.}(2024){Speranza}, {Ramos Almeida},
  {Acosta-Pulido}, {Audibert}, {Holden}, {Tadhunter}, {Lapi},
  {Gonz{\'a}lez-Mart{\'\i}n}, {Brusa}, {L{\'o}pez}, {Musiimenta}, \&
  {Shankar}}]{speranza24}
{Speranza}, G., {Ramos Almeida}, C., {Acosta-Pulido}, J.~A., {et~al.} 2024,
  \aap, 681, A63

\bibitem[{{Speranza} {et~al.}(2022){Speranza}, {Ramos Almeida},
  {Acosta-Pulido}, {Riffel}, {Tadhunter}, {Pierce}, {Rodr{\'\i}guez-Ardila},
  {Coloma Puga}, {Brusa}, {Musiimenta}, {Alexander}, {Lapi}, {Shankar}, \&
  {Villforth}}]{speranza22}
{Speranza}, G., {Ramos Almeida}, C., {Acosta-Pulido}, J.~A., {et~al.} 2022,
  \aap, 665, A55

\bibitem[{{Stanghellini} {et~al.}(2005){Stanghellini}, {O'Dea}, {Dallacasa},
  {Cassaro}, {Baum}, {Fanti}, \& {Fanti}}]{stanghe05}
{Stanghellini}, C., {O'Dea}, C.~P., {Dallacasa}, D., {et~al.} 2005, \aap, 443,
  891

\bibitem[{{Sun} {et~al.}(2014){Sun}, {Greene}, {Zakamska}, \&
  {Nesvadba}}]{sun14}
{Sun}, A.-L., {Greene}, J.~E., {Zakamska}, N.~L., \& {Nesvadba}, N. P.~H. 2014,
  \apj, 790, 160

\bibitem[{{Tacconi} {et~al.}(2013){Tacconi}, {Neri}, {Genzel}, {Combes},
  {Bolatto}, {Cooper}, {Wuyts}, {Bournaud}, {Burkert}, {Comerford}, {Cox},
  {Davis}, {F{\"o}rster Schreiber}, {Garc{\'{\i}}a-Burillo}, {Gracia-Carpio},
  {Lutz}, {Naab}, {Newman}, {Omont}, {Saintonge}, {Shapiro Griffin}, {Shapley},
  {Sternberg}, \& {Weiner}}]{tacconi13}
{Tacconi}, L.~J., {Neri}, R., {Genzel}, R., {et~al.} 2013, \apj, 768, 74

\bibitem[{{Tadhunter} {et~al.}(2014){Tadhunter}, {Morganti}, {Rose}, {Oonk}, \&
  {Oosterloo}}]{tadhunter14}
{Tadhunter}, C., {Morganti}, R., {Rose}, M., {Oonk}, J.~B.~R., \& {Oosterloo},
  T. 2014, \nat, 511, 440

\bibitem[{{Talbot} {et~al.}(2022){Talbot}, {Sijacki}, \& {Bourne}}]{talbot22}
{Talbot}, R.~Y., {Sijacki}, D., \& {Bourne}, M.~A. 2022, \mnras, 514, 4535

\bibitem[{{Ulivi} {et~al.}(2024){Ulivi}, {Venturi}, {Cresci}, {Marconi},
  {Marconcini}, {Amiri}, {Belfiore}, {Bertola}, {Carniani}, {D'Amato}, {Di
  Teodoro}, {Ginolfi}, {Girdhar}, {Harrison}, {Maiolino}, {Mannucci},
  {Mingozzi}, {Perna}, {Scialpi}, {Tomicic}, {Tozzi}, \& {Treister}}]{ulivi24}
{Ulivi}, L., {Venturi}, G., {Cresci}, G., {et~al.} 2024, \aap, 685, A122

\bibitem[{{Valentini} {et~al.}(2020){Valentini}, {Murante}, {Borgani},
  {Granato}, {Monaco}, {Brighenti}, {Tornatore}, {Bressan}, \&
  {Lapi}}]{valentini20}
{Valentini}, M., {Murante}, G., {Borgani}, S., {et~al.} 2020, \mnras, 491, 2779

\bibitem[{{Vallini} {et~al.}(2019){Vallini}, {Tielens}, {Pallottini},
  {Gallerani}, {Gruppioni}, {Carniani}, {Pozzi}, \& {Talia}}]{vallini19}
{Vallini}, L., {Tielens}, A.~G.~G.~M., {Pallottini}, A., {et~al.} 2019, \mnras,
  490, 4502

\bibitem[{{van der Tak} {et~al.}(2007){van der Tak}, {Black}, {Sch{\"o}ier},
  {Jansen}, \& {van Dishoeck}}]{radex}
{van der Tak}, F.~F.~S., {Black}, J.~H., {Sch{\"o}ier}, F.~L., {Jansen}, D.~J.,
  \& {van Dishoeck}, E.~F. 2007, \aap, 468, 627

\bibitem[{{van der Werf} {et~al.}(2010){van der Werf}, {Isaak}, {Meijerink},
  {Spaans}, {Rykala}, {Fulton}, {Loenen}, {Walter}, {Wei{\ss}}, {Armus},
  {Fischer}, {Israel}, {Harris}, {Veilleux}, {Henkel}, {Savini}, {Lord},
  {Smith}, {Gonz{\'a}lez-Alfonso}, {Naylor}, {Aalto}, {Charmandaris}, {Dasyra},
  {Evans}, {Gao}, {Greve}, {G{\"u}sten}, {Kramer}, {Mart{\'\i}n-Pintado},
  {Mazzarella}, {Papadopoulos}, {Sanders}, {Spinoglio}, {Stacey}, {Vlahakis},
  {Wiedner}, \& {Xilouris}}]{vanderwerf10}
{van der Werf}, P.~P., {Isaak}, K.~G., {Meijerink}, R., {et~al.} 2010, \aap,
  518, L42

\bibitem[{{Veilleux} {et~al.}(2020){Veilleux}, {Maiolino}, {Bolatto}, \&
  {Aalto}}]{veilleux20}
{Veilleux}, S., {Maiolino}, R., {Bolatto}, A.~D., \& {Aalto}, S. 2020, \aapr,
  28, 2

\bibitem[{{Venturi} {et~al.}(2021){Venturi}, {Cresci}, {Marconi}, {Mingozzi},
  {Nardini}, {Carniani}, {Mannucci}, {Marasco}, {Maiolino}, {Perna},
  {Treister}, {Bland-Hawthorn}, \& {Gallimore}}]{venturi21}
{Venturi}, G., {Cresci}, G., {Marconi}, A., {et~al.} 2021, \aap, 648, A17

\bibitem[{{Venturi} {et~al.}(2023){Venturi}, {Treister}, {Finlez}, {D'Ago},
  {Bauer}, {Harrison}, {Ramos Almeida}, {Revalski}, {Ricci}, {Sartori},
  {Girdhar}, {Keel}, \& {Tub{\'\i}n}}]{venturi23}
{Venturi}, G., {Treister}, E., {Finlez}, C., {et~al.} 2023, \aap, 678, A127

\bibitem[{{Villar-Mart{\'\i}n} {et~al.}(2017){Villar-Mart{\'\i}n}, {Emonts},
  {Cabrera Lavers}, {Tadhunter}, {Mukherjee}, {Humphrey}, {Rodr{\'\i}guez
  Zaur{\'\i}n}, {Ramos Almeida}, {P{\'e}rez Torres}, \& {Bessiere}}]{vm17}
{Villar-Mart{\'\i}n}, M., {Emonts}, B., {Cabrera Lavers}, A., {et~al.} 2017,
  \mnras, 472, 4659

\bibitem[{{Viti} {et~al.}(2014){Viti}, {Garc{\'\i}a-Burillo}, {Fuente}, {Hunt},
  {Usero}, {Henkel}, {Eckart}, {Martin}, {Spaans}, {Muller}, {Combes}, {Krips},
  {Schinnerer}, {Casasola}, {Costagliola}, {Marquez}, {Planesas}, {van der
  Werf}, {Aalto}, {Baker}, {Boone}, \& {Tacconi}}]{viti14}
{Viti}, S., {Garc{\'\i}a-Burillo}, S., {Fuente}, A., {et~al.} 2014, \aap, 570,
  A28

\bibitem[{{Ward} {et~al.}(2024){Ward}, {Costa}, {Harrison}, \&
  {Mainieri}}]{Ward24}
{Ward}, S.~R., {Costa}, T., {Harrison}, C.~M., \& {Mainieri}, V. 2024, \mnras,
  533, 1733

\bibitem[{{Whitmore} {et~al.}(2014){Whitmore}, {Brogan}, {Chandar}, {Evans},
  {Hibbard}, {Johnson}, {Leroy}, {Privon}, {Remijan}, \& {Sheth}}]{whitmore14}
{Whitmore}, B.~C., {Brogan}, C., {Chandar}, R., {et~al.} 2014, \apj, 795, 156

\bibitem[{{Ye} {et~al.}(2023){Ye}, {Sweijen}, {van Weeren}, {Williams}, {de
  Jong}, {Morabito}, {Rottgering}, {Shimwell}, {Best}, {Bondi}, {Br{\"u}ggen},
  {de Gasperin}, \& {Tasse}}]{ye23}
{Ye}, H., {Sweijen}, F., {van Weeren}, R., {et~al.} 2023, arXiv e-prints,
  arXiv:2309.16560

\bibitem[{{Zakamska} \& {Greene}(2014)}]{zakamska14}
{Zakamska}, N.~L. \& {Greene}, J.~E. 2014, \mnras, 442, 784

\bibitem[{{Zanchettin} {et~al.}(2021){Zanchettin}, {Feruglio}, {Bischetti},
  {Malizia}, {Molina}, {Bongiorno}, {Dadina}, {Gruppioni}, {Piconcelli},
  {Tombesi}, {Travascio}, \& {Fiore}}]{mavi21}
{Zanchettin}, M.~V., {Feruglio}, C., {Bischetti}, M., {et~al.} 2021, \aap, 655,
  A25

\bibitem[{{Zanchettin} {et~al.}(2025){Zanchettin}, {Ramos Almeida}, {Audibert},
  {Acosta-Pulido}, {Cezar}, {Hicks}, {Lapi}, \& {Mullaney}}]{mavi25}
{Zanchettin}, M.~V., {Ramos Almeida}, C., {Audibert}, A., {et~al.} 2025, \aap,
  695, A185

\bibitem[{{Zubovas} \& {King}(2012)}]{zubovas12}
{Zubovas}, K. \& {King}, A. 2012, \apjl, 745, L34

\end{thebibliography}

\begin{appendix}

\section{Radio observations}\label{sec:radio}

We describe the main properties of the radio continuum observations used in this work. In Table~\ref{tab:radio} we list the telescope and observed frequency, the synthesized beam size, rms and project name. We note that in the case of VLA observations, we only report the values for the HR configuration from \citet{jarvis19}, which were the mostly used in this work.

\begin{table*}
\caption{Properties of the radio observations.}
\centering
\begin{tabular}{lcccccc}
\hline
\hline
\multirow{ 2}{*}{Name} 	  &   telescope & frequency  &    beam     &    rms  & project  & Reference   \\
    &         & (GHz)     &  (\arcsec $\times$ \arcsec) &   ($\mu$Jy/beam)     & ID & \\
\hline
J1010   & VLA         & 6   &  0.25$\times$0.22      &  79    &  13B-127 & (1) \\
J1100   & VLA         & 6   &  0.23$\times$0.22      &  48    & 13B-127 & (1) \\
J1100   & e-MERLIN    & 1.5 &  0.30$\times$0.18      &  300   & CY2217 & (1) \\
J1347   & VLBA        & 15  & 1.25mas$\times$0.6mas  &  48    & 2cm survey$^\dagger$& (2) \\
J1356   & VLA         & 6   &  0.37$\times$0.29      &  100   &  13B-127 & (1) \\
J1430   & VLA         & 6   &  0.23$\times$0.23      &  23    &  13B-127 & (1) \\
\hline	  
\end{tabular}	
\tablefoot{Columns indicate the main observational properties of the sample: name, telescope, frequency, beam size, rms, project ID, and reference for the works that describe the data reduction in detail. References: (1) \citet{jarvis19} (2): \citet{kellermann04}. 
$^\dagger$ \url{https://www.cv.nrao.edu/2cmsurvey/}. Projects ID 13B-127 and CY2217 PI: C. Harrison. }
\label{tab:radio}
\end{table*}

\section{\barolo~modelling}\label{sec:baloro}

\begin{figure}
    \centering
    \includegraphics[width=\linewidth]{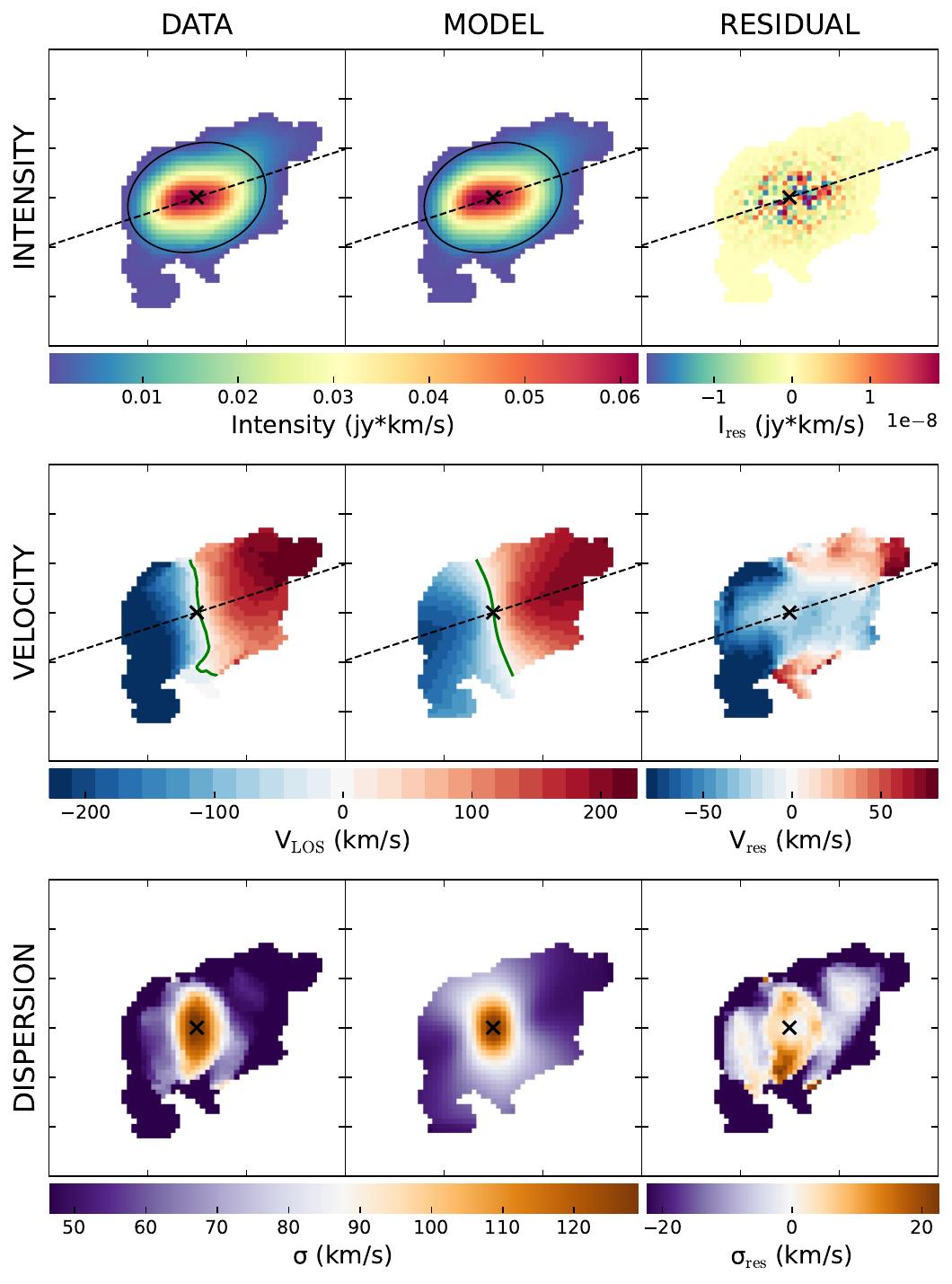}
    \caption{\barolo~kinematic maps of the CO(3-2) emission of J1010. The panels on the left show the data, the middle panels show the best \barolo~model, and the residuals (data-model) are shown on the right panels for the intensity (top row), velocity field (middle row), and velocity dispersion (bottom row). The cross represents the continuum peak at 200\,GHz, and the PA of the kinematic major axis of the galaxy is indicated with dashed lines.}
    \label{fig:bar1010}
\end{figure}

\begin{figure}
    \centering
    \includegraphics[width=\linewidth]{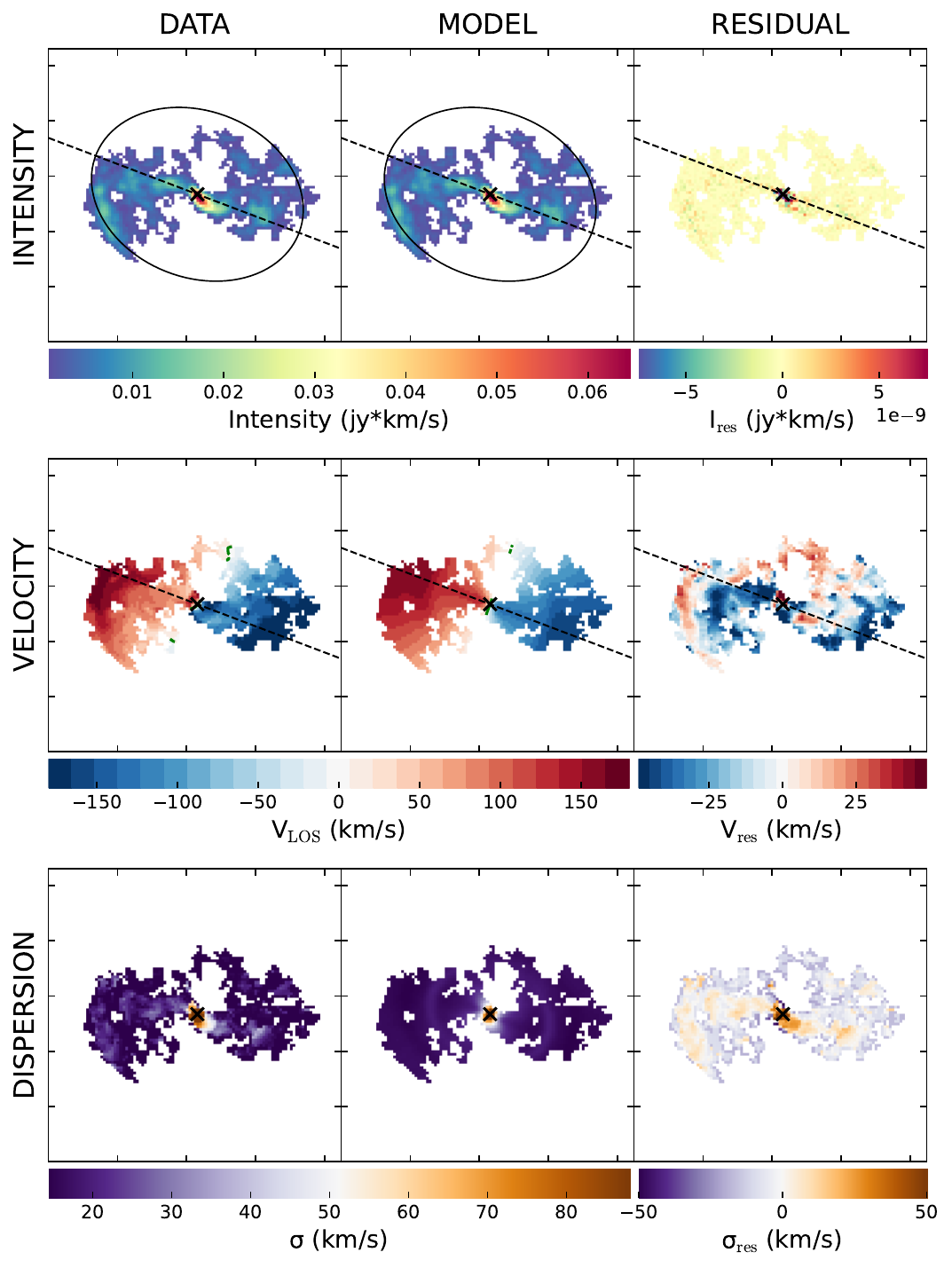}
    \caption{Same as Figure~\ref{fig:bar1010}, but for J1100.}
    \label{fig:bar1100}
\end{figure}

\begin{figure}
    \centering
    \includegraphics[width=1\linewidth]{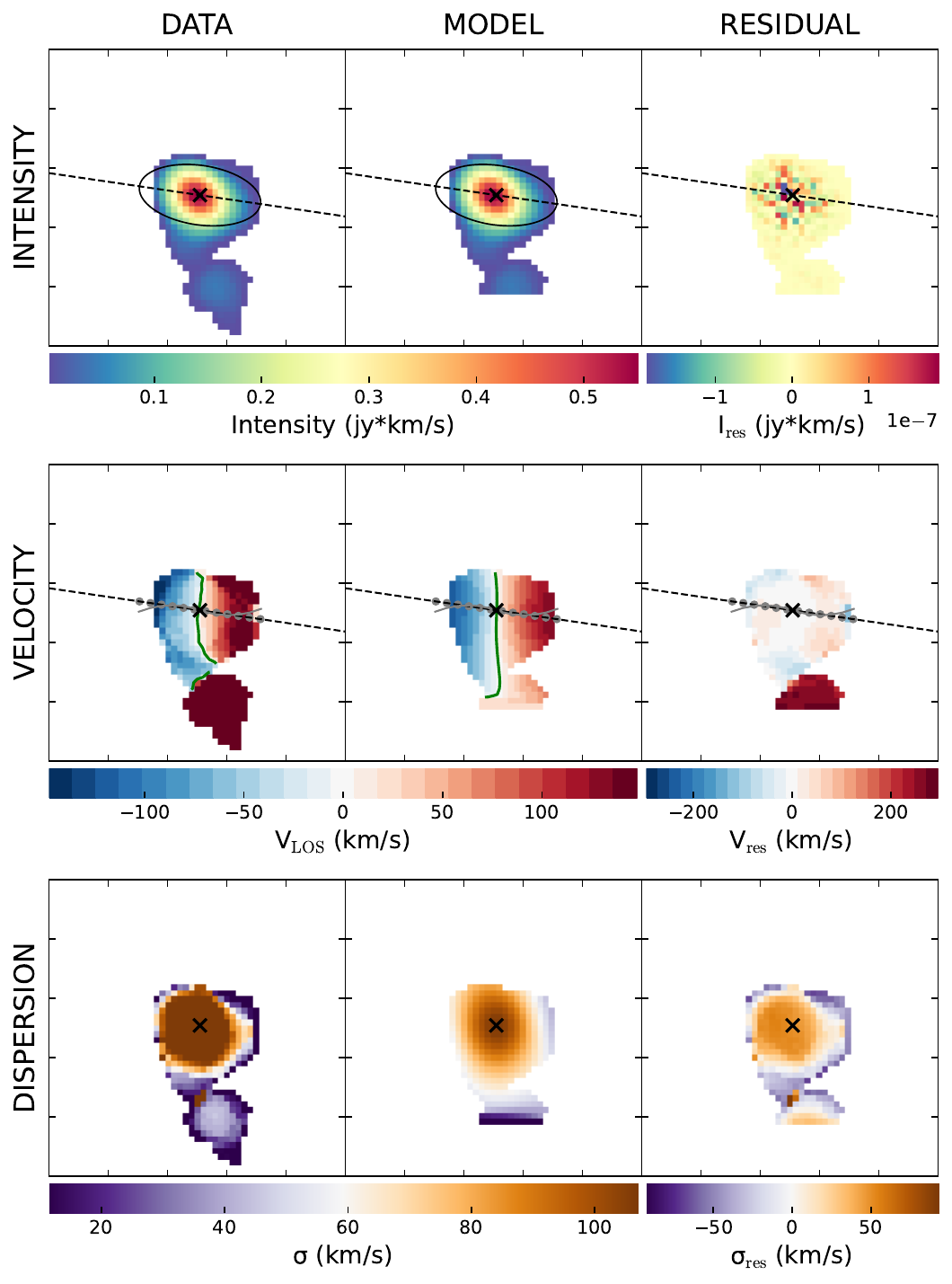}
    \caption{Same as Figure~\ref{fig:bar1010}, but for J1347.}
    \label{fig:bar1347}
\end{figure}

\begin{figure}
    \centering
    \includegraphics[width=\linewidth]{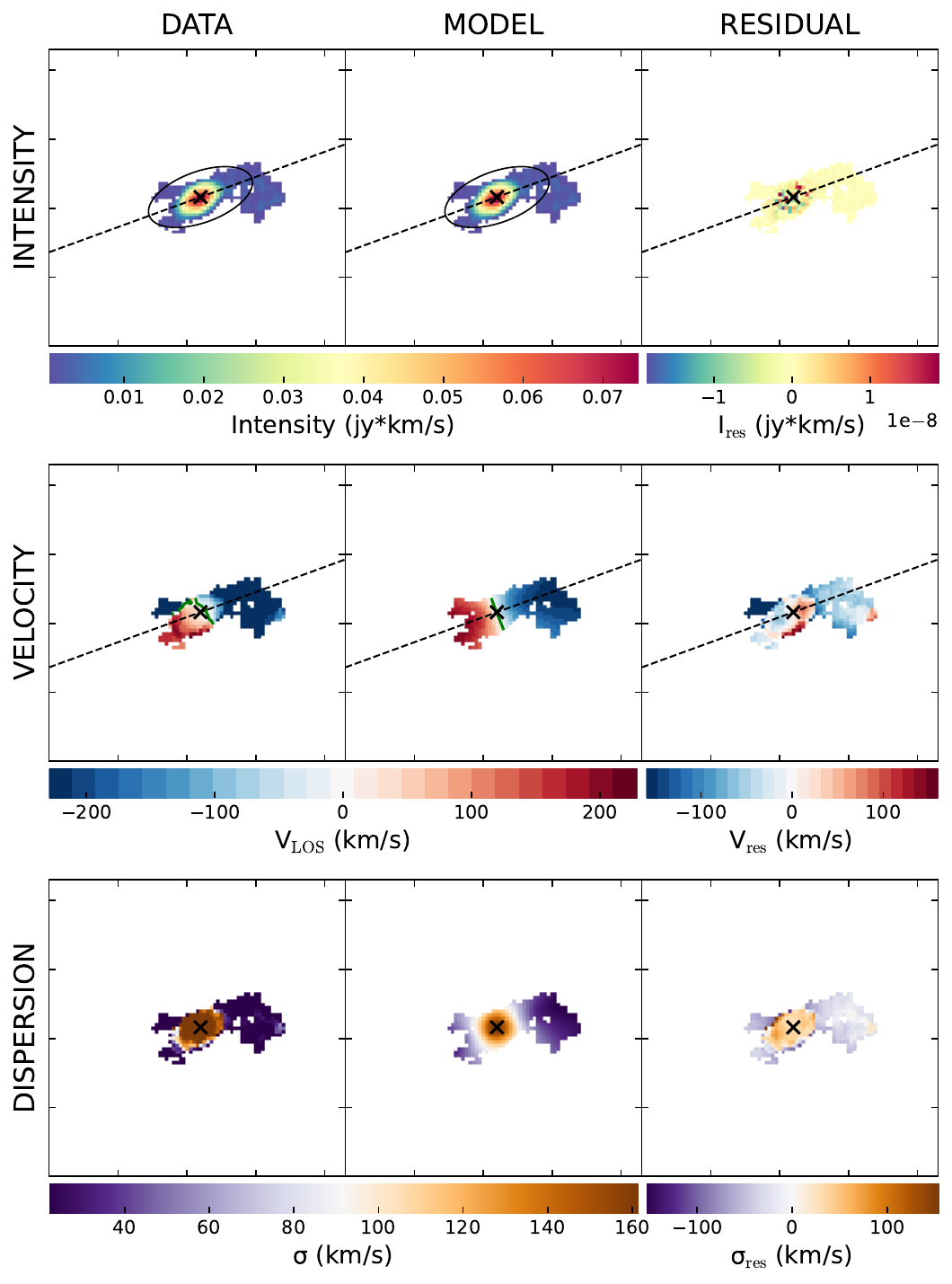}
    \caption{Same as Figure~\ref{fig:bar1010}, but for J1356.}
    \label{fig:bar1356}
\end{figure}

\begin{figure}
    \centering
    \includegraphics[width=1\linewidth]{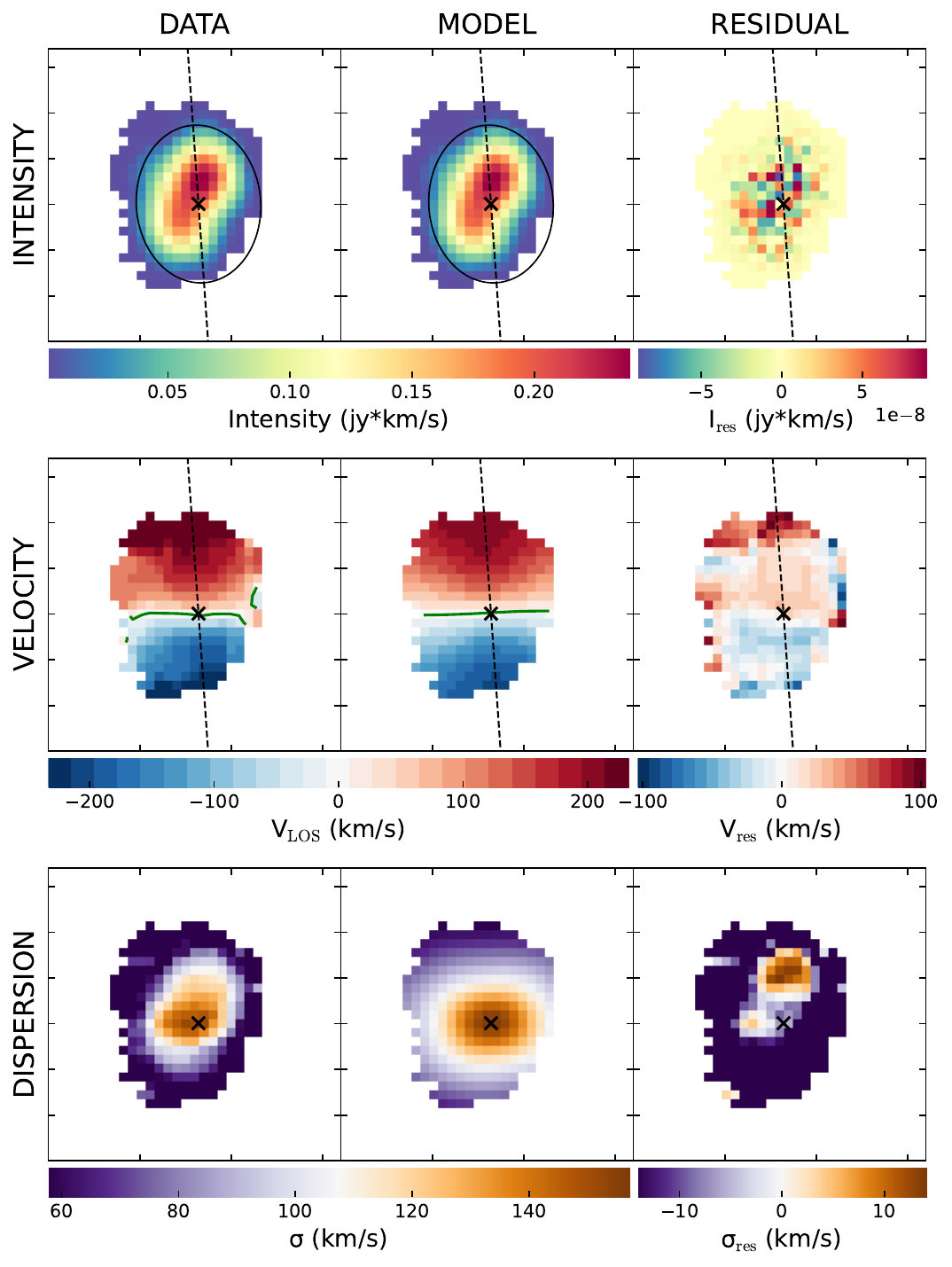}
    \caption{Same as Figure~\ref{fig:bar1010}, but for J1430.}
    \label{fig:bar1430}
\end{figure}

The results of the kinematic modelling using \barolo~are shown in Figures~\ref{fig:bar1010}-\ref{fig:bar1430} for the CO(3-2) emission. The main parameters derived from the best fit of the CO(3-2) data and of the CO(2-1) datacubes degraded to the resolution of CO(3-2) are shown in Table~\ref{tab:barolo}. All parameters are in agreement with the values obtained from the kinematic modelling presented in \cra.

\begin{table}[h]
\caption{\barolo~ best fits for the CO(2-1) and CO(3-2) data.}
\centering
\begin{tabular}{llccc}
\hline
\hline
\multirow{2}{*}{Name} 	 &    CO     & $i$       &  PA  &   $v_{\rm sys}$ \\
         & (J$\rightarrow$J-1) &  (\degree) & (\degree)    & (\kms) \\
\hline \noalign{\smallskip}
\multirow{2}{*}{J1010}    &   (2-1)    & 37$\pm$1 & 287$\pm$8 & -35  \\
                          &   (3-2)    & 36$\pm$1 & 288$\pm$4 & -35  \\
\noalign{\smallskip}
\multirow{2}{*}{J1100}    &   (2-1)    & 38$\pm$1 & 68$\pm$3 & 36  \\
                          &   (3-2)    & 38$\pm$2 & 67$\pm$5 & 36 \\
\noalign{\smallskip}
\multirow{2}{*}{J1347}    &   (2-1)    & 60$\pm$1 & 252$\pm$4 & 17   \\
                          &   (3-2)    & 61$\pm$2 & 262$\pm$7 & -50
 \\
\noalign{\smallskip}        
\multirow{2}{*}{J1356}    &   (2-1)    & 54$\pm$1 & 107$\pm$11 & 5   \\
                          &   (3-2)    & 62$\pm$0 & 110$\pm$10 & 5 \\
\noalign{\smallskip}
\multirow{2}{*}{J1430}    &   (2-1)    &  38$\pm$2 & 3$\pm$1 & 0   \\
                          &   (3-2)    &  41$\pm$2 & 4$\pm$3 & 0   \\
J1509 & (2-1) & 43$\pm$1 & 266$\pm$4 & 0 \\
\noalign{\smallskip}
\hline	   					 			    					 	
\end{tabular}	
\label{tab:barolo}
\end{table}

\section{Dispersion and rotation dominated systems}\label{sec:vsig}

\begin{figure*}
\centering
\includegraphics[width=17cm]{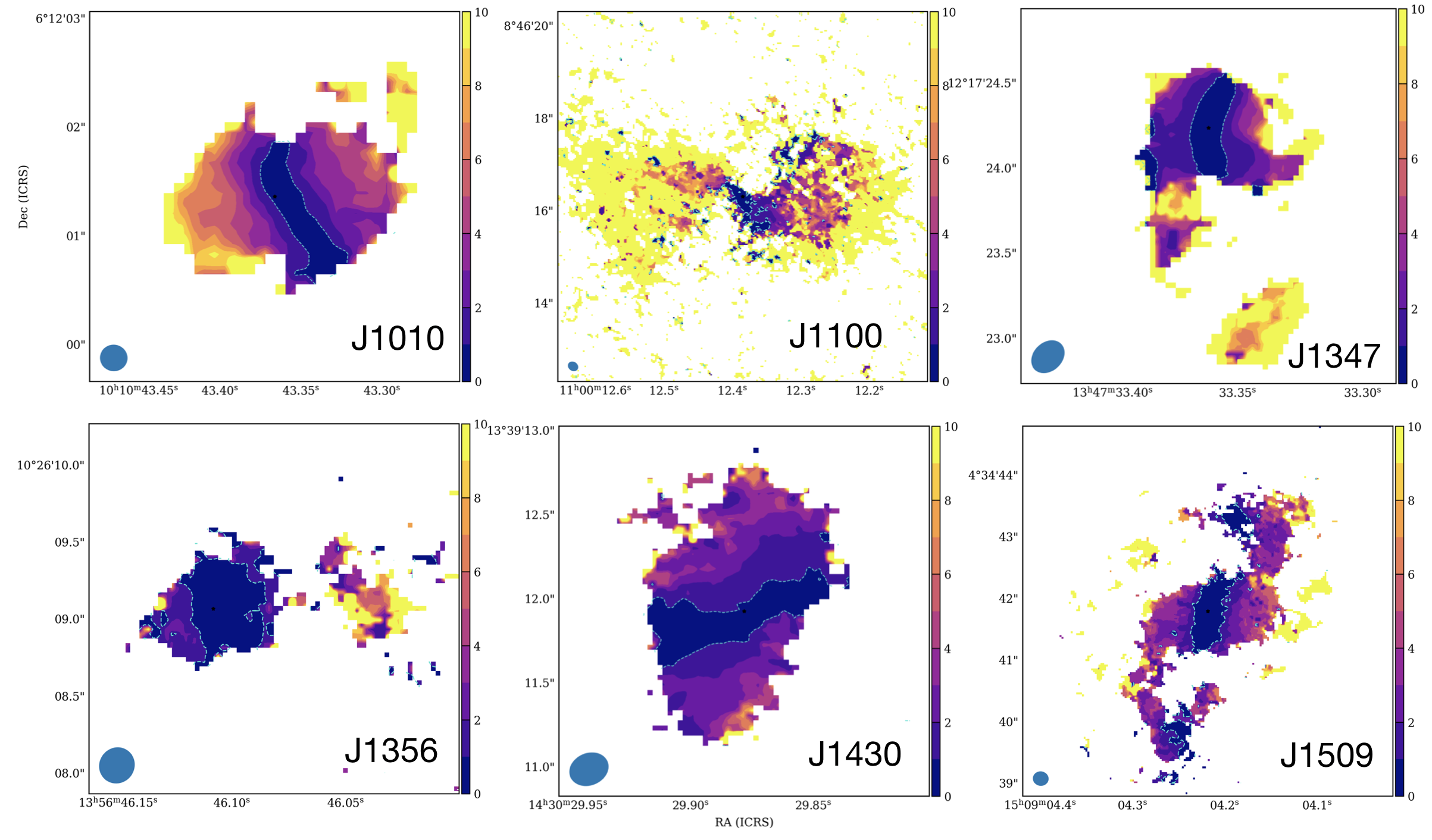}
\caption{CO(2-1) velocity over velocity dispersion ($v/\sigma$) maps of the QSO2s. In the case of J1010 (top left panel) CO(3-2) has been used instead. The color bars indicate $v/\sigma$ ranging from 0 to 10 and the dotted blue line shows the $v/\sigma$=1 contour. The beam sizes are indicated by the blue ellipses at the bottom-left corner of each panel.}
\label{fig:vsig}
\end{figure*}

We present the velocity over the velocity dispersion maps, $v/\sigma$, for the six QSO2s studied in this work in Figure~\ref{fig:vsig}. To create the  $v/\sigma$ maps we used the higher resolution CO(2-1) observations, except for J1010, for which we used the higher spatial resolution CO(3-2) data.

All but one systems appear to be rotation-dominated in most of the emitting regions according to the criterion $v/\sigma>$1, highlighted by the dotted blue lines showing the $v/\sigma$=1 contour. The regions showing 
$v/\sigma<$1 correspond to the line of zeros in the velocity field, as expected for rotation along the minor axis of the galaxies. The exception is J1356, where high velocity dispersion values ($v/\sigma<$1) dominate the main gas concentration (i.e. J1356N), while the western arc 
shows lower velocity dispersion, as shown by the moment maps in Figure  \ref{fig:co32moms}. Therefore, the determination of the rotation curve for this galaxy is not accurate and therefore we did not derive scenarios \textsc{i} and \textsc{iii} for the outflows, which rely on a correct determination of gas rotation.

\end{appendix}

\end{document}